\documentclass[10pt,aps,prstab,reprint,twocolumn,amsmath,superscriptaddress,preprintnumbers,floatfix,nofootinbib,showpacs]{revtex4-1}
\usepackage{siunitx}
\usepackage{graphicx}
\usepackage[dvipsnames]{xcolor}
\usepackage[bookmarks]{hyperref} 
\usepackage{tikz}
\usepackage{booktabs}
\usepackage{mathtools}
\usepackage{braket}
\usepackage[utf8]{inputenc}
\usepackage[T1]{fontenc}
\usepackage{natbib}
\usepackage{amssymb,latexsym}
\usepackage{wasysym}
\usepackage{soul}

\newcommand{\matr}[1]{\mathbf{#1}}
\newcommand{\dd}{\text{d}}
\DeclareMathOperator{\Tr}{Tr}
\begin{document}
\preprint{Draft for PR ST AB \today}
\title{Spin tune mapping as a novel tool to probe the spin dynamics in storage rings}
%
%
\author{A.~Saleev}
\affiliation{Institut f\"ur Kernphysik, Forschungszentrum J\"ulich, 52425 J\"ulich, Germany}
\affiliation{Samara National Research University, 443086 Samara, Russia}
\author{N.N. Nikolaev}
\affiliation{L.D. Landau Institute for Theoretical Physics, 142432 Chernogolovka, Russia}
\affiliation{Moscow Institute for Physics and Technology, 141700 Dolgoprudny, Russia}
\author{F.~Rathmann}
\affiliation{Institut f\"ur Kernphysik, Forschungszentrum J\"ulich, 52425 J\"ulich, Germany}
\author{W.~Augustyniak}
\affiliation{Department of Nuclear Physics, National Centre for Nuclear Research, 00681 Warsaw, Poland}
\author{Z.~Bagdasarian}
\affiliation{High Energy Physics Institute, Tbilisi State University, 0186 Tbilisi, Georgia}
\affiliation{Institut f\"ur Kernphysik, Forschungszentrum J\"ulich, 52425 J\"ulich, Germany}
\author{M.~Bai}
\affiliation{Institut f\"ur Kernphysik, Forschungszentrum J\"ulich, 52425 J\"ulich, Germany}
\affiliation{JARA--FAME (Forces and Matter Experiments), Forschungszentrum J\"ulich and RWTH Aachen University, Germany}
\author{L.~Barion}
\affiliation{University of Ferrara and INFN, 44100 Ferrara, Italy}
\author{M.~Berz}
\affiliation{Department of Physics and Astronomy, Michigan State University,  East Lansing, Michigan 48824, USA}
\author{S.~Chekmenev}
\affiliation{III. Physikalisches Institut B, RWTH Aachen University, 52056 Aachen, Germany}
\author{G.~Ciullo}
\affiliation{University of Ferrara and INFN, 44100 Ferrara, Italy}
\author{S.~Dymov}
\affiliation{Institut f\"ur Kernphysik, Forschungszentrum J\"ulich, 52425 J\"ulich, Germany}
\affiliation{Laboratory of Nuclear Problems, Joint Institute for Nuclear Research, 141980 Dubna, Russia}
\author{D. Eversmann}
\affiliation{III. Physikalisches Institut B, RWTH Aachen University, 52056 Aachen, Germany}
\author{M.~Gaisser}
\affiliation{Center for Axion and Precision Physics Research, Institute for Basic Science, 291 Daehak-ro, Yuseong-gu, Daejeon 305-701, Republic of Korea}
\affiliation{III. Physikalisches Institut B, RWTH Aachen University, 52056 Aachen, Germany}

\author{R.~Gebel}
\affiliation{Institut f\"ur Kernphysik, Forschungszentrum J\"ulich, 52425 J\"ulich, Germany}
\author{K.~Grigoryev}
\affiliation{III. Physikalisches Institut B, RWTH Aachen University, 52056 Aachen, Germany}
\author{D.~Grzonka}
\affiliation{Institut f\"ur Kernphysik, Forschungszentrum J\"ulich, 52425 J\"ulich, Germany}
\author{G.~Guidoboni}
\affiliation{University of Ferrara and INFN, 44100 Ferrara, Italy}
\author{D.~Heberling}
\affiliation{Institut f\"ur Hochfrequenztechnik, RWTH Aachen University, 52056 Aachen, Germany}
\affiliation{JARA--FAME (Forces and Matter Experiments), Forschungszentrum J\"ulich and RWTH Aachen University, Germany}
\author{V.~Hejny}
\affiliation{Institut f\"ur Kernphysik, Forschungszentrum J\"ulich, 52425 J\"ulich, Germany}
\author{N.~Hempelmann}
\affiliation{III. Physikalisches Institut B, RWTH Aachen University, 52056 Aachen, Germany}
\author{J.~Hetzel}
\affiliation{Institut f\"ur Kernphysik, Forschungszentrum J\"ulich, 52425 J\"ulich, Germany}
\author{F.~Hinder}
\affiliation{III. Physikalisches Institut B, RWTH Aachen University, 52056 Aachen, Germany}
\affiliation{Institut f\"ur Kernphysik, Forschungszentrum J\"ulich, 52425 J\"ulich, Germany}
\author{A.~Kacharava}
\affiliation{Institut f\"ur Kernphysik, Forschungszentrum J\"ulich, 52425 J\"ulich, Germany}
\author{V.~Kamerdzhiev}
\affiliation{Institut f\"ur Kernphysik, Forschungszentrum J\"ulich, 52425 J\"ulich, Germany}
\author{I.~Keshelashvili}
\affiliation{Institut f\"ur Kernphysik, Forschungszentrum J\"ulich, 52425 J\"ulich, Germany}
\author{I.~Koop}
\affiliation{Budker Institute of Nuclear Physics, 630090 Novosibirsk, Russia}
\author{A.~Kulikov}
\affiliation{Laboratory of Nuclear Problems, Joint Institute for Nuclear Research, 141980 Dubna, Russia}
\author{A.~Lehrach}
\affiliation{Institut f\"ur Kernphysik, Forschungszentrum J\"ulich, 52425 J\"ulich, Germany}
\affiliation{JARA--FAME (Forces and Matter Experiments), Forschungszentrum J\"ulich and RWTH Aachen University, Germany}
\author{P.~Lenisa}
\affiliation{University of Ferrara and INFN, 44100 Ferrara, Italy}
\author{N.~Lomidze}
\affiliation{High Energy Physics Institute, Tbilisi State University, 0186 Tbilisi, Georgia}
\author{B.~Lorentz}
\affiliation{Institut f\"ur Kernphysik, Forschungszentrum J\"ulich, 52425 J\"ulich, Germany}
\author{P.~Maanen}
\affiliation{III. Physikalisches Institut B, RWTH Aachen University, 52056 Aachen, Germany}
\author{G.~Macharashvili}
\affiliation{High Energy Physics Institute, Tbilisi State University, 0186 Tbilisi, Georgia}
\affiliation{Laboratory of Nuclear Problems, Joint Institute for Nuclear Research, 141980 Dubna, Russia}
\author{A.~Magiera}
\affiliation{Institute of Physics, Jagiellonian University, 30348 Cracow, Poland}
\author{D.~Mchedlishvili}
\affiliation{High Energy Physics Institute, Tbilisi State University, 0186 Tbilisi, Georgia}
\affiliation{Institut f\"ur Kernphysik, Forschungszentrum J\"ulich, 52425 J\"ulich, Germany}
\author{S.~Mey}
\affiliation{III. Physikalisches Institut B, RWTH Aachen University, 52056 Aachen, Germany}
\affiliation{Institut f\"ur Kernphysik, Forschungszentrum J\"ulich, 52425 J\"ulich, Germany}
\author{F.~M\"uller}
\affiliation{Institut f\"ur Kernphysik, Forschungszentrum J\"ulich, 52425 J\"ulich, Germany}
\author{A.~Nass}
\affiliation{Institut f\"ur Kernphysik, Forschungszentrum J\"ulich, 52425 J\"ulich, Germany}
\author{A.~Pesce}
\affiliation{University of Ferrara and INFN, 44100 Ferrara, Italy}
\author{D.~Prasuhn}
\affiliation{Institut f\"ur Kernphysik, Forschungszentrum J\"ulich, 52425 J\"ulich, Germany}
\author{J.~Pretz}
\affiliation{III. Physikalisches Institut B, RWTH Aachen University, 52056 Aachen, Germany}
\affiliation{JARA--FAME (Forces and Matter Experiments), Forschungszentrum J\"ulich and RWTH Aachen University, Germany}
\author{M.~Rosenthal}
\affiliation{III. Physikalisches Institut B, RWTH Aachen University, 52056 Aachen, Germany}
\affiliation{Institut f\"ur Kernphysik, Forschungszentrum J\"ulich, 52425 J\"ulich, Germany}
\author{V.~Schmidt}
\affiliation{III. Physikalisches Institut B, RWTH Aachen University, 52056 Aachen, Germany}
\affiliation{Institut f\"ur Kernphysik, Forschungszentrum J\"ulich, 52425 J\"ulich, Germany}
\author{Y.~Semertzidis}
\affiliation{Center for Axion and Precision Physics Research, Institute for Basic Science, 291 Daehak-ro, Yuseong-gu, Daejeon 305-701, Republic of Korea}
\author{Y.~Senichev}
\affiliation{Institut f\"ur Kernphysik, Forschungszentrum J\"ulich, 52425 J\"ulich, Germany}
\author{V.~Shmakova}
\affiliation{Laboratory of Nuclear Problems, Joint Institute for Nuclear Research, 141980 Dubna, Russia}
\author{A.~Silenko}
\affiliation{Research Institute for Nuclear Problems, Belarusian State University, 220030 Minsk, Belarus}
\affiliation{Bogoliubov Laboratory of Theoretical Physics, Joint Institute for Nuclear Research, 141980 Dubna, Russia}
\author{J.~Slim}
\affiliation{Institut f\"ur Hochfrequenztechnik, RWTH Aachen University, 52056 Aachen, Germany}
\author{H.~Soltner}
\affiliation{Zentralinstitut f\"ur Engineering, Elektronik und Analytik (ZEA-1), Forschungszentrum J\"ulich, 52425 J\"ulich, Germany}
\author{A.~Stahl}
\affiliation{III. Physikalisches Institut B, RWTH Aachen University, 52056 Aachen, Germany}%
\affiliation{JARA--FAME (Forces and Matter Experiments), Forschungszentrum J\"ulich and RWTH Aachen University, Germany}
\author{R.~Stassen}
\affiliation{Institut f\"ur Kernphysik, Forschungszentrum J\"ulich, 52425 J\"ulich, Germany}
\author{E.~Stephenson}
\affiliation{Indiana University Center for Spacetime Symmetries, Bloomington,  Indiana 47405, USA}
\author{H.~Stockhorst}
\affiliation{Institut f\"ur Kernphysik, Forschungszentrum J\"ulich, 52425 J\"ulich, Germany}
\author{H.~Str\"oher}
\affiliation{Institut f\"ur Kernphysik, Forschungszentrum J\"ulich, 52425 J\"ulich, Germany}
\affiliation{JARA--FAME (Forces and Matter Experiments), Forschungszentrum J\"ulich and RWTH Aachen University, Germany}
\author{M.~Tabidze}
\affiliation{High Energy Physics Institute, Tbilisi State University, 0186 Tbilisi, Georgia}
\author{G.~Tagliente}
\affiliation{INFN, 70125 Bari, Italy}
\author{R.~Talman}
\affiliation{Cornell University, Ithaca,  New York 14850, USA}
\author{P.~Th\"orngren Engblom}
\affiliation{Department of Physics, KTH Royal Institute of Technology, SE-10691 Stockholm, Sweden}
%
\author{F.~Trinkel}
\affiliation{III. Physikalisches Institut B, RWTH Aachen University, 52056 Aachen, Germany}
\affiliation{Institut f\"ur Kernphysik, Forschungszentrum J\"ulich, 52425 J\"ulich, Germany}
\author{Yu.~Uzikov}
\affiliation{Laboratory of Nuclear Problems, Joint Institute for Nuclear Research, 141980 Dubna, Russia}
\author{Yu.~Valdau}
\affiliation{Helmholtz-Institut f\"ur Strahlen- und Kernphysik, Universit\"at Bonn, 53115 Bonn, Germany}
\affiliation{Petersburg Nuclear Physics Institute, 188300 Gatchina, Russia}
\author{E.~Valetov}
\affiliation{Department of Physics and Astronomy, Michigan State University,  East Lansing, Michigan 48824, USA}
\author{A.~Vassiliev}
\affiliation{Petersburg Nuclear Physics Institute, 188300 Gatchina, Russia}
\author{C.~Weidemann}
\affiliation{University of Ferrara and INFN, 44100 Ferrara, Italy}
\author{A.~Wro\'{n}ska}
\affiliation{Institute of Physics, Jagiellonian University, 30348 Cracow, Poland}
\author{P.~W\"ustner}
\affiliation{Zentralinstitut f\"ur Engineering, Elektronik und Analytik (ZEA-2), Forschungszentrum J\"ulich, 52425 J\"ulich, Germany}
\author{P.~Zupra\'nski}
\affiliation{Department of Nuclear Physics, National Centre for Nuclear Research, 00681 Warsaw, Poland}
\author{M.~Zurek}
\affiliation{Institut f\"ur Kernphysik, Forschungszentrum J\"ulich, 52425 J\"ulich, Germany}
\collaboration{JEDI collaboration}

\begin{abstract}
Precision experiments, such as the search for electric dipole moments of charged particles using storage rings, demand for an understanding of the spin dynamics with unprecedented accuracy. The ultimate aim is to measure the electric dipole moments with a sensitivity up to 15 orders in magnitude better than the magnetic dipole moment of the stored particles. This formidable task requires an understanding of the background to the signal of the electric dipole  from rotations of the spins in the spurious magnetic fields of a storage ring. One of the observables, especially sensitive to the imperfection magnetic fields in the ring is the angular orientation of stable spin axis. Up to now, the stable spin axis has never been determined experimentally, and in addition, the JEDI collaboration for the first time succeeded to quantify the background signals that stem from false rotations of the magnetic dipole moments in the horizontal and longitudinal imperfection magnetic fields of the storage ring. To this end, we developed a new method based on the spin tune response of a machine to artificially applied longitudinal magnetic fields. This novel technique, called \textit{spin tune mapping}, emerges as a very powerful tool to probe the spin dynamics in storage rings. The technique was experimentally tested in 2014 at the cooler synchrotron COSY, and for the first time, the angular orientation of the stable spin axis at two different locations in the ring has been determined to an unprecedented accuracy of better than $\SI{2.8}{\micro.rad}$. 
\end{abstract}
\pacs{13.40.Em, 11.30.Er, 29.20.Dh, 29.27.Hj}
\maketitle
\tableofcontents
\section{Introduction}
Our very existence hinges on the net baryonic content of the Universe. In the Big Bang  paradigm, the baryon asymmetry of the Universe is generated  during the off-equilibrium expansion of the Universe due to baryon number and $CP$ non-conserving processes\,\cite{Sakharov:1967dj}. The Standard Model (SM) possesses a topological baryon number violation\,\cite{Kuzmin:1985mm}  and the $CP$-violation can be successfully parameterized by the non-vanishing phase of the CKM mixing parameters\,\cite{Olive:2016xmw}.The baryonic abundance predicted by the SM, however, is some nine orders of magnitude smaller than the experimentally observed one\,\cite{Rubakov:1996vz,Riotto:1999yt,Engel201321}. That clearly calls for  $CP$-violating mechanisms beyond the CKM parameterization within the SM (for a discussion of alternative approaches to the matter-antimatter asymmetry, see\,\cite{Kusenko:2015nta} and references therein).

Electric dipole moments (EDMs) become only possible when parity $P$ and time-reversal
invariance $T$ (and $CP$ by virtue of the $CPT$ theorem) are broken. Hence the search for
EDMs of hadrons and leptons constitutes an important window toward new physics beyond the
SM. An EDM would precess in the electric field precisely as the magnetic dipole moment (MDM)
does in a magnetic field. The nuclear magneton $\mu_N = e\hbar/2m_N c \approx \SI{e-14}{e.cm}$ sets a natural scale for the MDM of nucleons and light nuclei. The EDM calls for
a $P$ violating weak interaction, the price for which to pay is a factor of $\sim
\num{e-7}$, and one pays extra a factor of $\sim \num{e-3}$ for $CP$ violation\
\cite{Khriplovich:1997ga}. Hence the natural scale for the EDM of nucleons is given by 
\begin{equation}
d_N \sim \num{e-3} \times  \num{e-7} \times  \mu_N \sim \SI{e-24}{e.cm}\,.
\label{eq:1.1}
\end{equation}
In the SM the $CP$ symmetry is violated due to flavor changing transitions. To generate a flavor-neutral EDM one has to change the flavor back invoking the weak interaction once again, which entails an exceedingly small lower bound on the nucleon EDM from the SM of
\begin{equation}
d_N^{\text{SM}} \sim \num{e-7} \times d_N \sim \SI{e-31}{e.cm}\,.
\label{eq:1.2}
\end{equation}
 
So far stringent upper bounds have been set on the EDM of neutral atoms, molecules and neutrons, which can readily be subjected to strong electric fields still being at rest. In these investigations one usually looks for a shift of the spin precession frequency caused by an electric $E$-field parallel or anti-parallel 
to the $B$-field (for a review see\,\cite{Lamoreaux:1900zz}). For the neutron ($n$) EDM, an upper bound of $d_n < \SI{2e-26}{e.cm}$ has been reached\,\cite{Afach:2015sja,Serebrov:2013tba,Serebrov:2015gga}. The ultimate sensitivity anticipated in the present neutron EDM experiments may reach $d_n \sim \SI{e-27}{e.cm}$.  

The parallel fields approach does not work for charged particles, such as protons ($p$), deuterons ($d$) and other nuclei though. Here the electric field must be part of what confines charged particles on a closed orbit in a storage ring.  On a purely statistical basis, the sensitivity to the proton and deuteron EDMs can be higher than that of the neutron. In addition, the existing bound on the neutron EDM does not preclude much larger proton, deuteron  and helion ($^3\text{He}$) EDMs (for a comprehensive discussion, see\,\cite{Bsaisou:2014zwa}). The principal point is that there are no model-independent sum rules relating EDMs for $n$, $p$, $d$ and $^3\text{He}$ -- they all probe different aspects of generic mechanisms of $CP$ violation.

The present study, carried out by the JEDI Collaboration (J\"ulich Electric Dipole moment Investigations) \cite{jedi-collaboration} in September 2014 at COSY, is motivated by ideas on the search for EDMs of protons and deuterons using a storage ring\,\cite{srEDM-collaboration,jedi-collaboration}. It is part of an extensive world-wide effort to push further the frontiers of precision spin dynamics of polarized particles in storage rings. We developed a new method to map out the spin tune response of a machine with respect to artificially introduced magnetic field imperfections. The theoretical background to this method and its experimental vindication are prerequisites to the planned precursor EDM experiments at COSY\,\cite{jedi-collaboration}, and will also have an impact on the design of future dedicated EDM storage rings.

The present investigation is part of the preparations for the search for the deuteron EDM at COSY, using a radio-frequency (RF) Wien filter (WF)\,\cite{Slim:2016pim}. The idea is to look for an EDM-driven resonant rotation of the stored deuteron spins from the horizontal to vertical direction and vice versa,  generated by the RF Wien filter at the deuteron spin precession frequency. The RF Wien filter \textit{per se} is transparent to the EDM of the particle, its net effect is a frequency modulation of the spin tune. This modulation couples to the EDM precession in the static motional $E$-field of the ring, and generates an EDM-driven up-down oscillation of the polarization\,\cite{PhysRevSTAB.16.114001}. 

On the other hand, the EDM interaction with the horizontal motional electric field tilts the vertical stable spin axis inwards or outwards the ring. This tilt constitutes another static EDM observable, dual to the EDM-driven resonant spin rotation. Any offset and misalignment of magnetic elements in the ring produces horizontal and/or longitudinal imperfection magnetic fields as well. A rotation of the MDM in these magnetic imperfections is indistinguishable from that of the EDM in the horizontal motional electric field. In practice, those imperfection magnetic fields cannot readily be compensated for element by element and thus emerge as a principal background to the search for the EDM using an RF Wien filter.  

Recently, the JEDI collaboration has developed a method to measure the spin tune of deuterons to a relative precision of nine decimal places  in 100 s cycles \,\cite{Eversmann:2015jnk}. This very high precision can be applied to provide a diagnostics tool to quantify the magnetic ring properties. Specifically, the imperfections are known to affect the spin tune\,\cite{lee1997spin,Mane:2005xh}. The new technique is based on the introduction of artificial imperfections in the ring and to study the spin tune as a function of the spin kick in these artificial imperfections. Such a mapping of the spin tune response enables one to determine the orientation of the stable spin axis at the location of the artificial imperfections, and we report here about the first ever direct measurement of the stable spin axis in a storage ring. Preliminary results are reported in\,\cite{Saleev:2015ghs}. In the present experiment the two electron cooler solenoids, placed in the opposite straight sections of COSY, have been utilized  as makeshift artificial imperfections. Remarkably, such a two-solenoid setup with pure longitudinal magnetic fields allows one to deduce both longitudinal and horizontal components of the stable spin axis at two positions in the ring. 

The further presentation is organized as follows. In Sec.\,\ref{sec:II}, we present a brief theoretical introduction to the experimental investigations. The principal results of the exploratory study using COSY are reported in Sec.\,\ref{sec:III}. They do fully confirm the principal theoretical expectations on the impact of magnetic imperfections on the spin tune. We have shown for the first time, that the angular orientation of the stable spin axis can be controlled to an accuracy of about $\SI{2.8}{\micro.rad}$.  The experimental data exhibit certain systematic effects that have been uncovered in the course of the data analysis, those stemming from beam-orbit distortions by the misaligned solenoids are discussed in Sec.\,\ref{sec:IV}. In the analysis of systematic effects, we invoked simulations based on the orbit- and spin-tracking code COSY-Infinity\,\cite{COSY-Infinity}. The interpretation of the experimental findings and possible applications of the spin-tune mapping technique are reviewed in Sec.\,\ref{sec:V}. A summary is given in Sec.\,\ref{sec:VI}, where we emphasize our points on the utility of the spin tune as a probe to characterize the MDM background in searches for the EDMs of charged particles. The Appendices (\ref{sec:appendixB}, \ref{sec:appendixA}, \ref{sec:appendixF}, \ref{sec:appendixC}, \ref{sec:appendixG}, and \ref{sec:appendixE}) are reserved for technical aspects on the statistical and the systematic accuracy of the spin tune determination and on the theoretical background behind the spin tune mapping. 

\section{Background of the EDM signal from magnetic imperfection fields}
\label{sec:II}
\subsection{Spin dynamics with EDM} 

The spin dynamics in a storage ring is governed by the Frenkel-Thomas-Barmann-Michel-Telegdi (FT-BMT) 
equation\,\cite{Frenkel:1926zz, Thomas:1926dy,Thomas:1927yu, PhysRevLett.2.435}\, extended to include the EDM effects\,\cite{PhysRevLett.2.492, Fukuyama:2013ioa}. We start with an ideal storage ring with static vertical magnetic field $\vec{B} = B\vec{e}_y$,  and horizontal electric field $\vec{E} = E \vec{e}_x$, so that $(\vec{\beta}\cdot \vec{E}) = (\vec{\beta}\cdot \vec{B})=0$, where $\vec{\beta} = \beta \vec e_z$ is the particle velocity in units of the velocity $c$ of light  [($\vec e_x,\vec e_y,\vec e_z$) form a right-handed coordinate system]. We use the system of units  $\hbar=c=1$. Let the stored particle of mass $m$ and  of electric change $q$ have a non-vanishing EDM, 
\begin{equation}
d = \eta \frac{q}{2m}.
 \end{equation}
Here $\eta$ plays for the EDM the same role as the $g$-factor does for the MDM, $\mu = g q/{2m}$. With allowance for an EDM, the FT-BMT equation for the spin precession takes the form\,\cite{PhysRevLett.2.492, Fukuyama:2013ioa}
\begin{equation}
\frac{d\vec{S} }{ dt } = \vec{\Omega}_s \times \vec{S}\, , 
\label{eq:spin-precession}
\end{equation}
where the spin precession angular velocity is given by
\begin{equation}
\begin{split}
\vec{\Omega}_s = - &\frac{q}{m} \left[
\underbrace{
G{\vec{B}} +\left(\frac{1}{\beta^2} -1 -G\right) \vec{\beta}\times\vec{E} }_{\text{MDM}}\right.\\
 & \quad\quad + \left. \underbrace{\frac{\eta}{2} (\vec{E}+\vec{\beta}\times\vec{B})}_{\text{EDM}} \right] \, .
\end{split}
\label{eq:Omega_s}
\end{equation}
Here $G = (g-2)/2$ describes  the magnetic anomaly. The EDM part in $\vec{\Omega}_s$ is proportional to the Lorentz force,
\begin{equation}
\frac{d\vec{p}}{dt} = q \left(\vec{E} + \vec{\beta} \times \vec{B}\right)\, , \label{eq:Lorentz-force}
\end{equation}
while the MDM part receives a contribution from the motional magnetic field $\propto \vec{\beta}\times\vec{E}$.

In the standard spinor formalism\,\cite{lee1997spin,Mane:2005xh}, the spin transfer matrix per turn in a ring $\text{R}$ equals 
\begin{equation}
\begin{split}
\matr{t}_{\text{R}} =  \exp \left(-i\pi\nu_s \vec{\sigma}\cdot \vec{c}\right)
         = &  \cos \left(\pi\nu_s \right) \\
           & -i \left(\vec{\sigma} \cdot \vec{c} \right) \sin(\pi\nu_s) \, ,
         \label{eq:2A.3}
\end{split}
\end{equation}
where $\vec{\sigma}$ stands for the Pauli matrices and $\vec{c}$ is a unit vector pointing along the local spin precession axis. The angular velocity of the spin precession equals
\begin{equation}
\vec{\Omega}_s = 2\pi f_s \vec c = 2\pi f_{\text{R}} \nu_s \vec{c}\,, 
\label{eq:Omega_s-2} 
\end{equation}
where $f_{\text{R}}$ is the revolution frequency of the particles in the ring, and $\nu_s$ the spin tune, \textit{i.e.}, the number of spin revolutions per turn. The EDM produces two important effects. Firstly, it tilts the stable spin axis (also called spin-closed orbit) away from the vertical direction in a plane perpendicular to the particle velocity, described by
\begin{equation}
\vec{c} = \vec{e}_x\sin\xi_{\text{EDM}} + \vec{e}_y\cos\xi_{\text{EDM}}\,, \label{eq:tilt-of-spin-axis}
\end{equation}
where 
\begin{eqnarray}
\tan\xi_{\text{EDM}}=\frac{\eta\beta}{G} \, . \label{eq:2A.4}
\end{eqnarray}
 Secondly, besides this tilt, the EDM interaction also modifies the spin tune from the canonical $\nu_s = G\gamma$ to
\begin{equation}
\nu_s^0=\frac{G\gamma}{\cos\xi_{\text{EDM}}} \, . \label{eq:2A.5}
\end{equation}

\subsection{Radio-frequency and static approaches to EDM measurements in ideal storage rings}
\subsubsection{Radio-frequency driven EDM signal}
The early discussion of signals of the EDM focused on the EDM-driven resonance rotation of the spin from the horizontal to the vertical direction or vice versa by employing an RF Wien filter with a horizontal $\vec{E}$-field ($\vec E_{\text{WF}}(t) = \vec e_x E_{\text{WF}} \cos(2\pi f_ {\text {WF}}t +\Delta_{\text {WF}}$) and a vertical $\vec B$-field ($\vec B_{\text{WF}}(t) = \vec e_y B_{\text{WF}}\cos(2\pi f_ {\text {WF}}t +\Delta_{\text {WF}}$). According to the FT-BMT equation, such a Wien filter with vanishing Lorentz force, 
\begin{equation}
\vec F_{\text{L}}(t) = \vec{E}_{\text{WF}} (t) + \vec{\beta} \times \vec{B}_{\text{WF}} (t) = 0\,,
\label{eq:vanishing-lorentz-force}
\end{equation}
exerted on the beam, is entirely EDM-transparent. 

Nevertheless, the MDM interaction with the vertical RF magnetic field [see the MDM component of $\vec \Omega_s$ in Eq.\,(\ref{eq:Omega_s})], yields the precession around the $y$-axis with the angular velocity
\begin{equation}
 \vec{\Omega}_{\text{WF}}(t) = -\frac{q}{m}\cdot \frac{1+G}{\gamma^2} \vec B_{\text{WF}}(t)\, .
\label{eq:WFprecession}
\end{equation}
The resulting spin kick in the WF causes an RF modulation of the spin tune. As Morse, Orlov and  Semertzidis showed\,\cite{PhysRevSTAB.16.114001}, when the RF WF frequency is locked to the spin precession frequency ($f_{\text{WF}} = f_s$), the RF modulation of the spin tune couples to the EDM interaction with the static motional $\vec{E}$-field $\propto \vec{\beta}\times \vec{B}$ and generates an up-down rotation of the particle spins.

The strength of such an EDM-driven spin resonance is given by (see the detailed discussion in Appendix\,\ref{sec:appendixB})
\begin{equation}
\epsilon = \frac{1}{2} \chi_{\text{WF}} \left| \vec{c}\times \vec{w} \right|  \, .
\label{eq:2B.1}
\end{equation}
Hereafter, $\vec{c}$ denotes the stable spin axis of the ring [$\vec{c}$ is a static quantity, defined at the location of the RF WF, \textit{before} the RF was activated, see also Eq.\,(\ref{eq:2A.3})], $\chi_{\text{WF}}$ the spin kick in the WF, and $\vec{w}$ the magnetic field axis of the WF. 

For an ideal WF, $\vec{w} =\vec{e}_y$ and $\left| \vec{c}\times \vec{w} \right|=\sin\xi_{\text{EDM}}$. The EDM resonance strength 
\begin{equation}
\epsilon  =\frac{1}{2} \chi_{\text{WF}}\sin\xi_{\text{EDM}} 
\label{eq:epsilon}
\end{equation}
manifestly vanishes if $\xi_{\text{EDM}} \propto d =0$. A full derivation of the on-resonance case [Eq.\,(\ref{eq:2B.1})] is given in Appendix\,\ref{sec:appendixB-1}, the off-resonance case is treated in Appendix\,\ref{sec:appendixB-2}.

\subsubsection{Orientation of the stable spin axis as a static EDM signal}
The second option, elaborated in more detail in the subsequent Sec.\,\ref{sec:II.C}, is to measure directly the angular orientation of the stable spin axis [see Eq.\,(\ref{eq:2A.4})]. If it were possible, measuring this static quantity may prove more advantageous than measuring the resonance strength $\epsilon$, which is suppressed  by the small factor $\chi_{\text{WF}} \ll 1$ [see Eq.\,(\ref{eq:epsilon})]. The issue is false EDM signals, which are of major concern throughout the present study.

\begin{figure}[tb]
\includegraphics[width=\columnwidth]{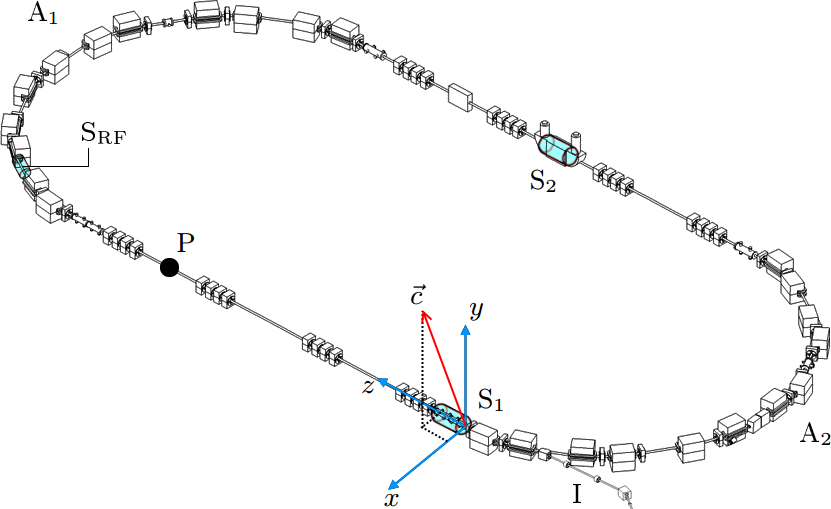}
\caption{\label{fig:ring-sketch} Sketch of the experimental setup with two solenoids $\text{S}_1$ and $\text{S}_2$ located in the opposite straight sections of the COSY ring. The vector $\vec c$  indicates the spin closed orbit before solenoid $\text{S}_1$, when $\text{S}_1$ and $\text{S}_2$ are switched off. The two arcs are denoted by $\text{A}_1$ and $\text{A}_2$, $\text{P}$ shows the location of the polarimeter, $\text{S}_{\text{RF}}$ the location of the RF solenoid,  and $\text{I}$ indicates the injection. The beam orbits in clockwise direction in the machine.}
\end{figure}
\subsection{Imperfections and spin tune mapping approach to the determination of the stable spin axis} 
\label{sec:II.C}

Realistic all-magnetic storage rings are laden with in-plane imperfection magnetic fields, induced by misalignments, rolls and offsets of magnetic elements. The interaction of the MDM with such imperfection fields therefore also contributes to the tilt of the stable spin axis, which, to first order, is given by
\begin{equation}
\vec{c}  = c_y \vec{e}_y + \Big[c_x(\text{MDM}) + \sin\xi_{\text{EDM}}\Big] \vec{e}_x + c_z(\text{MDM}) \vec{e}_z\,,
\label{eq:2B.2}
\end{equation}

Thus, imperfection magnetic fields provide \textit{the} major background to the EDM signal. This point about the false EDM signal from imperfections has already been raised in the discussion of the bound on the muon EDM from the BNL muon $g-2$ experiment \cite{Bennett:2008dy}.  Understanding the imperfection content of a storage ring is therefore among the top priorities for an EDM measurement using a magnetic machine, and this was precisely the principal task of the JEDI experiment at COSY.

\subsubsection{Spin tune mapping in a  ring with a single artificial imperfection}
An extremely precise observable at our disposal is the spin tune\,\cite{Eversmann:2015jnk}, which is prone to the imperfection magnetic fields. In order to apply the precise measurement of the spin tune as a tool to probe the imperfection magnetic fields, two  artificial solenoidal magnetic imperfections, $\text{S}_1$ and $\text{S}_2$, were activated  in the ring (see Fig.\,\ref{fig:ring-sketch}).

In the following, the idea of spin tune mapping using a single, artificially introduced imperfection is exposed. In anticipation of the modification of the spin tune by artificial imperfections, we define the spin tune for a ring without artificial imperfections by
\begin{equation}
\begin{split}
\matr{t}_{\text{R}} =   \exp\left(-i\pi\nu_s^0 \vec{\sigma}\cdot \vec{c}\right)
         =  & \cos \left(\pi\nu_s^0 \right) \\ 
            & - i(\vec{\sigma} \cdot \vec{c}) \sin(\pi\nu_s^0) \, ,
        \label{eq:spin-tune-without-solenoid}
\end{split}
\end{equation}
where $\nu_s^0$ denotes the unperturbed spin tune.

In such a situation, the spin transfer matrix of the artificial imperfection (AI) is given by
\begin{equation}
\matr{t}_{\text{AI}}= \cos\left(\frac{1}{2}\chi_{\text{AI}}\right) -i \left(\vec \sigma
 \cdot \vec k \right)\sin\left(\frac{1}{2}\chi_{\text{AI}}\right)  \, ,\label{eq:2C.1}
\end{equation}
where $\chi_{\text{AI}}$ denotes the spin rotation angle of the imperfection, and $\vec k $ its spin rotation axis. The total spin transfer matrix of the ring in the presence of the AI is given by the product 
\begin{equation}
\begin{split}
\matr{T} = &\matr{t}_{\text{R}} \matr{t}_{\text{AI}}\\
=  & \cos \left[\pi\nu_s(\chi_{\text{AI}}) \right] 
             - i\left[\vec{\sigma} \cdot \vec{c}(\chi_{\text{AI}})\right] \sin\left[\pi\nu_s(\chi_{\text{AI}})\right] \, ,
\label{eq:T}
\end{split}
\end{equation} 
where by definition $\nu_s(\chi_{\text{AI}}=0)= \nu_s^0$ and $\vec{c}(\chi_{\text{AI}}=0)=\vec{c}$, thus 
\begin{equation} 
\begin{split}
 \cos \left[ \pi \nu_s(\chi_{\text{AI}}) \right] = 
 & \cos \left( \pi \left[\nu_s^0 +\Delta\nu_s(\chi_{\text{AI}})\right]\right)  \\
 = \frac{1}{2} \Tr \matr{T}  
= &\cos \left( \pi \nu_s^0\right) \cos \left( \frac{1}{2}\chi_{\text{AI}} \right) \\  & -\sin \left( \pi\nu_s^0\right) \sin\left( \frac{1}{2}\chi_{\text{AI}}\right) \left(\vec{c} \cdot \vec{k}\right) \, . 
\end{split}
\label{eq:2C.2}
\end{equation}
Here, $\Delta \nu_s(\chi_{\text{AI}})$ denotes the change in spin tune from the unperturbed value $\nu_s^0$ when the artificial imperfection is activated.

For the sake of illustration of the idea of spin tune mapping, take the perturbative expansion
\begin{widetext}
\begin{equation}
\begin{split}
 \cos \left( \pi\nu_s^0\right) - \cos \left( \pi \left[\nu_s^0 + \Delta\nu_s(\chi_{\text{AI}})\right]\right)  = & \, \cos \left(\pi \nu_s^0 \right)  \left[1 - \cos \left( \frac{1}{2} \chi_{\text{AI}}\right)\right] + (\vec c \cdot \vec k) \sin \left( \pi \nu_s^0 \right) \sin \left(\frac{1}{2} \chi_{\text{AI}} \right)\\
 \simeq  & \, \frac{1}{8}\cos \left( \pi \nu_s^0\right) \left\{ \left(\chi_{\text{AI}}  + 2 \left(\vec{c} \cdot \vec{k}\right) \tan \left( \pi\nu_s^0 \right) \right)^2   
- 4 (\vec{c} \cdot \vec{k})^2 \tan^2 \left( \pi\nu_s^0 \right) \right\} \\
 \simeq & \, \pi \sin \left( \pi \nu_s^0 \right) \Delta \nu_s(\chi_{\text{AI}})\, , 
\end{split}
\label{eq:2C.3}
\end{equation} 
\end{widetext}
which is a quadratic function of $\chi_{\text{AI}}$. In case the spin rotation axis of the artificial imperfection is in the ring plane, then $(\vec{c} \cdot \vec{k}) = c_x k_x + c_z k_z$. Mapping the spin tune as function of  $\chi_{\text{AI}}$ and the orientation of $\vec{k}$  would enable one to determine both projections of $c_x$ and $c_z$ of $\vec c$ (see Fig.\,\ref{fig:ring-sketch}).

However, the determination of $c_x$ calls for an AI with a horizontal magnetic field which would cause unwanted vertical collective beam orbit excursions. In the above idealized example, the induced orbit excursions have been ignored. In principle, distortion-free AIs using a static Wien filter could be realized, but may require inaccessibly large electric fields to ensure vanishing Lorentz forces (estimates are given in Appendix\,\ref{sec:appendixG}). Fortunately enough, there exists a simple solution with pure longitudinal magnetic fields, which is free of orbit distortions.

\subsubsection{Spin tune mapping in a ring with two solenoids}
\label{sec:II.C.2}
Specifically, using two solenoids $\text{S}_1$ and $\text{S}_2$ as AIs in the ring (as shown in Fig.\,\ref{fig:ring-sketch}), well apart in opposite straight sections, constitutes the simplest option. Let $\matr{t}_{\text{A}_{1,2}}$ and $\matr{t}_{\text{S}_{1,2}}$ be the spin transfer matrices of the two arcs and of the two solenoids. The spin transfer matrix $\matr{T}$ for the full ring then reads
\begin{equation}
\matr{T} = \matr{t}_{\text{A}_2} \matr{t}_{\text{S}_2}   \matr{t}_{\text{A}_1} \matr{t}_{\text{S}_1} =  \matr{t}_{\text{A}_2} \matr{t}_{\text{A}_1} \matr{t}_{\text{A}_1}^{-1} \matr{t}_{\text{S}_2}   \matr{t}_{\text{A}_1} \matr{t}_{\text{S}_1} \, . 
\label{eq:2C.4}
\end{equation} 
In the absence of imperfections, $\matr{t}_{\text{A}_1}^{-1} \matr{t}_{\text{S}_2}   \matr{t}_{\text{A}_1} \matr{t}_{\text{S}_1} = 1$, and $\matr{T}=\matr{t}_{\text{R}} = \matr{t}_{\text{A}_2} \matr{t}_{\text{A}_1}$. Spin-wise this amounts to the apparent transport of the second imperfection downstream of the first one, generating one combined local AI, given by
\begin{equation}
\matr{t}_{\text{AI}} = \matr{t}_{\text{A}_1}^{-1} \matr{t}_{\text{S}_2} \matr{t}_{\text{A}_1} \matr{t}_{\text{S}_1}\,. 
\label{eq:t_AI}
\end{equation}
Let the spin transfer matrices in the arcs  $\text{A}_j$ ($j=1,2$) be,
\begin{equation}
\matr{t}_{\text{A}_j} = \exp \left\{-\frac{i}{2}\theta_j (\vec{\sigma} \cdot \vec{m_j})\right\}\,,
\end{equation}
where $\theta_j \simeq \pi \nu_s^0$ is the spin rotation angle in arc $\text{A}_j$ around the direction of $\vec{m_j} \simeq \vec{e}_y$. The spin transfer matrices in the two solenoids $\text{S}_{j}$ are given by
\begin{equation}
\matr{t}_{\text{S}_j} = \exp \left\{-\frac{i}{2} \chi_j(\vec{\sigma} \cdot \vec{n}_j) \right\}
\end{equation}
with $\vec{n}_j \simeq \vec{e}_z$. 

Upon the above apparent transport of the imperfection, one finds 
\begin{equation}
\matr{t}_{\text{A}_1}^{-1} \matr{t}_{\text{S}_2}   \matr{t}_{\text{A}_1} = \exp\left\{-\frac{i}{2} \chi_2(\vec{\sigma} \cdot {\vec{n}_2}^{\,\text{r}})\right\} \,, 
\end{equation}
where the spin rotation axis is transformed from $\vec n_2$ to 
\begin{equation}
 \begin{split}
{\vec{n}_2}^{\,\text{r}} & =  \cos\theta_1 \vec{n}_2 +\sin \theta_1 \left[\vec{n}_2 \times \vec{m}_1\right] \\
& \phantom{=} + (1 - \cos \theta_1)\left(\vec{m}_1 \cdot \vec{n}_2 \right)\vec{m}_1 \\
& \simeq  \cos \left( \pi\nu_s^0\right) \vec{e}_z -\sin \left( \pi \nu_s^0\right)  \vec{e}_x\, . 
\end{split}
\label{eq:2C.5}
\end{equation}
The last line of the above equation is an approximation that holds when $\vec m_1$ $(\simeq \vec e_y)$ and $n_2$ $(\simeq \vec e_z)$ are orthogonal, and when $\left[\vec{n}_2 \times \vec{m}_1\right] \simeq -\vec e_x$. Consequently, this apparent transport amounts to a rotation of the axis of solenoid $\text{S}_2$ by an angle $\theta_1\simeq \pi \nu_s^0$. This rotation is denoted by the upper index $\,\text{r}$ in Eq.\,(\ref{eq:2C.5}). We thus managed to generate a local artificial imperfection with an apparent horizontal component of the magnetic field \textit{without} excitation of transverse beam excursions.

Thus, the spin transfer matrix of the combined artificial imperfection is given by  
\begin{equation}
\begin{split}
\matr{t}_{\text{AI}}  =   & \cos \left( \frac{1}{2}\chi_1\right) \cos \left( \frac{1}{2}\chi_2 \right)  \\
                          & - \left({\vec{n}_2}^{\,\text{r}} \cdot \vec{n}_1 \right) \sin \left(\frac{1}  {2}\chi_1\right)  \sin \left( \frac{1}{2}\chi_2 \right)  \\
                          & -i\left( \vec{\sigma} \cdot \vec{k}_{\text{AI}} \right)\, ,
\end{split}
\label{eq:2C.6}
\end{equation}
where 
\begin{widetext}
\begin{equation}
\begin{split}
\vec{k}_{\text{AI}} =  \cos\left( \frac{1}{2}\chi_1 \right)  \sin \left(\frac{1}{2}\chi_2\right) {\vec{n}_2}^{\,\text{r}} 
                       + \cos \left( \frac{1}{2}\chi_2\right)  \sin \left( \frac{1}{2}\chi_1 \right) \vec{n}_1 
                       + \sin \left( \frac{1}{2}\chi_1 \right)  \sin \left( \frac{1}{2}\chi_2 \right) \left[{\vec{n}_2}^{\,\text{r}} \times \vec{n}_1 \right]\, . 
\end{split}
\label{eq:2C.7}
\end{equation} 
\end{widetext}

Finally, to an accuracy adequate for the purposes of the present investigation (see Appendix\,\ref{sec:appendixF}), 
\begin{widetext}
\begin{equation}
\begin{split}
 \cos\left(\pi \nu_s^0\right) - \cos\left( \pi \left[\nu_s^0 + \Delta\nu_s(\chi_1,\chi_2)\right]\right) 
 = & \left[ 1 + \cos\left(\pi \nu_s^0\right) \right] \sin^2 \left(\frac{1}{2}\chi_+ \right) - \left[ 1-\cos \left(\pi \nu_s^0\right)\right] \sin^2\left(\frac{1}{2}\chi_-\right) \\
 & - \frac{1}{2}a_+ \sin(\pi \nu_s^0)\sin \chi_+ +\frac{1}{2}a_-\sin(\pi \nu_s^0)\sin \chi_- \,,
\end{split}
\label{eq:2C.8}
\end{equation}
\end{widetext}
where the spin kick angles $\chi_\pm$  and the imperfection parameters $a_\pm$ are given by
\begin{equation}
 \chi_\pm =\frac{1}{2} \left(\chi_1\pm \chi_2 \right) \quad \text{and} \quad a_\pm =  \left(\vec{c} \cdot {\vec{n}_2}^{\,\text{r}}\right) \pm \left( \vec{c} \cdot \vec{n}_1 \right)\,. 
\label{eq:2C.9}
\end{equation}
Consequently, the determination of $a_\pm$ amounts to the determination of the projections of the stable spin axis $\vec c$ onto a plane spanned by the vectors $\vec n_1$ and ${\vec{n}_2}^{\,\text{r}}$.

Note the different status of the four terms in Eq.\,(\ref{eq:2C.8}). The first two terms, proportional to  $\sin^2 \left(\frac{1}{2}\chi_\pm \right)$, are uniquely predicted with absolute normalization. The last two terms, proportional to $\sin \chi_\pm $, enter with unknown coefficients $a_\pm$, to be determined experimentally.

For weak AIs, the left-hand side of Eq.\,(\ref{eq:2C.8}) can be further approximated as  $ \cos(\pi \nu_s^0) - \cos(\pi [\nu_s^0 + \Delta\nu_s(\chi_1,\chi_2)]) \simeq  \pi \Delta\nu_s(\chi_1,\chi_2) \sin\pi \nu_s^0$. Then, the right-hand side of Eq.\,(\ref{eq:2C.8}) entails a saddle point of $\Delta\nu_s(\chi_+,\chi_-)$ in the $(\chi^+,\chi^-)$-plane. Simple algebra yields the location of the saddle point (sp) $\chi_\pm^{\,\text{sp}}$ at  
\begin{eqnarray}
\tan \chi_{\pm}^{\,\text{sp}} = \frac{ a_\pm \sin \left( \pi \nu_s^0\right)}{1\pm \cos \left( \pi\nu_s^0 \right) }\, ,
\label{eq:2C.10}
\end{eqnarray}
so that the determination of the location of the saddle point amounts to a measurement of the imperfection parameters $a_\pm$. 

It should be noted that we could equally have applied the above described trick of Eq.\,(\ref{eq:2C.4}) to the 
apparent transport of the imperfection S$_1$ to the location of S$_2$. That would not have changed anything apart from the interchange of subscripts  1 and 2 in Eqs.\,(\ref{eq:2C.5}) to (\ref{eq:2C.9}). Consequently, modulo to this interchange, our findings for $a_\pm$ are applicable to the orientation of the stable spin axis at the location of both solenoids S$_1$ and S$_2$.

\section{Exploring magnetic imperfections of the COSY ring}
\label{sec:III}
\subsection{Experimental setup and data taking}
One of the goals of the investigations at COSY (carried out in September 2014) was to explore the spin closed orbit by introducing AIs, as exposed in the previous section. For that purpose the drift solenoid of the $\SI{2}{MeV}$ electron cooler (solenoid $\text{S}_1$ in Fig.\,\ref{fig:ring-sketch}, see\,\cite{Dietrich:2011tu} for details), and the difference of fields of drift and compensation solenoids of the $\SI{120}{keV}$ electron cooler (solenoid $\text{S}_2$ in Fig.\,\ref{fig:ring-sketch}, see\,\cite{6620982} for details) have been used as makeshift AIs. They are located in the opposite straight sections and the longitudinal artificial imperfection magnetic fields were adjusted by two separate power supplies. 

At first, the vertically polarized deuteron beam was injected and accelerated to the kinetic energy of $T=\SI{270}{MeV}$. Subsequently, the beam was prepared for $\SI{75}{s}$ by cooling and bunching. Afterwards the beam was extracted onto the carbon target. Then the initial vertical polarization of the particle ensemble was flipped into the horizontal plane by means of a resonant RF solenoid $\text{S}_{\text{RF}}$\,\cite{PhysRevLett.100.054801} (see Fig.\ref{fig:ring-sketch}). Subsequently, the particle spins perform an idle precession around the vertical axis in the horizontal plane of the machine with a frequency
\begin{equation}
f_s = |\nu_s^0 |f_{\text{R}} \simeq \SI{120}{kHz} \, , \label{eq:3A.1}
\end{equation}
where $\nu_s^0$ denotes the spin tune and $f_{\text{R}}$ the revolution frequency of the particle bunch. The initial vertical deuteron vector polarization provided by an atomic beam source was alternated from up to down states. One run typically contained 6 cycles, in order to allow us to estimate the fluctuations due to instabilities of COSY.

\begin{figure}[tb]
\includegraphics[width=\columnwidth]{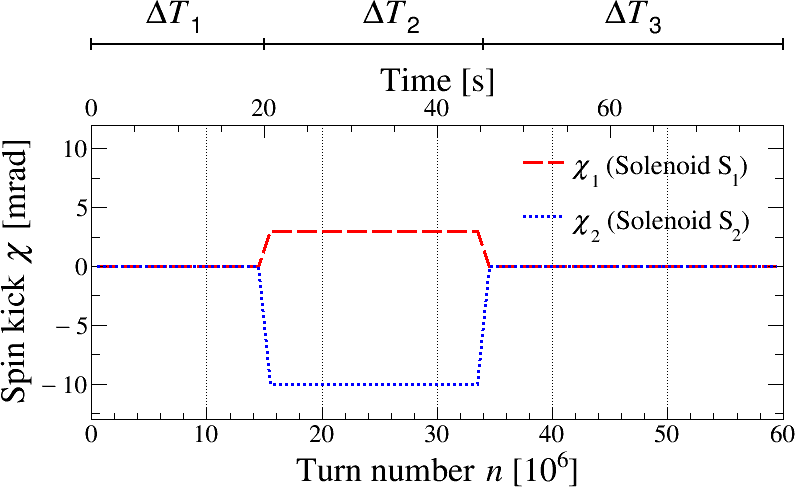}
\caption{\label{fig:fstep} Timing of spin kicks $\chi_{1,2}$ (see Eq.\,[\ref{eq:3A.2}]) of the two solenoids $\text{S}_1$ and $\text{S}_2$ during a measurement cycle. The achievable field integrals and spin kicks are listed in Table\,\ref{tab:solenoid-ramping}.} 
\end{figure}
The experimental scheme is illustrated in Fig.\,\ref{fig:fstep}. The first baseline spin tune measurement interval lasted for $\Delta T_1 = \SI{20}{s}$ after the spins of the particle ensemble had been flipped into the horizontal plane. Then the current of both solenoids $\text{S}_1$ and $\text{S}_2$ were ramped during a short time interval of $2-\SI{3}{s}$ to the specified values (see also Fig.\,\ref{fig:phasejump} in Appendix\,\ref{sec:appendixC}). The solenoids remained switched on for a longer time period $\Delta T_2 = \SI{25}{s}$ in order to obtain approximately the same statistical accuracy for the determination of the spin tune compared to the first time interval $\Delta T_1$. Afterwards, the solenoids were ramped down for a still longer $\Delta T_3 = \SI{35}{s}$ to provide a second baseline measurement. A comparison of the two baseline spin tune measurements allows one to keep track of potential spin tune drifts within each cycle.

The spin kicks $\chi_i$ ($i=1,2$) induced by the currents $I_i$ of the two solenoids are given by 
\begin{equation}
\chi_i = \frac{(1+G)}{B\rho} \int B_{i,z} \dd z -\chi_i^0= \frac{(1+G)}{B\rho} F_i (I_i - I_i^0) \, ,\label{eq:3A.2}
\end{equation}
where $B\rho$ denotes the magnetic rigidity of the ring, and $F_i$ the corresponding calibration factors, which in free space are given by Ampere's law in terms of the coil winding numbers\,\cite{Soltner:2014doa}. For the drift solenoid $\text{S}_1$ of the $\SI{2}{MeV}$ electron cooler $I_1^0 =0$. For $\text{S}_2$, the nominal current $I_2^0$ corresponds to the normal operation regime of compensating the longitudinal field integrals from the main drift solenoid, the toroids and the two compensation solenoids. In our study the drift and toroid solenoids and the corresponding steerers were run at the nominal current.  In $\text{S}_2$ the AI was generated by ramping the currents of the two compensation solenoids away from the nominal $I_2^0$ and then back to $I_2^0$ at the end of $\Delta T_2$. The ranges of the applied field integrals using the two solenoids $\text{S}_1$ and $\text{S}_2$ are listed in Table\,\ref{tab:solenoid-ramping}.
\begin{table}[tb]
\renewcommand{\arraystretch}{1.1}
\begin{ruledtabular}
\begin{tabular}{c|cccc}
  &  \multicolumn{2}{c}{Field integral} & \multicolumn{2}{c}{Spin kick angle} \\
  &  \multicolumn{2}{c}{$[\si{T.mm}]$} & \multicolumn{2}{c}{$[\si{mrad}]$}\\
                       & min   & max  & min  & max \\\hline
 Solenoid $\text{S}_1$ & $-33$ & $+33$ &  $-\phantom{1}8.787$  & $+8.787$ \\
 Solenoid $\text{S}_2$ & $-49$ & $+14$ &  $-12.978$ & $+3.708$\\
\end{tabular}
\end{ruledtabular}
 \caption{\label{tab:solenoid-ramping} Minimum and maximum of the field integrals $\int B_1 \dd z$ and $\int B_2 \dd z$ applied by the solenoids $\text{S}_1$ and $\text{S}_2$, and the corresponding minimal and maximal spin rotation angles $\chi_1$ and $\chi_2$.}
\end{table}

One necessary requirement to determine the spin tune in each time interval $\Delta T_i$ ($i = 1,2,3$)  to high accuracy is a long horizontal polarization lifetime. This was achieved by tuning the sextupole magnets in the ring to correct for decoherence effects like emittance and momentum spread of the beam\,\cite{Guidoboni:2016bdn}.

\subsection{Analysis method}
The method to unfold the fast spin precession in the horizontal plane and thus to determine the spin tune is described in the previous JEDI publication\,\cite{Eversmann:2015jnk} and is outlined in Appendix\,\ref{sec:appendixC}. The  EDDA detector is operated as a polarimeter to measure count rates in each of the four detector quadrants (up, right, down, left)\,\cite{PhysRevSTAB.17.052803}. The beam particles are brought into interaction with the carbon target of the polarimeter by stochastic heating of the beam.

Six measurement cycles with alternating polarization states (up, down) were taken for each solenoid setting, each measurement taking about $\Delta T_1 + \Delta T_2 + \Delta T_3 = \SI{80}{s}$.  In each of the three time intervals ($i=1,2,3$), the spin tunes $\nu_{s_i}$ were determined, and subsequently two spin tune jumps
\begin{equation}
\begin{split}
\Delta\nu_{s_1} & = \nu_{s_2}(n^{\text{ON}}) - \nu_{s_1}  \, , \text{and}\\
\Delta\nu_{s_2} & = \nu_{s_2}(n^{\text{OFF}}) - \nu_{s_3}  \, ,\label{eq:3B.1}
\end{split}
\end{equation}
were determined, where $n^{\text{ON}}$ denotes the turn number when the solenoids are switched on, and $n^{\text{OFF}}$ the turn number when the solenoids are switched off (see Fig.\,\ref{fig:phasejump} in Appendix\,\ref{sec:appendixC}). Measurements containing six-cycle runs were repeated and spin tune jumps were measured on a mesh of spin kicks $\chi_1$ versus $\chi_2$, and this procedure is referred to as \textit{spin tune mapping}.

In a stable ring with perfectly stable solenoid power supplies, within each cycle the two baseline spin tunes  $\nu_{s_1}$ and $\nu_{s_3}$, and the corresponding spin tune jumps $\Delta \nu_{s_1}$ and  $\Delta\nu_{s_2}$ must coincide. This is not quite the case with COSY as is. A drift of the spin tune within each cycle, from cycle to cycle of the same run, and from run to run was already observed during our previous experiment\,\cite{Eversmann:2015jnk}. This drift could arise from a walk of the solenoid currents $I_{1,2}$, from a temperature dependence of the magnetic fields, or from hysteresis effects in the main dipole magnets causing a continuous displacement of the beam orbit and a resulting change of the beam axis with respect to the magnetic axes of the solenoids. 

\begin{figure}[tb]
\includegraphics[width=\columnwidth]{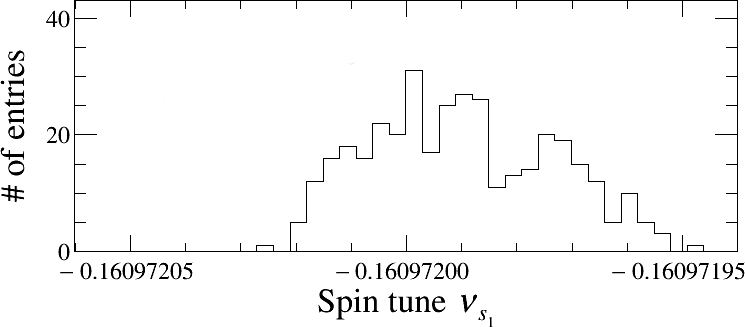}
\caption{ \label{fig:nu1dist}  Distribution of the baseline spin tunes $\nu_{s_1}$ in the time interval $\Delta T_1$ for all 359 measurement cycles, exhibiting a mean value of $\braket{\nu_{s_1}} = -(\num{16097199.0} \pm \num{1.6})\cdot \num{e-8}$.}
\end{figure}

The cycle-to-cycle variations in the machine are clearly demonstrated by the graph of unperturbed spin tune $\nu_{s_1}$, shown in Fig.\,\ref{fig:nu1dist}. The RMS of this distribution must be regarded as a cycle-to-cycle systematic uncertainty of the baseline spin tune, which amounts to $\delta\nu_{s_1}^{\text{syst}} = \num{1.6e-8}$. The cycle-to-cycle statistical uncertainty of the baseline spin tune is evaluated in Appendix\,\ref{sec:appendixC}, and the values are given in Table\,\ref{tab:all-abc}. For the first time interval $\Delta T_1$ it amounts to $\delta\nu_{s_1}^{\text{stat}} = (7.1\pm 1.1)\cdot \num{e-10}$.

Similarly, we  regard the difference of the baseline spin tunes $\nu_{s_1} -\nu_{s_3}$ (see Fig.\,\ref{fig:dnu13} in Appendix\,\ref{sec:appendixC}) and the difference of the two spin tune jumps $\Delta \nu_s{_1} - \Delta \nu_s{_2}$ within a cycle  as a systematic error due to ring instabilities. This difference comes out much larger than the statistical accuracy to which the spin tunes in the three time intervals of the same cycle can be determined. 

The best estimate for the statistical, systematic and quadratically combined errors of the spin tune jumps $\delta \Delta\nu_s$ per cycle is derived in Appendix\,\ref{sec:appendixC}, and amounts to 
\begin{equation}
\begin{split}
\delta \Delta\nu_s^{\text{stat}} & = \num{0.70e-9} \,, \\ 
\delta \Delta\nu_s^{\text{syst}} & = \num{3.23e-9} \,, \\ 
\delta \Delta\nu_s               & = \num{3.30e-9} \, .
\end{split}
\label{eq:3B.2}
\end{equation}

Since, depending on the measurement scheme, the above given statistical error can be made small, in the subsequent data analyses and simulations the systematic error $\delta \Delta\nu_s^{\text{syst}}$ is used. Remarkably, the within-the-cycle walk of the spin tune $\nu_{s_1} -\nu_{s_3}$ (shown in Fig.\,\ref{fig:dnu13} in Appendix\,\ref{sec:appendixC}) is almost an order of magnitude smaller than the cycle-to-cycle walk (shown in Fig.\,\ref{fig:nu1dist}).

One serendipitous finding of a systematic effect was that the operation of the COSY ionization beam profile monitor (IPM)\,\cite{PhysRevSTAB.18.020101} causes spin tune jumps as large as $\sim \num{e-6}$. This finding indicates the high sensitivity of the spin tune to the seemingly small electromagnetic perturbations in the ring. Since the observed IPM effects are large, all cycles with IPM ON were excluded from the data analysis.

\begin{figure*}[htb]
\includegraphics[width=\textwidth]{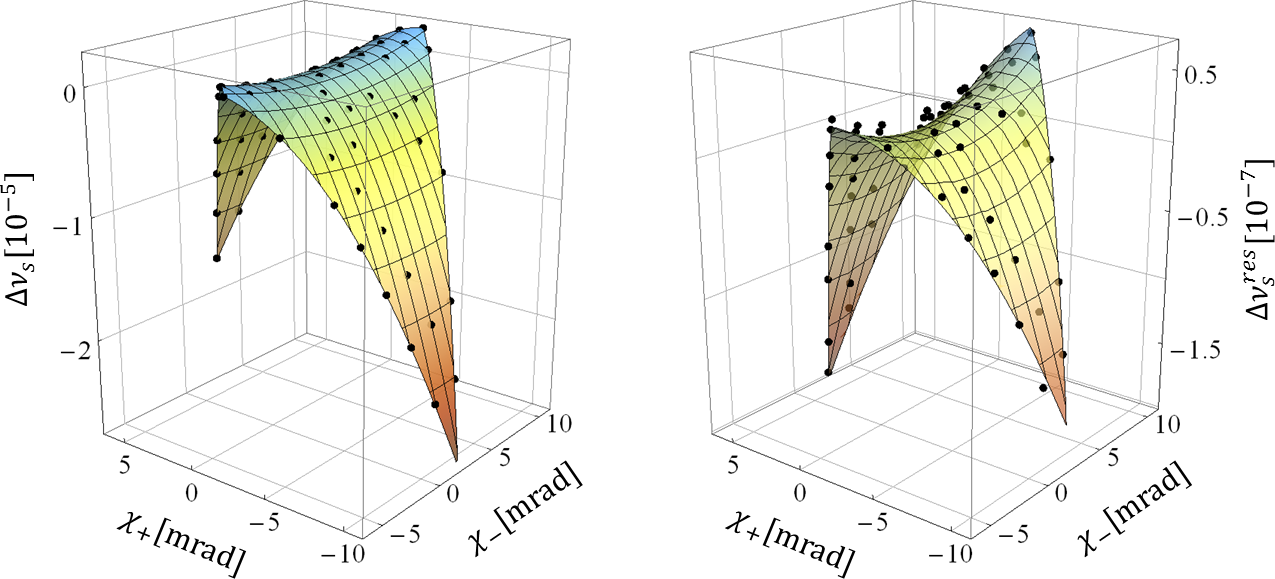}
\caption{\label{fig:map12} Left panel: Map 2 shows the results of the measurement of spin tune jumps $\Delta\nu_s(\chi_+,\chi_-)$. Each point represents a single measurement. The error bars are smaller than the size of the symbols. A surface is fit to the data as described in the text, the location of the saddle point is given in Eq.\,(\ref{eq:saddlepoint}). Right panel: Residuals for Map 2, showing $\Delta\nu_s^{\text{res}} = \Delta{\nu_s} - \Delta\nu_s^{\text{fit}}$. As described in the text, using Eq.\,(\ref{eq:3C.3}) a surface is fit to $\Delta\nu_s^{\text{res}}(\chi_+,\chi_-)$. Note the difference of the vertical scales for the two graphs.}
\end{figure*}

As we had no \textit{a priori} idea about the strength of the imperfection fields, a first exploratory map (Map\,1) was recorded using a coarse mesh. Later on, during about $\SI{33}{\hour}$ a second map (Map\,2) was recorded with twice smaller mesh spacing. Initially, Map\,2 contained $9\times 9 = 81$ data points. After runs with IPM switched ON had been discarded, and one row of measurements was not recorded properly, Map\,2 altogether contained 60 data points. While Maps\,1 and 2 are fully consistent with each other, in view of the higher statistics in the following the experimentally observed data for Map\,2 are considered, and are depicted in the left panel of Fig.\,\ref{fig:map12}.

\subsection{Confirmation of the saddle point of the spin tune map}
\label{sec:3.C}
The observed spin tune map shown in Fig.\,\ref{fig:map12} clearly confirms the theoretically expected saddle point property. The graph shows  
\begin{equation}
\begin{split}
& \Delta\nu_s(\chi_+,\chi_-)  \\
&  \simeq \frac{\cos( \pi \nu_s^0) - \cos(\pi [\nu_s^0 + \Delta\nu_s(\chi_+,\chi_-)])}{\pi \sin(\pi \nu_s^0 )} \\
                            & = f(\nu_s^0,\chi_+, \chi_-) \, ,
\label{eq:3C.1}
\end{split}
\end{equation}
where the shape of the surface of the spin tune map is produced by the numerator, given by Eq.\,(\ref{eq:2C.8}). For deuterons $\sin \left( \pi \nu_s^0 \right)  < 0$, and according to Eq.\,(\ref{eq:2C.8}), $f(\nu_s^0,\chi_+, \chi_-)$ is a sum of the convex function of $\chi_+$ and the concave function of $\chi_-$. 

Each data point has been assigned a quadratically combined error bar, given by Eq.\,(\ref{eq:3B.2}). The principal fitted parameters are the ring imperfections $a_+$ and $a_-$. By virtue of Eq.\,(\ref{eq:2C.8}), the missing cross terms assure that these parameters are basically uncorrelated. 

\subsubsection{Validation of the fitting procedure}
\label{sec:3C.1}
The employed fitting procedure is illustrated by a simulation using the spin-tracking code COSY-Infinity\,(\cite{COSY-Infinity}, for applications to spin tracking at COSY, see\,\cite{Rosenthal:2016zbf}). The simulations assume a single particle with nominal momentum orbiting on the closed orbit. We used a model for the spin-tune jump described by Eq.\,(\ref{eq:2C.2}). We assumed an ideal ring and vanishing EDM, so that at every point along the orbit the stable spin axis is precisely oriented along the $y$-axis. An additional $\SI{5}{\tesla.mm}$ solenoid is placed in arc A$_1$ (same location as $\text{S}_{\text{RF}}$ in Fig.\,\ref{fig:ring-sketch}). According to the COSY-Infinity simulations, this solenoid  generates an imperfection which produces a $c_z$ different from zero of 
\begin{equation}
c_z = -0.001323429 
\label{eq:input-c}
\end{equation}
at the location of solenoid $\text{S}_1$. Now, we want to use a single solenoid in the simulation to determine the value of $c_z$ using spin tune mapping. To this end, we produced a set of 53 spin tune jumps, based on Eq.\,(\ref{eq:2C.2}) with uncertainties given by Eq.\,(\ref{eq:3B.2}), equally spaced in $\chi_{\text{AI}}$. Then we fit the resulting set of data points using Eq.\,(\ref{eq:2C.2}). As expected, the resulting fit is of good quality, yielding a $\chi^2/N_{\text{dof}} = 50.27/52$, and the input value for $c_z$ [given in Eq.\,(\ref{eq:input-c})] and the fitted value of $c_z^{\text{fit}}$ are perfectly consistent with each other,
\begin{equation}
c_z - c_z^{\text{fit}} = (3.99 \pm 3.43)\cdot \num{e-7} \,.
\end{equation}

If it  were not for the systematic errors to be discussed below, it would have been possible to determine $c_z$ to an accuracy of $\delta c_z \approx 3.5\cdot 10^{-7}$. To put this number into perspective, supposing a similar accuracy of the determination of $c_x$ would entail a resolution of the angle of the stable spin axis $\delta \xi_{\text{EDM}} \sim  3.5\cdot 10^{-7}$. Then, in the absence of machine imperfections at $T = \SI{270}{MeV}$, Eq.\,(\ref{eq:2A.4}), corresponds to a resolution of the deuteron EDM of about 
\begin{equation}
\begin{split}
\sigma(d) & \sim \frac{ q}{2m_d}\delta \eta  \\
          & = \frac{G q}{\beta_d m_d} \delta \xi_{\text{EDM}} \sim \SI{2e-21}{e.cm}\,. 
\end{split}
\label{bound-on-EDM}
\end{equation}

\subsubsection{Fitting the map of residuals}
\label{sec:fitting-the-map-of-residuals}
As we have seen above, there exists a run-to-run variation of $\nu_s^0$. We account for that by evaluating the theoretically expected spin tune jump function $f(\nu_s^0,\chi_+, \chi_-)$ [Eq.\,(\ref{eq:3C.1})] at the average $\nu_s^0$ as measured in the corresponding run. 

The fit to the spin tune jump Map 2 with $a_\pm$ as free parameters yields 
\begin{equation}
 \begin{split}
   a_+ & = (\num{50172.1} \pm 5.9) \cdot \num{e-7}\,,\, \text{and} \\
   a_- & = (-\num{4452.5} \pm 5.7) \cdot \num{e-7}\,,
 \end{split}
\end{equation}
with an enormous $\chi^2/N_{\text{dof}} =  \num{22017}/58$. According to Eq.\,(\ref{eq:2C.10}), the saddle point is located at
\begin{equation}
 \begin{split}
   \chi^{\text{sp}}_+ & = (-\num{1.29637} \pm \num{0.00015})\,\si{mrad},\, \text{and} \\
   \chi^{\text{sp}}_- & = (\num{0.11505} \pm \num{0.00015})\,\si{mrad}.
 \end{split}\label{eq:saddlepoint}
\end{equation}

In order to understand the reason for the large $\chi^2/N_{\text{dof}}$, we investigate the map of residuals, 
\begin{equation}
\Delta\nu_s^{\text{res}} = \Delta{\nu_s} - \Delta\nu_s^{\text{fit}} \,,
\label{eq:residuals}
\end{equation}
shown in Fig.\,\ref{fig:map12} (right panel). This map exhibits a similar saddle point pattern with an amplitude at the level of about one per cent of the observed spin tune jumps $\Delta\nu_s(\chi_+,\chi_-)$. 

The observed saddle point property hints at the possibility to fit the residuals by a function reminiscent of Eq.\,(\ref{eq:3C.1}), where we allow for an additional scaling of effects stemming from $\chi_+$ and $\chi_-$
\begin{widetext}
\begin{equation}
\begin{split}
\pi \sin(\pi \nu_s^0) \Delta\nu_s^{\text{res}}(\chi_+,\chi_-) = \;  A_+ & \left\{\left[ 1 + \cos\left( \pi \nu_s^0\right)\right]\sin^2 \left(\frac{1}{2}\chi_+ \right)-\frac{1}{2}b_+ \sin(\pi \nu_s^0)\sin \chi_+ \right\}\\
             -A_- &\left\{ \left[ 1 - \cos\left(\pi \nu_s^0\right)\right]\sin^2\left(\frac{1}{2}\chi_-\right) - \frac{1}{2}b_-\sin(\pi \nu_s^0)\sin \chi_- \right\}  \,.
\end{split}
\label{eq:3C.3}
\end{equation}
\end{widetext}
We assign to the residuals the error bars of the corresponding spin tune jumps. Such a fit yields, 
\begin{equation}
\begin{split}
    A_+ & =  (91.4 \pm 0.7) \cdot \num{e-4} \,, \\ 
    A_- & =  (21.9 \pm 1 )  \cdot \num{e-3} \,, \\
    b_+ & = (1022 \pm 0.7) \cdot \num{e-4} \,, \\
    b_- & =  -(15.9 \pm 3.9) \cdot \num{e-5} \,, 
\label{eq:fit-parameters-of-residuals}
\end{split}
\end{equation}
with a  $\chi^2/N_{\text{dof}} =235.5/55$, which improves by about a factor of $100$ the $\chi^2/N_{\text{dof}} =22017/56$ found for the simplified formalism, given in Eq.\,(\ref{eq:3C.1}). 

Now the full spin tune jump takes the form
\begin{widetext}
\begin{equation}
\begin{split}
\pi \sin(\pi \nu_s^0)\Delta\nu_s(\chi_+,\chi_-)  \simeq  & \, \pi \sin(\pi \nu_s^0)\left[\Delta\nu_s^{\text{fit}}(\chi_+,\chi_-) + \Delta\nu_s^{\text{res}}(\chi_+,\chi_-) \right]\\
  = & \, (1 + A_+)  \left[1 + \cos \left( \pi \nu_s^0\right)\right] \sin^2 \left( \frac{1}{2}\chi_+ \right)  -(1 + A_-) \left[1 - \cos\left( \pi \nu_s^0\right)\right] \sin^2 \left( \frac{1}{2}\chi_- \right) \\
    & \, -\frac{1}{2} \left(a_+ + A_+ b_+\right) \sin(\pi \nu_s^0)\sin \chi_+ -\frac{1}{2} \left(a_- + A_- b_-\right) \sin(\pi \nu_s^0)\sin \chi_- \\
  \simeq  & \, \left[1 + \cos \left( \pi \nu_s^0 \right) \right] \sin^2 \left( \frac{1}{2}k_+\chi_+\right) - \left[1 - \cos \left( \pi \nu_s^0\right)\right] \sin^2 \left( \frac{1}{2}k_-\chi_- \right) \\
    & \, -\frac{1}{2} a_+^* \sin(\pi \nu_s^0)\sin \left(k_+ \chi_+\right) -\frac{1}{2} a_-^* \sin(\pi \nu_s^0)\sin \left( k_-\chi_- \right) \,.
\end{split}
\label{eq:3C.4}
\end{equation}
\end{widetext}
This guess for the functional dependence of the map of residuals [Eq.\,(\ref{eq:2C.8})] suggests that the spin tune jumps can still be described by Eq.\,(\ref{eq:3C.1}) at the expense of rescaling the spin kick angles via
\begin{equation}
 \tilde\chi_\pm \rightarrow k_\pm \chi_\pm\,,
 \label{eq:rescaling-of-spin-kick-angles}
\end{equation}
where $k_{\pm}^2 = 1 + A_\pm$. The variables $\chi_\pm$ are somewhat obscure, because they mix the effects of the two solenoids. One may prefer to apply the rescaling to the \textit{individual} solenoids, described by 
\begin{equation}
 \tilde\chi_{1,2}\rightarrow k_{1,2}\chi_{1,2},.
 \label{eq:rescaling-of-spin-kick-angles_2}
\end{equation}
This empirical finding looks as if the actual spin kicks $\tilde\chi_{1,2}$ are different from what is given by Ampere's law applied to the readout currents of the solenoid power supplies\footnote{The power supply of the compensation solenoid (type SM 30-200) provides a current control stability of $\SI{100}{ppm}$ and a temperature coefficient of $\SI{60}{ppm \per \kelvin}$. The power supply of the drift solenoid of the $\SI{2}{MeV}$ electron cooler (type BPS SW MODULE PUISS BIP 30/22) has an absolute current calibration of $\SI{0.1}{\percent}$, an output stability of $\SI{20}{ppm}$, and a temperature coefficient of $\SI{5}{ppm \per \kelvin}$. Presently, for the two solenoids $\text{S}_{1,2}$, \SI{16}{bit} power supply controllers (type PSC-ETH) are used.}.

In the simplified formalism, given in Eq.\,(\ref{eq:2C.9}), the parameters $a_\pm$ were related to projections of the spin stable axis onto the AI axes $\vec n_1$ and $\vec n_2^{\,{\text{r}}}$. This interpretation is somewhat obscured by the yet unknown systematic effects behind the residuals, which also contribute to
\begin{equation}
a_\pm^*= a_{\pm}+ A_\pm b_\pm\,.
\label{eq:shift_of_apm}
\end{equation}

To summarize,  Eq.\,(\ref{eq:2C.8}), constrained by assuming an ideal alignment of solenoids S$_1$ and S$_2$, only contains two free parameters $a_\pm$. This approach obviously misses the experimental data on the spin tune jump by $\Delta\nu_s^{\text{res}}$ which numerically amounts to about 1\% of $\Delta\nu_s$ (see Fig.\,\ref{fig:map12}). However, in view of the achieved record-high precision of the spin tune determination, even this small mismatch becomes statistically very relevant. We must therefore conclude that Eq.\,(\ref{eq:2C.8}) does not account for certain substantial systematic effects. Specifically, we shall discuss in the following, whether an apparent rescaling of the spin kick angles, given in Eq.\,(\ref{eq:rescaling-of-spin-kick-angles}) is borne out by realistic physics mechanisms. A more detailed discussion of such a mechanism and the quantitative description in terms of a fit will be presented in Sec.\,\ref{sec:spin-tune-mapping-with-allowance-for-misalignment-of-solenoids}. 

\begin{figure*}[!]
\includegraphics[width=0.95\textwidth]{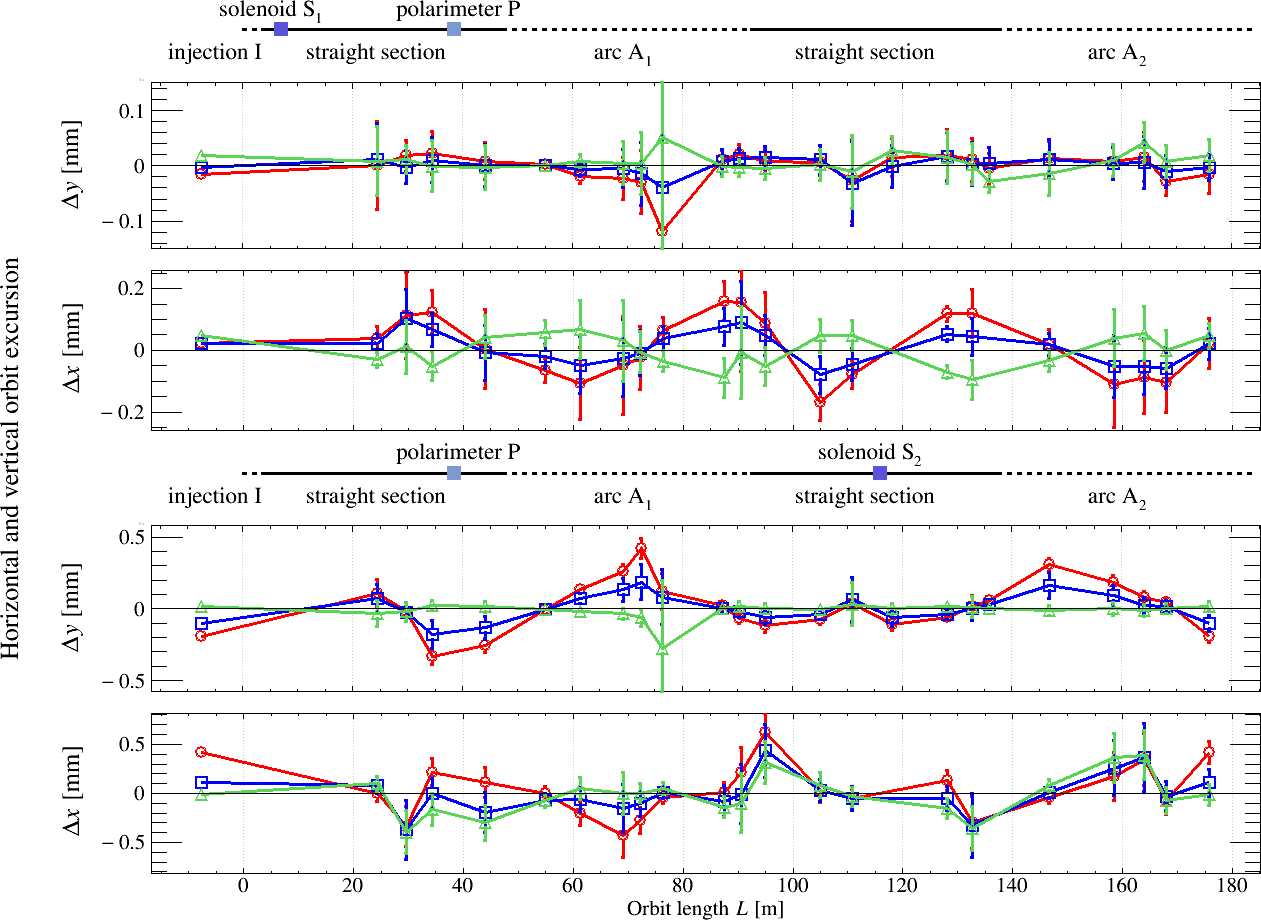}
\caption{\label{fig:allbpm-x1-and-x2} Upper pair of panels: the experimentally observed  horizontal and vertical orbit excursions along the ring for several spin kicks in solenoid $\text{S}_1$ ($\chi_1 = -8.79(\textcolor{red}{\Circle}), -4.39(\textcolor{blue}{\square}), 4.39(\textcolor{green}{\triangle}) $\,mrad), while solenoid $\text{S}_2$ is switched off. The positions of solenoids $\text{S}_1$ and $\text{S}_2$ in the ring are indicated. The last BPM reading at the end of arc A$_2$ is replicated as first point of straight section 1. The lines connecting the points are to guide eye. Bottom pair of panels: Measured horizontal and vertical orbit excursions along the ring for several spin kicks in solenoid $\text{S}_2$ ($\chi_2 = -12.98(\textcolor{red}{\Circle}), -7.42(\textcolor{blue}{\square}), 1.85(\textcolor{green}{\triangle}) $\,mrad), while solenoid $\text{S}_1$ is switched off.}
\end{figure*}

\section{Systematic limitations of spin tune mapping}
\label{sec:IV}
\subsection{Evidence for steering effect of solenoids}
One obvious source of systematics is the misalignment of the solenoid axes with respect to the beam trajectory. In such a case, the magnetic field of a solenoid exhibits vertical and horizontal field components which are proportional to the solenoid field $\chi_{\text{AI}}$ and the angles of rotation $\xi_{x,y}$ of the solenoid around the $x$- and $y$-axis, respectively. To a first approximation, a misaligned solenoid can be regarded as an ideal solenoid complemented by one horizontal and one vertical steerer dipole magnet. The steering effect of the misaligned solenoid, \textit{i.e.}, the momentum rotation angle $\vartheta$, is related to the solenoid spin kick $\chi_{\text{AI}}$ via [see Eq.\,(\ref{eq:3A.2})]
\begin{equation}
\vartheta_{x,y} = \frac{\xi_{y,x} \chi_{\text{AI}}}{1+G}\,. 
\label{eq:momentum-bending-in-solenoids}
\end{equation}

The transverse magnetic fields  affect the spin tune of the ring both directly and indirectly via excursions of the beam  which change the orbit length and also affect the magnetic imperfections acting on the spin in all magnetic elements of the ring. Even the sign of the impact on the spin tune cannot be readily predicted. Preliminary experimental findings on the effect of steerer magnets on the spin tune 
have been reported elsewhere\,\cite{StasBeijing}.

The drift solenoid $\text{S}_1$ of the \SI{2}{MeV} electron cooler is operated independently from the toroidal magnetic fields and the related steerers. The case of solenoid $\text{S}_2$ in the \SI{120}{keV} electron cooler is more complex\,\cite{Soltner:2014doa}. During standard operation, the longitudinal field integral  $\int B_z dz$ of the drift solenoid and the two toroidal magnets is compensated for by the field integral of the two compensation solenoids up- and downstream of the main electron cooler solenoid. During the spin tune mapping experiment, the two compensation solenoids were operated using an additional power supply, whereby the field integral of the electron cooler could be adjusted (see Table\,\ref{tab:solenoid-ramping}).

In S$_2$, the effect of the transverse toroid fields is compensated for by two sets of steerers upstream of the compensation solenoid and downstream of a family of quadrupoles (and vice versa for the second compensation solenoid at the downstream end of the drift solenoid). The principal point is that even the reference trajectory runs along the compensation solenoid at a finite angle, and operating the compensation solenoid at a current $I_2$, which differs from the nominal current $I_2^0$, inevitably generates unwanted transverse fields along with the uncompensated longitudinal AI field required for spin tune mapping. At the moment there are no spin tracking codes available which fully account for the fields of the toroidal magnets in the \SI{120}{keV} electron cooler.


In Fig.\,\ref{fig:allbpm-x1-and-x2} the experimentally observed orbit excursions along the ring are shown as function of the spin kicks $\chi_1$ or $\chi_2$ in each of the two solenoids when the second solenoid is switched off, \textit{i.e.} only single solenoids, either $\text{S}_1$ or $\text{S}_2$ are active. The observed beam excursions are quite substantial. Solenoid $\text{S}_1$ perturbs the orbit mostly in the horizontal plane, \textit{i.e.}, $\text{S}_1$ predominantly provides a vertical magnetic field. Solenoid $\text{S}_2$ shifts the orbit in both the horizontal and vertical plane. It should be noted that the vertical orbit displacement is approximately linear in both planes as function of $\chi_2$ (see bottom pair of panels in Fig.\,\ref{fig:allbpm-x1-and-x2}). Quantitative estimates for the solenoid rotation angles, derived from the simulations carried out with COSY-Infinity, will be discussed in more detail in Sec.\,\ref{sec:IV:B}). 

\begin{figure}[t]
\includegraphics[width=\columnwidth]{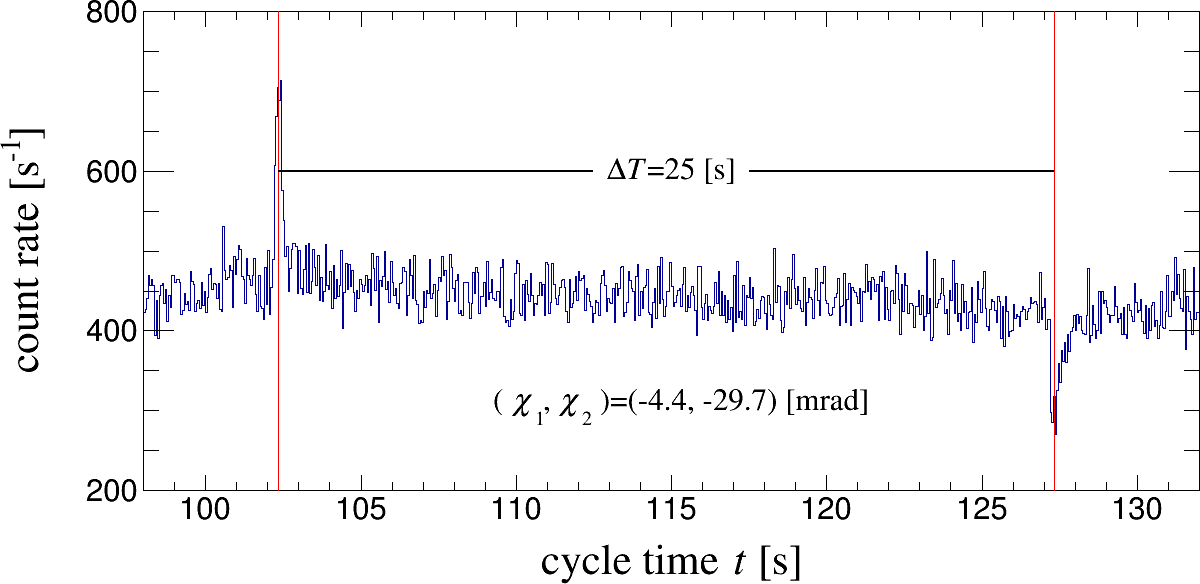}
\caption{\label{fig:rate3864} Spikes of the count rate in the polarimeter at the beginning of the time interval $\Delta T_2$ when the two solenoids $\text{S}_1$ and $\text{S}_2$ are switched ON (producing the indicated rotation angles $\chi_1$ and $\chi_2$) at $t\approx \SI{103}{s}$ and at the end at $t\approx \SI{127}{s}$ when $\text{S}_1$ and $\text{S}_2$ are turned OFF. }
\end{figure}
\begin{figure}[t]
\includegraphics[width=1\columnwidth]{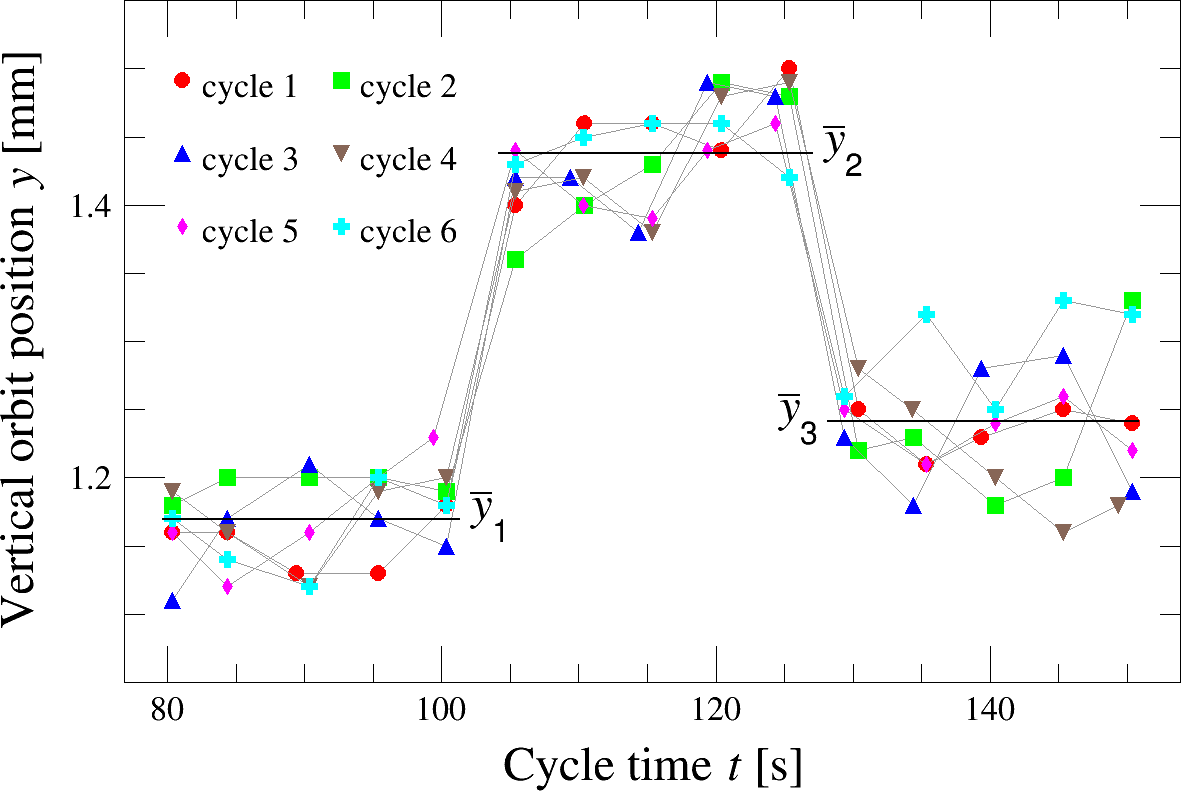}
\caption{\label{fig:bpmx13t} Measurements of the vertical beam positions in BPM13 during six cycles, each ranging in cycle time from 80 to $\SI{150}{s}$. The results shown were obtained with the same beam parameters and solenoid settings: $\text{S}_1 (\chi_1=0)$ and $\text{S}_2(\chi_2 = \SI{-11}{mrad})$. The  vertical orbit positions $\bar{y}_{i}$ ($i=1,2,3$), averaged over 6 cycles, are calculated for the time intervals $\Delta T_1$ and $\Delta T_3$ when $\text{S}_2$ was switched OFF, and for time interval $\Delta T_2$ when $\text{S}_2$ was switched ON. The numerical values of $\bar{y}_{i}$ are listed in Table\,\ref{tab:beam-positions}.}
\end{figure}

Since the particles are removed from the beam by the interaction with the target, collective orbit displacements should therefore also affect the count rate in the polarimeter. In Fig.\,\ref{fig:rate3864} the observed count rate for one particular cycle is shown. The first spike occurs when the solenoids $\text{S}_1$ and $\text{S}_2$ are switched on, and fades away because of the absorption of beam particles at the target. The effect should be reversible, and indeed a drop in the count rate is observed at the end of the time interval $T_2$ when the solenoids are turned off. Subsequently, due to continuous heating of the beam, the count rate approaches again the previous level.

After the AIs in the two solenoids $\text{S}_1$ and  $\text{S}_2$ are switched off, the spin tune jumps are not always perfectly reversible (see Appendix\,\ref{sec:appendixC}). Such effects accumulate and the spin tune drifts from injection to injection, as reported in\,\cite{Eversmann:2015jnk}.  There is an indication for a similar irreversibility with respect to the orbit position, and in Fig.\,\ref{fig:bpmx13t} one example is shown. From the average values of the vertical beam positions, listed in Table\,\ref{tab:beam-positions}, it is clearly seen that within errors, the vertical position of the beam is not recovered when solenoid $\text{S}_2$ is switched OFF at the end of the time interval $\Delta T_2$.


\subsection{Unraveling the steering effect of misaligned solenoids using simulations with COSY-Infinity \label{sec:IV:B}}
Here we shall analyze to which extent the residuals of the spin tune map, shown in Fig.\,\ref{fig:map12} (right panel), can be related to the steering effect of the misaligned solenoids. In order to evaluate the situation, we resort again to the spin- and orbit-tracking code COSY-Infinity\,\cite{COSY-Infinity}. For instance, when the solenoid is rotated around the $y$ axis by an angle $\xi_y$, then its axis is given by $\vec{k}= \cos \xi_y \vec{e}_z  +\sin\xi_y  \vec{e}_x$, and a horizontal magnetic field proportional to $ \sin\left( \xi_y \chi_{\text{AI}} \right) \simeq \xi_y \chi_{\text{AI}}$ will be generated, causing vertical excursions of the beam orbit.

Excursions of the orbit change the torque exerted on the spin by the magnetic elements in the ring. Thus, the simplified prediction of the spin tune jumps, given in Eq.\,(\ref{eq:2C.8}),  needs to be revised. Here we evaluate the salient features of the steering effect in a simplified model using only one solenoid as an AI, as described by Eq.\,(\ref{eq:2C.2}). 

Specifically, now the spin transfer matrix of the ring excluding the solenoid is given by
\begin{equation}
\begin{split}
  \matr{t}_{\text R}(\xi,\chi_{\text{AI}}) =  & \cos \left( \pi q_s^{\text{R}}(\xi,\chi_{\text{AI}})\right) \\
  &  - i \left[\vec \sigma \cdot \vec c(\xi,\chi_{\text{AI}})\right] \sin \left( \pi q_s^{\text{R}}(\xi,\chi_{\text{AI}}) \right)\,,
\end{split}
\end{equation}
where $2\pi q_s^{\text{R}}(\xi,\chi_{\text{AI}})$ denotes the spin-phase advance in the ring \textit{excluding} the solenoid (not to be confused with the spin tune $\nu_s(\xi,\chi_{\text{AI}})$, which is defined for the full ring \textit{including} the solenoid). The spin-transfer properties of the ring only depend on the beam orbit. An ideally aligned solenoid ($\xi=0$, $\chi \neq 0$) does not disturb the beam orbit, and leaves the spin advance in the ring unchanged from the canonical $2\pi \nu_s^0$,  therefore, $q_s^{\text{R}}(\xi = 0,\chi_{\text{AI}}) \equiv \nu_s^0$, \textit{i.e.}, equal to the spin tune of the ring when the solenoid is switched off.
\begin{table}[thb]
\renewcommand{\arraystretch}{1.2}
\begin{ruledtabular}
\begin{tabular}{ccc}
  Time interval &   Vertical beam position [mm] \\\hline
  $\Delta T_1$ & $\bar{y}_1 = 1.169 \pm 0.006$ \\
  $\Delta T_2$ & $\bar{y}_2 = 1.438 \pm 0.007$ \\
  $\Delta T_3$ & $\bar{y}_3 = 1.241 \pm 0.008$ \\
\end{tabular}
\end{ruledtabular}
 \caption{\label{tab:beam-positions} Average vertical beam positions $\bar{y}_i$ ($i=1,2,3$) for one run composed of six cycles, as shown in Fig.\,\ref{fig:bpmx13t}.}
\end{table}

Now, we proceed to the decomposition
\begin{equation}
\begin{split}
   \cos \left( \pi \nu_s^0 \right) & - \cos \left( \pi \nu_s(\xi,\chi_{\text{AI}}) \right) = \\
    & \cos \left( \pi \nu_s^0 \right) - \cos \left(  \pi q^{\text{R}}_s(\xi,\chi_{\text{AI}}) \right) \\
    & + \cos \left( \pi q_s^{\text{R}}(\xi,\chi_{\text{AI}})\right)   - \cos \left(\pi \nu_s(\xi,\chi_{\text{AI}})\right)\,.
\end{split}
 \label{spin-tune-decomposition}
\end{equation}
Here the last two terms describe the change of the spin transfer properties of the ring that are caused by the orbit excursions, 
\begin{equation}
\begin{split}
 \cos \left( \pi \nu_s^0 \right) - \cos \left( \pi q^{\text{R}}_s(\xi,\chi_{\text{AI}})\right) 
 \simeq \\ 
 \pi \sin  \left(\pi \nu_s^0\right) \Delta q^{\text{R}}_s(\xi,\chi_{\text{AI}})\,,
\end{split}
\end{equation}
where $\Delta q^{\text{R}}_s(\xi,\chi_{\text{AI}}) = q^{\text{R}}_s(\xi,\chi_{\text{AI}}) - \nu_s^0$ is the first systematic effect, which changes the spin phase advance per turn
by $2\pi \Delta q^{\text{R}}_s(\xi,\chi_{\text{AI}})$. 

The last two terms of Eq.\,(\ref{spin-tune-decomposition}) can be regarded as an extension of Eq.\,(\ref{eq:2C.3}), which leads to 
\begin{widetext}
\begin{equation}
 \begin{split}
    \cos \left( \pi q_s^{\text{R}}(\xi,\chi_{\text{AI}})\right) & - \cos \left( \pi \nu_s(\xi,\chi_{\text{AI}}) \right) \\
    =  & \cos \left( \pi q^{\text{R}}_s(\xi,\chi_{\text{AI}})\right) \cdot \left[ 1 - \cos \left( \frac{1}{2} \chi_{\text{AI}}\right) \right] 
     + \left( \vec c(\xi,\chi_{\text{AI}}) \cdot \vec k\right) \sin \left( \pi q^{\text{R}}_s(\xi,\chi_{\text{AI}})\right) \sin \left( \frac{1}{2} \chi_{\text{AI}}\right) \\
    \simeq & \cos \left( \pi \nu_s^0\right)  \left[1 - \cos \left( \frac{1}{2} \chi_{\text{AI}}\right)\right]  
     + \left(\vec c(\xi,\chi_{\text{AI}}) \cdot \vec k\right) \sin \left( \pi \nu_s^0\right)  \sin \left( \frac{1}{2} \chi_{\text{AI}}\right) \\
     & + \left(\Delta \vec c(\xi,\chi_{\text{AI}}) \cdot \vec k \right) \sin \left( \pi \nu_s^0\right)  \sin \left(  \frac{1}{2} \chi_{\text{AI}}\right) \,.
 \end{split}
 \label{eq:decomposition-with-misaligned-solenoids}
\end{equation}
\end{widetext}
Here emerges yet another systematic effect in the form of $\Delta \vec c(\xi,\chi_{\text{AI}})$, which denotes the change of the spin rotation axis in the ring caused by the orbit excursions. The approximation in the final form of Eq.\,(\ref{eq:decomposition-with-misaligned-solenoids}) is valid to a quadratic accuracy, which allows us to put $q_s^{\text{R}}(\xi,\chi_{\text{AI}}) \simeq \nu_s^0$. 

Then the first two terms in the last line of Eq.\,(\ref{eq:decomposition-with-misaligned-solenoids}) precisely reproduce the result for an ideally aligned solenoid, given earlier in Eq.\,(\ref{eq:2C.3}). The final result can be written as
\begin{widetext}
\begin{equation}
 \begin{split}
    \cos \left(\pi \nu_s^0\right) - \cos \left(\pi \nu_s(\xi,\chi_{\text{AI}})\right) \simeq & \cos \left( \pi \nu_s^0 \right) - \cos \left( \pi q_s^{\text{R}} (\xi=0,\chi_{\text{AI}}) \right)\\
    & + \pi \sin \left( \pi \nu_s^0 \right)  \Delta q^{\text{R}}_s(\xi,\chi_{\text{AI}}) \\
    & + \left(\Delta \vec c(\xi,\chi_{\text{AI}}) \cdot \vec k\right) \sin \left( \pi \nu_s^0 \right) \sin \left( \frac{1}{2} \chi_{\text{AI}} \right)\,,
     \end{split}
\end{equation}
\end{widetext}
where the last two terms can precisely be interpreted as the residuals, defined in Eq.\,(\ref{eq:residuals}),
\begin{equation}
\begin{split}
  \Delta \nu_s^{\text{res}} (\xi, \chi_{\text{AI}}) =  & \, \Delta q^{\text{R}}_s(\xi,\chi_{\text{AI}}) \\
  & + \frac{1}{\pi} \left(\Delta \vec c(\xi,\chi_{\text{AI}}) \cdot \vec k\right) \sin \left( \frac{1}{2} \chi_{\text{AI}}\right)\,.
 \end{split}
 \label{eq:residuals-of-spin-tune-map}
\end{equation}

COSY-Infinity allows us to evaluate the spin-transfer matrix per turn with full allowance of the beam orbit excursions,
\begin{equation}
\matr{T}(\xi,\chi_{\text{AI}}) = \matr{t}_{\text{R}} \Big(q_s^{\text{R}}(\xi,\chi_{\text{AI}}),\vec{c}(\xi,\chi_{\text{AI}})\Big) \matr{t}_{\text{S}}(\xi,\chi_{\text{AI}})\, . \label{eq:4B.1}
\end{equation}
Now we can evaluate the desired spin-transfer matrix of the ring without solenoid, 
\begin{equation}
\matr{t}_{\text{R}} \Big(q_s^{\text{R}}(\xi,\chi_{\text{AI}}),\vec{c}(\xi,\chi_{\text{AI}})\Big) = \matr{T}(\xi,\chi_{\text{AI}}) \matr{t}_{\text{S}}^{-1}(\xi,\chi_{\text{AI}})\,. \label{eq:4B.2}
\end{equation}
Next we compare the so-determined $\matr{t}_{\text{R}}$ to the COSY-Infinity result for the unperturbed ring matrix $\matr{t}_{\text{R}}(\nu_s^0,\vec{c})$, evaluated with the solenoid switched off. This allows us to estimate the effect of beam steering on the spin tune via 
\begin{equation}
  q_s^{\text{R}}(\xi,\chi_{\text{AI}})  = \underbrace{q_s^{\text{R}}(\xi=0,\chi_{\text{AI}})}_{= \nu_s^0} +  \Delta q_s^{\text{R}}(\xi,\chi_{\text{AI}})\,,
	\label{eq:definition-of-residuals-1}
\end{equation}
and on the orientation of the stable spin axis via
\begin{equation}
  \vec{c}(\xi,\chi_{\text{AI}})  = \vec{c}(0,\chi_{\text{AI}}) + \Delta \vec{c}(\xi,\chi_{\text{AI}})\,.  
\label{eq:definition-of-residuals-2}
\end{equation}

\begin{figure*}[htb]
\includegraphics[width=0.9\textwidth]{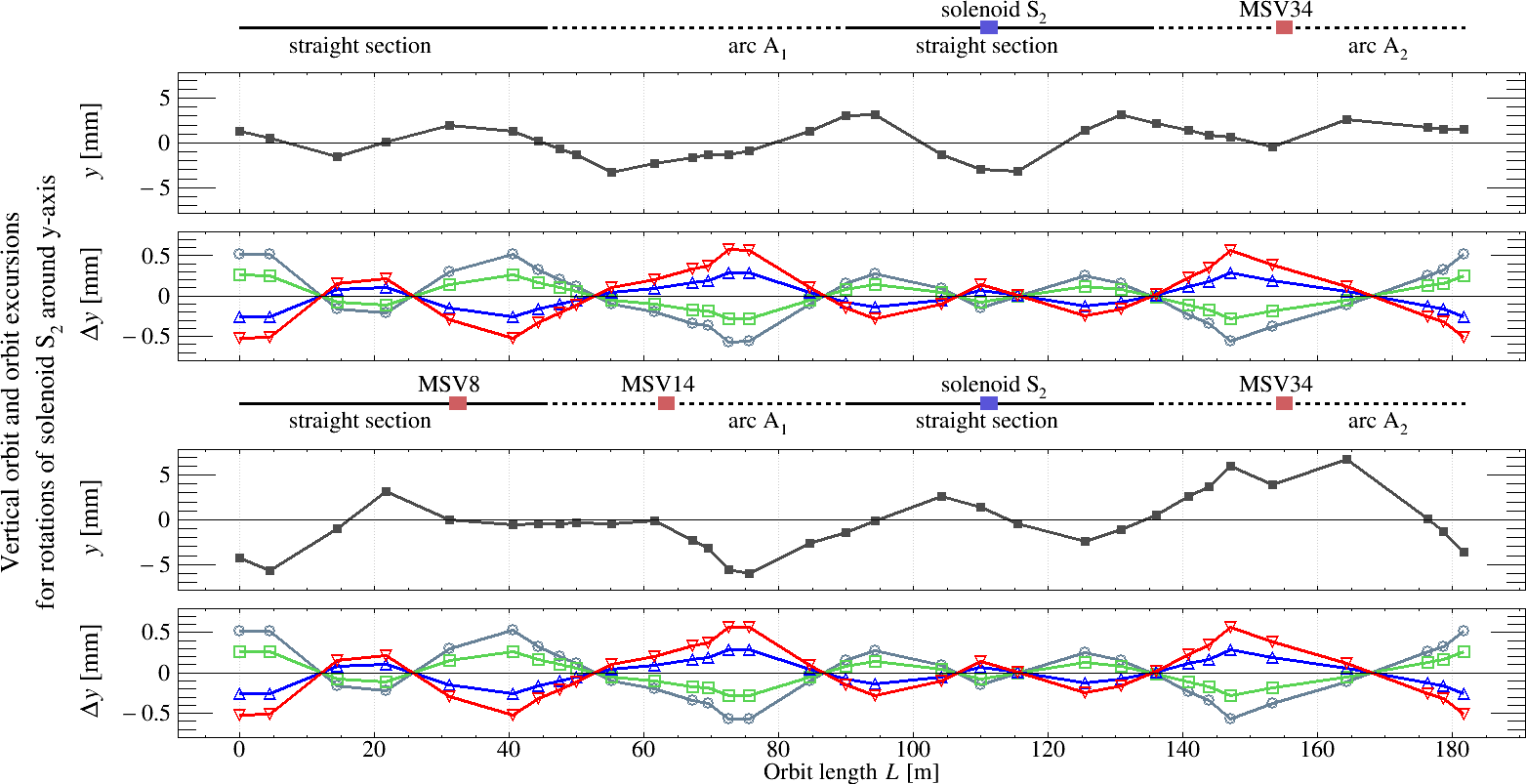}\vspace{0.2cm}
\includegraphics[width=0.9\textwidth]{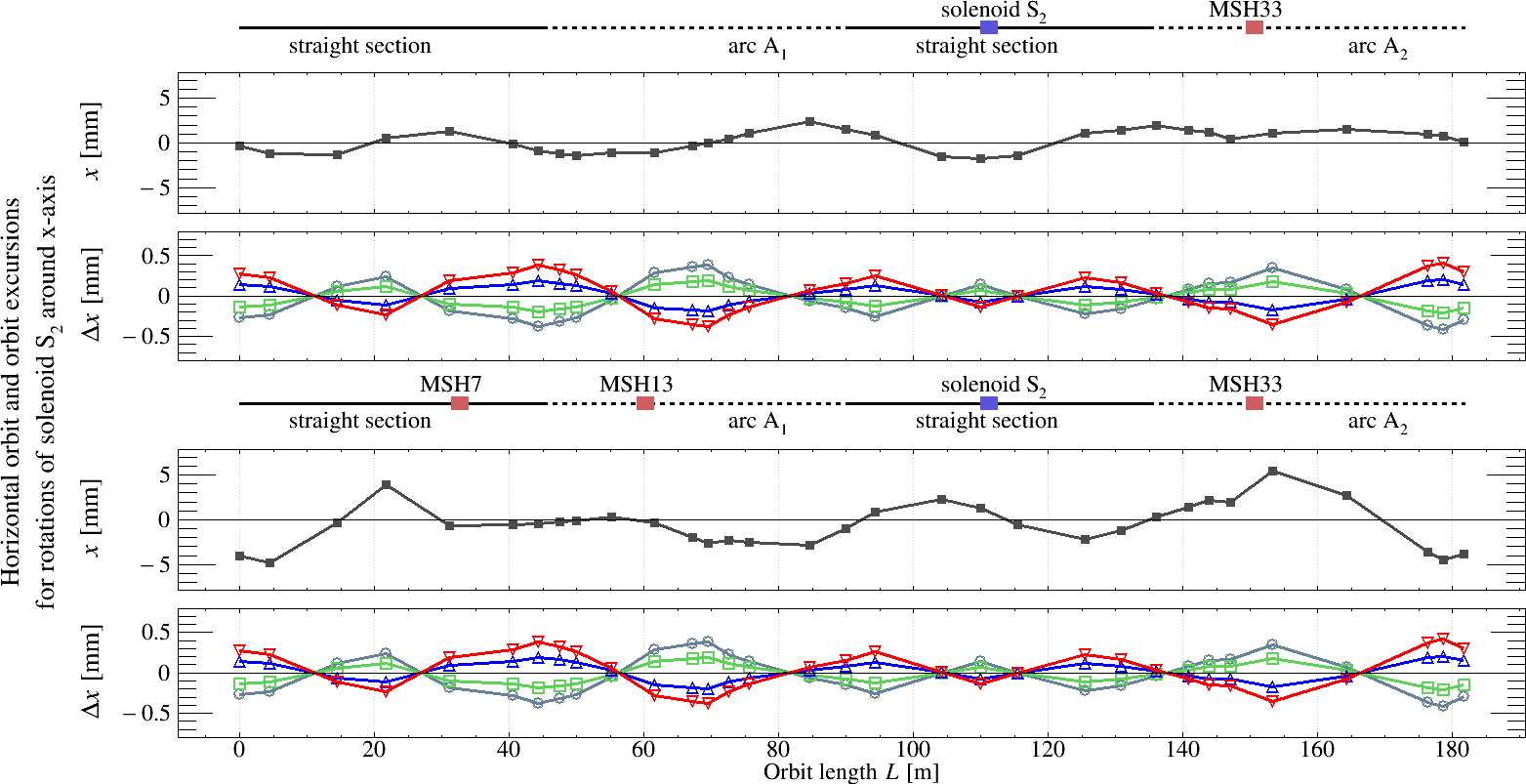}
\caption{\label{fig:simorb1} Top four panels: absolute beam positions $y\,[\text{mm}]$ indicate the vertical orbits for the vertical steerer sets 1 and 2 (Table\,\ref{tab:steerer-settings}), when solenoid $\text{S}_2$ is switched off. The panel for $\Delta y\,[\text{mm}]$ show the vertical orbit excursions with respect to the absolute beam positions  when solenoid $\text{S}_2$ is switched on with a spin kick angle $\chi_{\text{AI}}=\SI{12.98}{mrad}$, and rotated around the $y$-axis by $\xi_y = -8 (\textcolor{CadetBlue}{\Circle}), -4 (\textcolor{Green}{\square}), 4 (\textcolor{blue}{\triangle}), \text{ and } 8 (\textcolor{red}{\bigtriangledown})$\,mrad. The observed excursions are linear in $\xi_y$. Bottom four panels: Same as top four panels, but here for rotations of solenoid $\text{S}_2$ around the horizontal $x$-axis with $\xi_x = -8 (\textcolor{CadetBlue}{\Circle}), -4 (\textcolor{Green}{\square}), 4 (\textcolor{blue}{\triangle}), \text{ and } 8 (\textcolor{red}{\bigtriangledown})$\,mrad for the steerer sets 3 and 4. 
}
\end{figure*}

\begin{table*}[tb]
\renewcommand{\arraystretch}{1.3}
 \begin{ruledtabular}
  \begin{tabular}{clllr}
Orbit set  & Direction of beam excursion & Activated steerer  & Location           & Angle kick $\si{[mrad]}$\\ \hline 
    1      & vertical                    & MSV$_{34}$         & Arc $\text{A}_2$   & \SI{0.5}{mrad}  \\\hline
           & vertical                    & MSV$_8$            & Injection straight & \SI{0.5}{mrad}\\
    2      & vertical                    & MSV$_{14}$         & Arc A$_1$          & \SI{-0.5}{mrad}\\
           & vertical                    & MSV$_{34}$         & Arc A$_2$          & \SI{0.5}{mrad} \\\hline
    3      & horizontal                  & MSH$_{33}$         & Arc $\text{A}_2$   & \SI{0.5}{mrad} \\\hline
           & horizontal                  & MSH$_7$            & Injection straight & \SI{0.5}{mrad}   \\
    4      & horizontal                  & MSH$_{13}$         & Arc A$_1$          & \SI{-0.5}{mrad}\\
           & horizontal                  & MSH$_{33}$         & Arc A$_2$          & \SI{0.5}{mrad}  
\end{tabular}
 \end{ruledtabular}
 \caption{\label{tab:steerer-settings} The four orbit sets $1, 2, 3, 4$ used for the simulations with COSY-Infinity to evaluate the sensitivity of different orbit settings on the spin-transfer parameters of the ring.}
\end{table*}

\begin{figure*}[t]
 \includegraphics[width=0.9\textwidth]{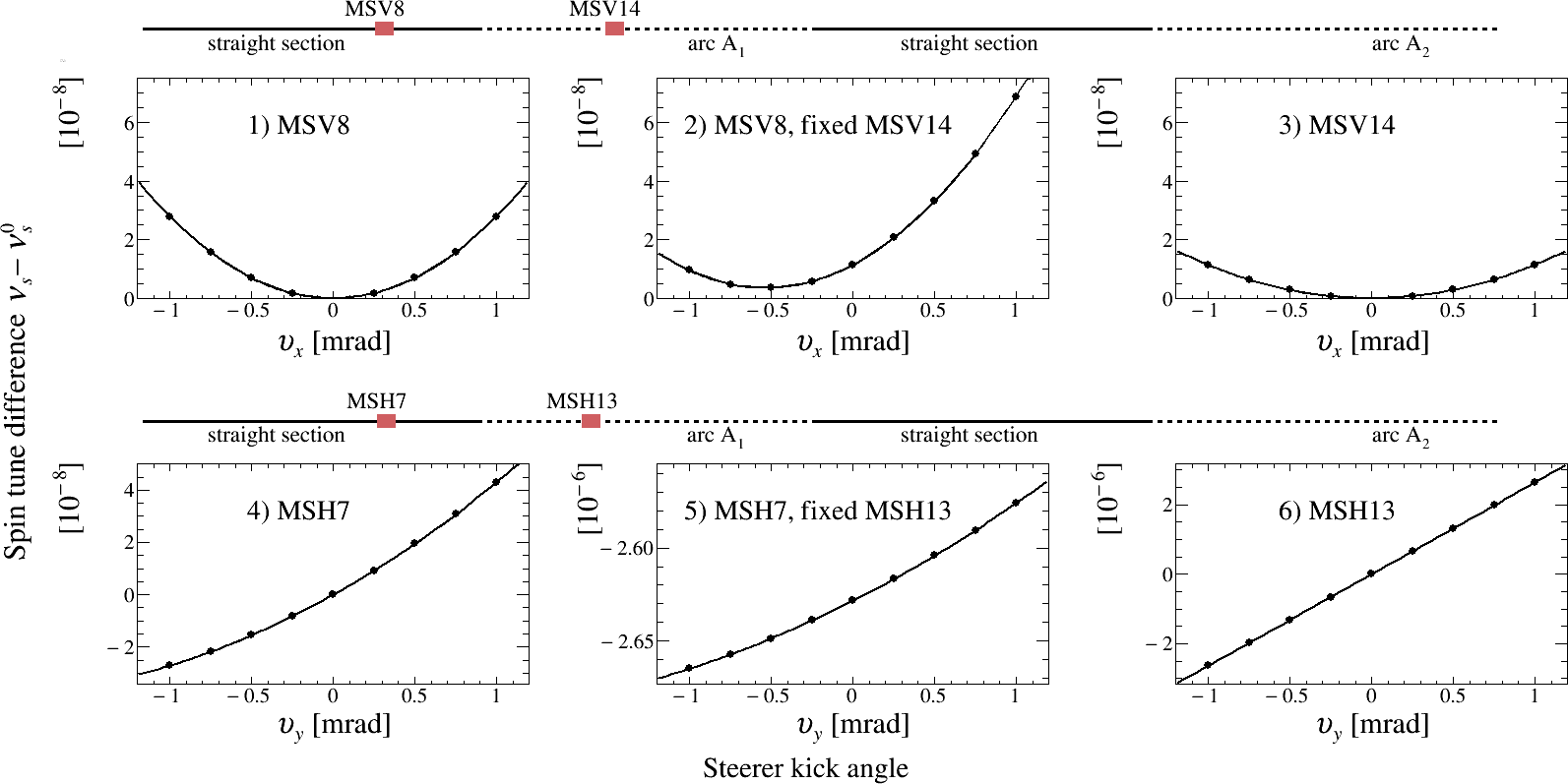}
\caption{\label{fig:steerer-simulations} 
Simulation of the spin tune difference $\nu_s - \nu_s^0$ using COSY-Infinity for various combinations of beam steerers in arc A$_1$ and in the injection straight section. The top (bottom) panel shows $\nu_s - \nu_s^0$ for vertical (horizontal) steerers as function of the steerer kick angle $\vartheta_y$  ($\vartheta_x$). The largest effect is observed when a horizontal steerer in the arc is activated, whereby $\nu_s - \nu_s^0$ is increased by about two orders in magnitude compared to vertical steerers anywhere or horizontal steerers in the straight sections.}
\end{figure*}

As soon as the misaligned solenoid is switched on, it will change the beam orbit all over the ring, the solenoid itself included. Because of the orbit excursions inside the \SI{120}{keV} electron cooler, its description as a simple solenoid $\text{S}_2$  already constitutes an approximation. This entails a {\it caveat} of Eq.\,(\ref{eq:4B.2}) -- the use of $\matr{t}_{\text{S}}^{-1}(\xi,\chi_{\text{AI}})$ evaluated assuming a fixed orientation of the solenoid axis $\vec k$ with respect to the beam axis.  Arguably, even so the above approximations provide a {\it qualitative} idea on $\Delta q_s^{\text{R}}(\xi,\chi_{\text{AI}})$ and on $\Delta \vec{c}(\xi,\chi_{\text{AI}})$. 
 
\subsection{Scaling properties of the orbit excursion effects on the spin transfer}
\subsubsection{Orbit settings}
\label{sec:orbit-settings}
First we need to understand to which extent the steering effect of the solenoid depends on the unknown orbit settings. The default orbit of COSY-Infinity corresponds to a situation when all steerers are turned off. The best we can do at this point is to compare the simulation results for different combinations of vertical and horizontal steerers. In the following, we use for the discussion the four specific orbit settings, listed in Table\,\ref{tab:steerer-settings}.



\subsubsection{Scaling properties of orbit excursions vs. solenoid misalignment}
\label{sec:scaling-properties-of-orbit-excursions}
In the top four panels of Fig.\,\ref{fig:simorb1}, we compare the initial beam orbits and the orbit excursions for the two orbit sets 1 and 2 (see Table\,\ref{tab:steerer-settings}) of the vertical steerers found using simulations with COSY-Infinity.  Solenoid $\text{S}_2$ with a spin kick of $\chi_2 = \SI{12.98}{mrad}$  is rotated around the $y$-axis in the angular range of $\SI{-8}{mrad} <\xi_y < \SI{8}{mrad}$ in steps of $\SI{4}{mrad}$. The steering effect of the rotated solenoid causes vertical orbit excursions.  The principal finding is that despite the striking difference between the two orbits, the corresponding excursions are identical, which evidently stems from the linear beam optics. 

The orbit excursions are also proportional to the strength of the AI induced by $\text{S}_2$, \textit{i.e.}, they scale as $\xi_y\chi_2$.  We also checked that the beam orbit excursions do not change when the beam rotation angles in individual steerers are varied. The horizontal orbit excursions from the vertical steering effect of the solenoid are at least two orders in magnitude smaller than the vertical ones and can safely be neglected.

The bottom four panels show the results when the horizontal steerer sets 3 and 4 (see Table\,\ref{tab:steerer-settings}) are activated, and the solenoid is rotated around the $x$-axis by an angle $\xi_x$. Again, the simulated orbit excursions exhibit a similar linear behavior of the orbit, as for the vertical steerer sets 1 and 2 -- the beam orbit excursions do not depend on the initial orbit setting. Compared to the horizontal orbit excursions, the vertical steering effect of the solenoid is negligibly small.

It should be noted that the pattern of variations of the observed orbit distortions along the ring, as depicted in Fig.\,\ref{fig:allbpm-x1-and-x2}, is in general quite consistent with the pattern exhibited by the simulated orbit distortions, shown in Fig.\,\ref{fig:simorb1}. The corresponding solenoid misalignment angles can be estimated as 
 \begin{align}
  \begin{aligned}
     {\xi_y}({\text{S}_1}) & \sim  1 - \SI{1.5}{mrad}, \\
     {\xi_y}({\text{S}_2}) & \sim  \SI{8}{mrad}\,,
  \end{aligned}   
  && 
  \begin{aligned}
  {\xi_x}({\text{S}_1})    & \sim \SI{6}{mrad}\,,\\
  {\xi_x}({\text{S}_2})    & \sim 6 - \SI{8}{mrad}\,.
  \end{aligned}
  \label{eq:solenoid-misalignment-angles}
 \end{align}
A full-fledged COSY-Infinity simulation of the complicated magnetic field structure of solenoid S$_2$ is not yet available, therefore we are limited to the semi-quantitative estimates, given in Eq.\,(\ref{eq:solenoid-misalignment-angles}).

\subsubsection{Ring steerers and baseline spin transfer parameters} 
\label{sec:ring-steerers}
As a prelude to the numerical simulations of the steering effect of a misaligned solenoid, we first look at the effect of ring steerers on the spin transfer properties of the ring. The COSY-Infinity simulations of the impact of steerers were performed starting with the default ideal orbit when all steerers are turned off, \textit{i.e.}, when $\nu_s^0 =G\gamma$. 

The strength of the ring steerers is conveniently defined by the momentum rotation angle (kick) $\vartheta_{x,y}$. Correspondingly, for a misaligned solenoid it is given by Eq.\,(\ref{eq:momentum-bending-in-solenoids}). We reiterate that horizontal steering is caused by momentum rotations by the angle $\vartheta_y$ around the vertical magnetic field $B_y$ of a steerer (and vertical steering by $\vartheta_x$ in the horizontal magnetic field $B_x$). 

Evidently, the non-commutation of spin rotations in the horizontal magnetic field of vertical steerers and in the vertical magnetic field of dipole magnets is similar to that in the solenoid and dipole fields. Therefore, vertical steerers would generate magnetic imperfections $c_{x,z}$ and spin tune shifts, similar to those generated by the solenoids [see Appendix\,\ref{sec:appendixA}, Eqs.\,(\ref{eq:spin-tune-formula}) and (\ref{eq:App-A.9})]. 

We demonstrate this property in panels 1 and 3 of Fig.\,\ref{fig:steerer-simulations}, where we show the spin tune shifts versus the momentum rotation angle $\vartheta_x$ for vertical steerer magnets activated in the injection straight section and arc A$_1$ (the corresponding spin rotation angles equal $\chi_x \simeq \nu_s^0 \vartheta_x$). We also show the results when $\vartheta_x$ using one steerer in a straight section is varied on the background of one steerer in arc A$_1$, operated at a fixed $\vartheta_x =  \SI{1}{mrad}$. The fixed steerer in the arc shifts the location  of the minimum of the spin tune shift vs. $\vartheta_x $, whereas the coefficient of the quadratic term in panels 1 and 2 is the same to better than 1\% accuracy.  

The case of the horizontal steerers is quite different. Here, the magnetic fields of the ring steerers and the dipole fields in the arcs are pointing along the same $y$ direction. The spin rotation in the horizontal steerer will closely follow the momentum rotation. A naive estimate for the shift of the spin tune 
\begin{equation}
\nu_s - \nu_s^0  \simeq \frac{\vartheta_y}{2\pi}\,. 
\label{eq:naive}
\end{equation}
is wrong for the reason that the momentum rotation in the steerer will be corrected for by the horizontal focusing fields, and because of the vanishing  dispersion in the straight section the orbit lengthening and corresponding energy shift by the horizontal steerer kick have a negligible impact on the spin tune. Indeed, the results from simulations with COSY-Infinity, shown in panel 6 of Fig.\,\ref{fig:steerer-simulations}, indicate a very strong suppression with respect to the naive Eq.\,(\ref{eq:naive}), 
\begin{equation}
\nu_s - \nu_s^0 \approx 1.5 \cdot 10^{-2} \frac{\vartheta_y }{2\pi}\, .
\end{equation}
For a steerer in the straight section, the effect computed using COSY-Infinity is about two orders of magnitude smaller than for a steerer in the arc, where the horizontal dispersion takes the largest value. We conclude that in contrast to the horizontal magnetic field $B_x$ of the vertical steerers, the vertical magnetic fields $B_y$ of the horizontal steerers do not affect the stable spin axis.

For the case of our interest, the steerer effect of misaligned solenoids, located in the straight sections, amounts to  $\vartheta_y \approx \xi_x\chi_{\text{AI}}/(1+G) < \SI{0.1}{mrad}$. Correspondingly, the results shown in panel 4 indicate the expected spin tune shift from solenoid rotations around the $x$-axis way below the uncertainty with which the spin tune jumps can be determined [see Eq.\,(\ref{eq:3B.2})].

\begin{figure*}[t]
 \includegraphics[width=1\textwidth]{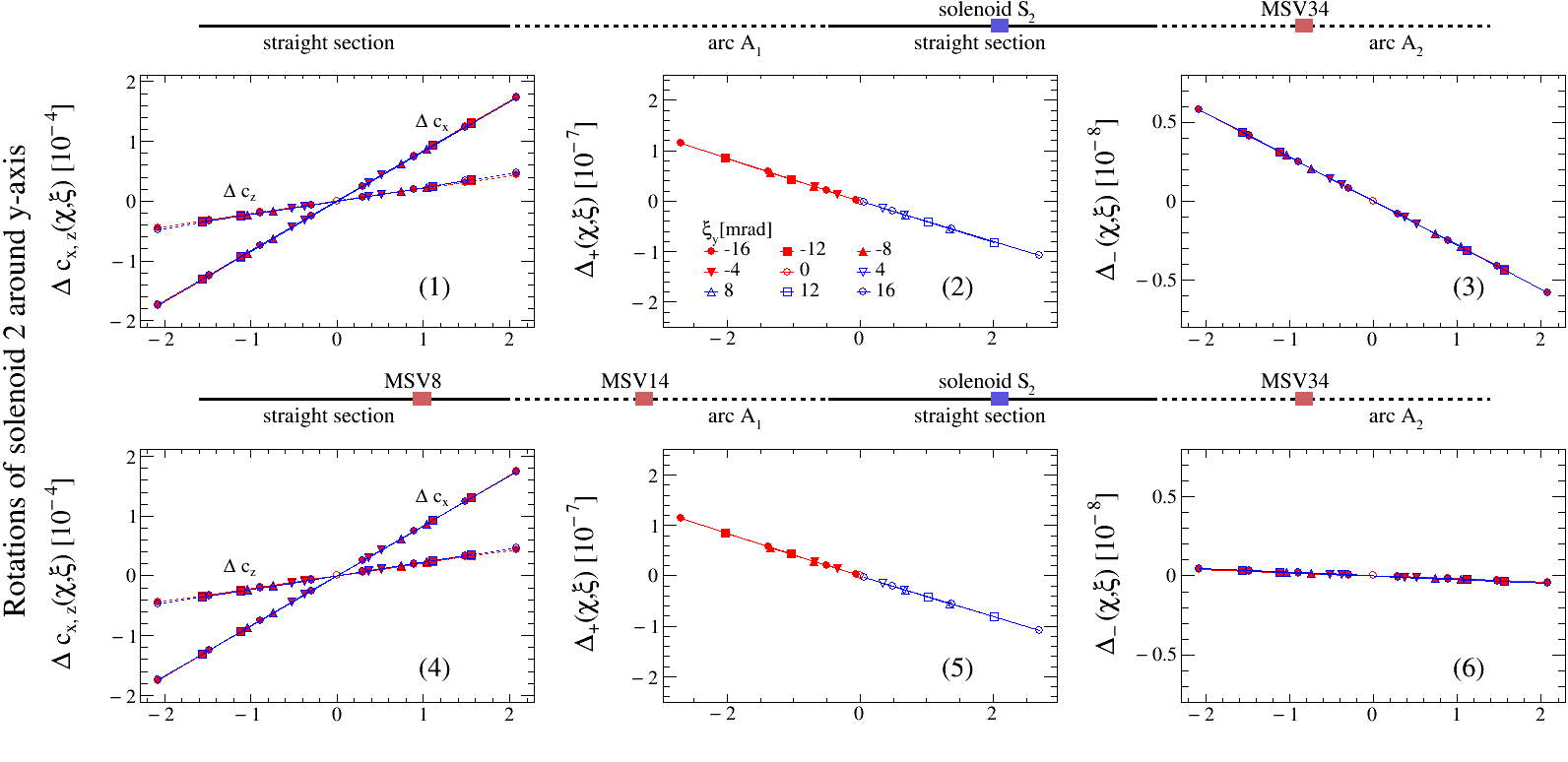}\vspace{0.2cm}
 \includegraphics[width=1\textwidth]{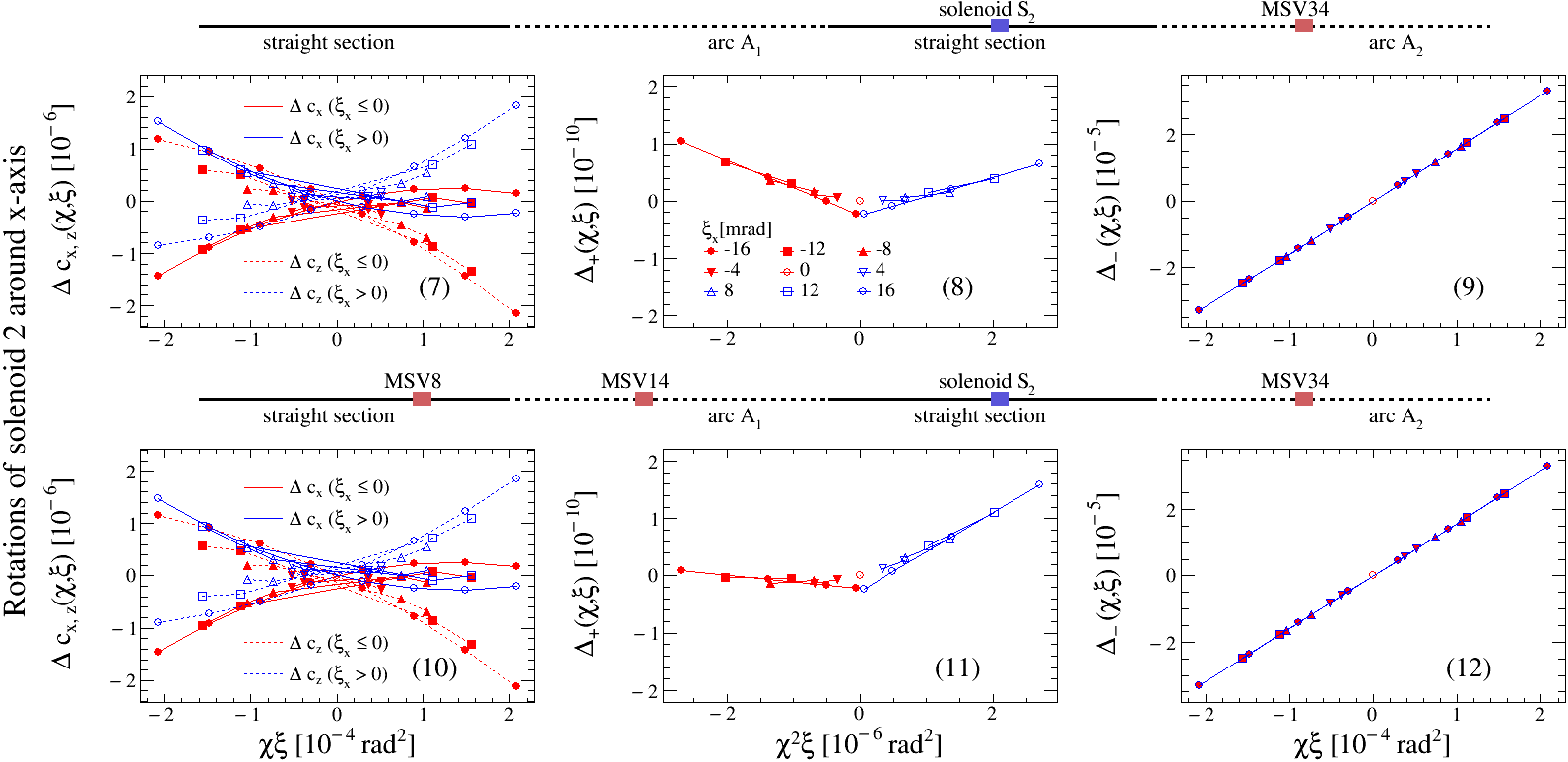}
\caption{\label{fig:fit2vecTwo_Ry-and-Rx} 
Perturbation of the spin transfer matrix $\matr{t}_{\text{R}}$ [see Eq.\,(\ref{eq:4B.2})] due to closed orbit excursions (see Fig.\,\ref{fig:simorb1}) by the steering fields of the misaligned solenoid $\text{S}_2$. Panels $1-6$ show the effect of rotations of solenoid S$_2$ around the $y$-axis. The first row of panels $1-3$ shows $\Delta c_{x,z}$ [Eq.\,(\ref{eq:slope-parameter-C})], and $\Delta_\pm$ [Eq.\,(\ref{eq:delta-plus-minus})] for orbit set 1 (Table\,\ref{tab:steerer-settings}). The second row of panels $4-6$ show the same parameters for orbit set 2. The corresponding situation for rotations of S$_2$ around the $x$-axis is depicted in panels $7-12$. Note that in order to emphasize the different scaling properties, $\Delta c_{x,z}$ (first column of panels) and $\Delta_-^{\,x,y}$ (third column) are plotted as a function of the variable $\chi\xi$, while $\Delta_+$ (second column) is plotted as a function of the variable $\chi^2\xi$.}
\end{figure*}

\subsubsection{Scaling properties of the stable spin axis $\vec c$ vs. solenoid misalignment}
\label{sec:scaling-properties-of-the-stable-spin-axis}

Next we turn to the second systematic effect, outlined in Eq.\,(\ref{eq:decomposition-with-misaligned-solenoids}), \textit{i.e.}, the change of the spin rotation axis in the ring caused by the orbit excursions. Specifically, in panels 1, 4, 7, and 10 of Fig.\,\ref{fig:fit2vecTwo_Ry-and-Rx}, we show the horizontal and longitudinal projections of $\Delta \vec{c}(\xi,\chi_{\text{AI}})$, determined from simulations using COSY-Infinity for  the misaligned solenoid and different steerer settings. The angles of solenoid rotation $\xi_y$ around the $y$-axis are indicated in panel 2, the ones for rotation around the $x$-axis ($\xi_x$) in panel 8. For the largest $\xi_{x,y}=\pm \SI{16}{mrad}$, we show the simulation results for nine equally spaced values of $\chi_{\text{AI}} \in [-12.98,12.98]\,\si{mrad}$. For smaller values of $\xi_{x,y}$, in order to avoid overlapping of points, we show only the four results for the outmost values of $\chi_{\text{AI}}$, two for $\chi_{\text{AI}}>0$ and two for $\chi_{\text{AI}}<0$. In principle, $\Delta c_{x,z}$ are functions of the two variables $\xi$ and $\chi_{\text{AI}}$. 

However, as illustrated in Fig.\,\ref{fig:fit2vecTwo_Ry-and-Rx}, the steering effect is indeed proportional to the product of the two, \textit{i.e.}, described by
\begin{equation}
 {\Delta\vec c}^{\,y} (\xi,\chi_{\text{AI}}) = \vec C^y \xi \chi_{\text{AI}}\,,
 \label{eq:slope-parameter-C}\end{equation}
where $\vec C^y = (C_x^{\,y},C_z^{\,y})$ describes the slope parameters. Close inspection of panels 1 and 4, which correspond to a rotation of the solenoid around the $y$-axis, shows that the slopes $C_{x,z}^{\, y}$ remain unchanged going from one orbit setting to another one. 

In Sec.\,(\ref{sec:ring-steerers}) we argued that the horizontal steerers should not affect the spin stable axis. Indeed,  numerically, ${\Delta\vec c}^{\,x} (\xi,\chi_{\text{AI}})$  shown in panels 7 and 10, are  found to be about two orders of magnitude smaller than ${\Delta\vec c}^{\,y} (\xi,\chi_{\text{AI}}) $, shown in panels 1 and 4. The origin of the very small ${\Delta\vec c}^{\,x} (\xi,\chi_{\text{AI}})$ can be attributed to the coupling of the horizontal and vertical betatron motion by the weak spiraling of the beam trajectory  in the misaligned solenoid [see Eq.\,(\ref{eq:momentum-bending-in-solenoids})].  We note that ${\Delta\vec c}^{\,x} (\xi,\chi_{\text{AI}})$ does not depend on the orbit setting, as evidenced by the comparison of panels 7 and 10. For the purposes of the subsequent analysis of the residuals of the spin tune map, the effect of ${\Delta\vec c}^{\,x} (\xi,\chi_{\text{AI}})$ can be neglected. 
 
\subsubsection{Scaling properties of the spin-phase advance in the ring vs. solenoid misalignment}
\label{sec:scaling-properties-of-spin-phase-advance}
It is convenient to define symmetric and anti-symmetric combinations of $\Delta q_s^{\text{R}}(\xi,\chi_{\text{AI}})$, introduced in Eq.\,(\ref{eq:definition-of-residuals-1}) via
\begin{equation}
 \Delta_{\pm}(\xi,\chi_{\text{AI}}) = \frac{1}{2} \left[ \Delta q_s^{\text{R}}(\xi,\chi_{\text{AI}}) \pm  \Delta q_s^{\text{R}}(\xi,-\chi_{\text{AI}}) \right]\,.
 \label{eq:delta-plus-minus}
\end{equation}
As we shall see, these functions exhibit different scaling properties as function of $\xi$  and $\chi_{\text{AI}}$. 

The simulation results for $\Delta_{+}(\xi,\chi_{\text{AI}})$, shown in panels 2 and 5 of Fig.\,\ref{fig:fit2vecTwo_Ry-and-Rx}, suggest the scaling law
\begin{equation}
 \Delta_{+}^{\, y}(\xi,\chi_{\text{AI}}) = D^{y} \xi {\chi_{\text{AI}}}^2\,.
  \label{eq:scaling-parameter-D}
\end{equation}
The slope $D^{y}$ is consistent with a constant, which is \textit{independent} of the orbit settings. This is evident from panels 2 and 5 for solenoid rotations around $y$. A summary of the simulation results for the slope parameter $D^{y}$ is found in Table \ref{tab:COSY-Infinity-fit-parameters}. 

The simulation results for the anti-symmetric combination $\Delta_{-}^{\, y}(\xi,\chi_{\text{AI}})$ are shown in panels 3 and 6 for solenoid rotations around the $y$-axis. These results suggest the scaling law 
\begin{equation}
 \Delta_{-}^{\, y}(\xi,\chi_{\text{AI}}) = E^y \xi_y \chi_{\text{AI}}\,.
 \label{eq:scaling-parameter-E}
\end{equation}
One notices, however, that $\Delta_{-}^{\, y}(\xi,\chi_{\text{AI}})$ is no longer independent of the orbit setting. 

\begin{table*}[tb]
\renewcommand{\arraystretch}{1.2}
 \begin{ruledtabular}
  \begin{tabular}{clcccrr}
Orbit set & Direction of beam excursion & Solenoid & \multicolumn{4}{c}{Parameters from COSY-Infinity}\\
          &            &          & $C_x^{\,y}$    & $C_z^{\, y}$     & $D^y$\,$[\num{e-2}]$  & $E^y$\,$[\num{e-6}]$ \\\hline
  1       & vertical   & S$_2$    & $0.8392$       & $0.2210$         & $\num{-4.134}$        & $\num{28}$  \\
  2       & vertical   & S$_2$    & $0.8389$       & $0.2210$         & $\num{-4.136}$        & $\num{2}$   \\
  3       & horizontal & S$_2$    & $0.8387$       & $0.2195$         & $\num{-4.136}$        & $0$            \\
  4       & horizontal & S$_2$    & $0.8393$       & $0.2205$         & $\num{-4.131}$        & $0$            \\\hline
  2       & vertical   & S$_1$    & $0.9455$       & $0.1475$         & $\num{-3.827}$        & $\num{-23}$ \\
  4       & horizontal & S$_1$    & $0.9441$       & $0.1464$         & $\num{-3.824}$        & $0$            \\
\end{tabular}
\end{ruledtabular}
 \caption{\label{tab:COSY-Infinity-fit-parameters} Summary of parameters of modifications of the spin transfer properties of the COSY ring by the orbit excursions induced by different steerer settings (listed in Table\,\ref{tab:steerer-settings}) as determined from the COSY-Infinity simulations for rotations of solenoids S$_1$ and S$_2$ around the vertical $y$-axis. Here $C^y_{x,z}$ are defined by Eq.\,(\ref{eq:slope-parameter-C}), $D^y$ is defined by Eq.\,(\ref{eq:scaling-parameter-D}), and $E^y$ is defined by Eq.\,(\ref{eq:scaling-parameter-E}). To emphasize the independence on the orbit setting, for solenoid S$_2$ we show results for all sets 1 to 4, while for solenoid S$_1$, we restricted ourselves to sets 2 and 4. }
\end{table*}

The case of rotations around the $x$-axis, shown in the two bottom rows of Fig.\,\ref{fig:fit2vecTwo_Ry-and-Rx}, is a special one. As mentioned in Sec.\,\ref{sec:ring-steerers} (and as shown in the upper panels of Fig.\,\ref{fig:steerer-simulations}), the horizontal steerers in the straight section have a very weak influence  on the spin tune. The spin kick in the vertical magnetic field of the misaligned solenoid, given by 
\begin{equation}
\chi^y = \frac{\xi_x\chi_{\text{AI}}}{2 \pi}\,,
\label{eq:chi_y-for-y-rotations}
\end{equation}
is to a high precision canceled by the combined action of the rest of the first straight section,  the arcs and the second straight section. Indeed, for rotations around the $x$-axis,  $|\Delta_+^{\,x}(\chi,\xi)|$ (shown in panels 8 and 11 of Fig.\,\ref{fig:fit2vecTwo_Ry-and-Rx}) is three orders of magnitude smaller than $|\Delta_{+}^{\, y}(\chi,\xi)|$ (shown in panels 2 and 5). Technically, the cancellation of the spin kick in the rest of the ring 
entails a fairly large slope of the antisymmetric $|\Delta_-^{\,x}(\chi,\xi)|$, 
\begin{equation}
E^x \approx \frac{1}{2\pi} \, ,
\label{eq:E-for-y-rotations}
\end{equation}
which is consistent with the numerical results shown in panels 9 and 12.

\subsection{Orbit excursion effects from solenoids S$_1$ vs. S$_2$}
\label{sec:orbit-excursions-from-solenoid-S1-vs-solenoid-S2}
Let the two solenoids S$_1$ and S$_2$ be located in the two straight sections exactly opposite to each other, and let the COSY ring be a precisely symmetric one. Then from the perspective of COSY-Infinity, simulations with activated S$_1$ would have differed from the case of activated S$_2$ only by the relative positions of the activated solenoid to the activated steerers, \textit{i.e.}, by the orbit setting. Consequently, in the case of perfect symmetry, the slopes $C_{x,z}$ and $D$, which were determined from the COSY-Infinity simulations for the activated solenoid S$_{2}$, would hold for solenoid S$_1$ as well. 

In COSY  the solenoids S$_1$ and S$_2$ are at asymmetric locations with different $\beta$-functions. Correspondingly, for the same values of $\xi\chi$, the beam orbit excursions could change from an activated S$_{1}$ to an activated S$_2$. Indeed, this is evident from a comparison of the patterns of beam orbit excursions along the ring, shown in the top and bottom panels of Fig.\,\ref{fig:allbpm-x1-and-x2}, where solenoids S$_1$ and S$_2$ were activated individually. Similarly, the steering effect in the spin transfer properties of the ring  from S$_{1}$ could be different than from  those induced by S$_2$. 

The findings for rotations of solenoids S$_1$ and S$_2$ around the $y$-axis are different. The last two entries in Table\,\ref{tab:COSY-Infinity-fit-parameters} list the results for two orbit settings, one with a vertical and another one with a  horizontal orbit steerer set. The scaling properties of the steering effect of solenoid S$_1$ exhibits the same independence on the orbit setting as the one observed for S$_2$. However,  the absolute values of the spin transfer parameters $C_{x,z}^y$ and $D^y$ for the two solenoids are different.  

\subsection{Spin tune mapping with allowance for misalignment of solenoids}
\label{sec:spin-tune-mapping-with-allowance-for-misalignment-of-solenoids}
Taking into account the considerations of Sec.\,\ref{sec:fitting-the-map-of-residuals} and the above Secs.\,\ref{sec:scaling-properties-of-orbit-excursions} to \ref{sec:scaling-properties-of-spin-phase-advance} for guidance, we proceed further with the analysis of the recorded spin-tune map by substituting in Eq.\,(\ref{eq:2C.8}) 
\begin{equation}
\chi_{\pm} \to \tilde\chi_\pm = \frac{1}{2}(k_1\chi_1 \pm k_2\chi_2)\,.
\label{eq:rescaling}
\end{equation}
The single-solenoid simulations using COSY-Infinity, which strongly suggest this rescaling, could have missed a cross talk of the two solenoids, caused by the orbit excursions, and in our final analysis, we fit the spin tune jump maps to the following function,
\begin{equation}
\begin{split}
 \cos(\pi &\nu_s^0) -  \cos\left(\pi \left[\nu_s^0 + \Delta\nu_s(\chi_1,\chi_2)\right]\right) \\ 
 = & \phantom{-} \,\left[1 + \cos\left(\pi \nu_s^0\right) \right] \sin^2 \left(\frac{1}{2} (k_1\chi_1 + k_2\chi_2)\right)\\
   &- \left[1-\cos \left(\pi \nu_s^0\right) \right]\sin^2\left(\frac{1}{2}(k_1\chi_1 - k_2\chi_2)\right)\\
   &-\frac{1}{2}a_+^* \sin(\pi \nu_s^0)\sin (k_1\chi_1 + k_2\chi_2)\\
   & +\frac{1}{2}a_-^*\sin(\pi \nu_s^0)\sin (k_1\chi_1 - k_2\chi_2) \\
   & + \frac{1}{4}F \chi_1\chi_2\,.
\end{split}
\label{eq:master-formula}
\end{equation}

Simple allowance for the above individual rescaling factors $k_{1,2}=1 + K_{1,2}$ without the cross-talk term $F=0$ already yields an acceptable Fit 1 with $\chi^2/N_{\text{dof}} = 226/56$ (see Table\,\ref{tab:rescaling-fit-parameters}). The allowance for the cross talk term $F\neq 0$ between the two solenoids S$_1$ and S$_{2}$ leads to a still better quality Fit 2 with $\chi^2/N_{\text{dof}} = 137/55$.
\begin{table}[tb]
\renewcommand{\arraystretch}{1.2}
\begin{ruledtabular}
\begin{tabular}{lccc}
Parameter                   & Unit                 & \multicolumn{2}{c}{Fit}                \\ 
                            &                      & ($1$)         & ($2$)                  \\ \hline
$\chi^2/N_{\text{dof}}$     &                      & $226.4/56$    & $137.5/55$             \\ 
$a_+^*$                     & $[\num{e-7}]$        & $49362\pm 9$  & $49385\pm 10.5$          \\                      
$a_-^*$                     & $[\num{e-7}]$        & $-4545 \pm 6$ &  $-4464 \pm 10$         \\
$K_1^{\text{fit}}$          & $[\num{e-5}]$        & $-384 \pm 7$  & $-356 \pm 8$           \\
$K_2^{\text{fit}}$          & $[\num{e-5}]$        & $-574 \pm 9$  & $-508  \pm 11$         \\
$F$                         & $[\num{e-5}]$        & $0$ (fixed)   & $-133 \pm 14$          \\                        
\end{tabular}
\end{ruledtabular}
 \caption{\label{tab:rescaling-fit-parameters} Summary of fits to the spin tune map with allowance for orbit distortion effects. Fit 1 uses Eq.\,(\ref{eq:master-formula}) with $F=0$, while Fit 2 allows in addition for a cross talks of solenoids S$_1$ and S$_2$.}
\end{table}

\subsection{Understanding the residuals $\Delta\nu_s^{\text{res}}$ of the spin-tune map
}
\label{sec:understanding-the-residuals}

Now, we are in the position to compare the residuals of the spin tune map, given by the empirical fits to the experimentally observed spin tune map, and the ones suggested by COSY-Infinity simulations. Lumping together the found scaling representation for the different contributions in Eq.\,(\ref{eq:residuals-of-spin-tune-map}), yields a simulation result for a single solenoid of the form
\begin{equation}
 \begin{split}
  \Delta\nu_s^{\text{res}}(\text{sim}) \simeq & D \xi {\chi_{\text{AI}}}^2 + E\xi \chi_{\text{AI}} + \frac{1}{2\pi} \left(\vec C \cdot \vec k \right) \xi {\chi_{\text{AI}}}^2\\
  \simeq & \left( D + \frac{C_z}{2\pi} \right)  \xi {\chi_{\text{AI}}}^2 + E \xi \chi_{\text{AI}}\,,
	\label{eq:SimulatedResidual}
 \end{split}
\end{equation}
where we have approximated $(\vec C \cdot \vec k) \simeq (\vec C \cdot \vec e_z) =C_z$. Now recall the point that in the basic Eq.\,(\ref{eq:2C.8}), only the two terms proportional to $\sin^2 \left(\frac{1}{2}\chi_\pm \right)$ come with uniquely prescribed coefficients, while the terms $\propto E$ renormalize the unknown $a_\pm^*$ [see also Eq.\,\ref{eq:shift_of_apm})]. For that reason, it only makes sense to compare the quadratic term in Eq.\,(\ref{eq:SimulatedResidual}) to the corresponding term in the residuals extracted from the fit function, given in Eq.\,(\ref{eq:master-formula})
\begin{equation}
 \begin{split}
  \Delta\nu_s^{\text{fit}}(\chi_1, \chi_2=0) & \simeq \frac{1}{4} K_1^{\text{fit}}\cos(\pi \nu_s^0)\chi_1^2 \,, \\
  \Delta\nu_s^{\text{fit}}(\chi_1=0, \chi_2) & \simeq \frac{1}{4} K_2^{\text{fit}} \cos(\pi \nu_s^0 )\chi_2^2\,.
	\end{split}
	\label{eq:calibration-fit}
	\end{equation}

We recall that contributions to the spin tune shifts from rotations of solenoids around the $x$-axis are negligibly small. Then the COSY-Infinity simulation results for $K_i$ for the two solenoids S$_i$ ($i=1,2$) will be dominated by the contribution from the solenoid rotations around the $y$-axis, described by
\begin{equation}
K_i^{\text{sim}}\approx \frac{4}{\cos(\pi \nu_s^0)} \left( D^y (\text{S}_i) + \frac{C_z^y (\text{S}_i)}{2\pi} \right) \xi_y(\text{S}_i)\, ,
\label{eq:calibration-simulation}
\end{equation}	
where S$_i$ as an argument denotes the parameters of the respective solenoid.

The evaluation of the vertical steering effect, using for the misalignment angles the estimates given in Eq.\,(\ref{eq:solenoid-misalignment-angles}), yields
\begin{equation}
\begin{split}
K_1^{\text{sim}} & \approx - 10 \cdot \num {e-5} \,, \\
K_2^{\text{sim}} & \approx - (17-23) \cdot \num{e-5}\, .
\label{eq:calibration-simulation-results}
\end{split}
\end{equation}
There are substantial cancellations between the contributions from $D^y$ and $C_z^y$ on the right-hand side of Eq.\,(\ref{eq:calibration-simulation}). The effect is larger for solenoid S$_2$. The simulation results have the same sign but are one order in magnitude smaller than the fitted $K_i^{\text{fit}}$, listed in Table\,\ref{tab:rescaling-fit-parameters}.

\subsection{Simulation of two tilted solenoids in an otherwise ideal ring}

Take a toy model consisting of two solenoids S$_1$ and S$_2$ that are embedded in a ring with ideally aligned magnetic elements and with all ring steerers turned off, so that along the complete orbit the stable spin axis $\vec c = \vec e_y$ and  $\nu_s^0 = G\gamma$. Next we assign to the two solenoids the rotation angles given in Eq.\,(\ref{eq:solenoid-misalignment-angles}), \textit{i.e.}, $\xi_y(\text{S}_1) = \SI{1.5}{mrad}$, $\xi_x(\text{S}_1) = \SI{6}{mrad}$, and $\xi_y(\text{S}_2) = \SI{8}{mrad}$, $\xi_x(\text{S}_2) = \SI{8}{mrad}$.

Now COSY-Infinity is used to generate a grid of $9 \times 9 = 81$ spin tune jumps $\Delta\nu_s$ as function of the same $\chi_1$ and $\chi_2$ as those used to produce Map\,2, and each of the generated spin tune jumps is associated to a statistical error of $\delta \Delta\nu_s^{\text{syst}} = \num{3.23e-9}$ [see Eq.\,(\ref{eq:3B.2})]. Fitting the so simulated spin-tune map by Eq.\,(\ref{eq:master-formula}) yields the results summarized in Table\,\ref{tab:toy-model}.
\begin{table}[tb]
\renewcommand{\arraystretch}{1.2}
\begin{ruledtabular}
\begin{tabular}{lccc}
Parameter                   & Unit                 & \multicolumn{2}{c}{Fit}                 \\ 
                            &                      & ($1$)         & ($2$)                   \\ \hline
$\chi^2/N_{\text{dof}}$     &                      & $78.4/77$    & $75.5/76$                \\ 
$a_+^*$                     & $[\num{e-7}]$        & $-6.4\pm 9.4$  & $-15.3\pm 10.8$         \\                      
$a_-^*$                     & $[\num{e-7}]$        & $26.6 \pm 6.5$ &  $12.1 \pm 10.8$     \\
$K_1^{\text{fit}}$          & $[\num{e-5}]$        & $-8 \pm 6.5$  & $-11 \pm 6.7$           \\
$K_2^{\text{fit}}$          & $[\num{e-5}]$        & $-53 \pm 8.4$  & $-68 \pm 12$           \\
$F$                         & $[\num{e-5}]$        & $0$ (fixed)   & $25 \pm 15$             \\                        
\end{tabular}
\end{ruledtabular}
  \caption{\label{tab:toy-model} Fit parameters of a COSY-Infinity simulation using Eq.\,(\ref{eq:master-formula}) of the toy model with two tilted solenoids S$_1$ and S$_2$ embedded in an otherwise ideal ring.}
\end{table}

Firstly, the results show that the fits are of the expected good quality. Secondly, the so obtained $K_{1,2}^{\text{fit}}$ have the same sign as the earlier estimates $K_{1,2}^{\text{sim}}$, given in Eq.\,(\ref{eq:calibration-simulation-results}). Thirdly, the present $K_{1}^{\text{fit}}$ and the earlier estimated  $K_{1}^{\text{sim}}$ are of the same magnitude, although we do not see obvious reasons why $K_{2}^{\text{fit}}$ is about $3$ times larger than the estimate for  $K_{2}^{\text{sim}}$ of Eq.\,(\ref{eq:calibration-simulation-results}). Fourthly, within the error bars the fitted $a_\pm^*$ are consistent with zero, as expected for an ideal ring. This can be taken as additional evidence for the credibility of estimates based on simulations with COSY-Infinity. 

Finally, it should be noted that the pseudo-experimental data discussed in this section were generated assuming the true solenoids S$_{1,2}$. Fits to these pseudo data assumed either a fixed cross-talk parameter $F=0$, or treated $F$ as a free parameter yield basically identical $\chi^2/N_{\text{dof}}$. In the latter case, the resulting $F$ differs from zero by a mere $1.7 \sigma$. In contrast to that, the  experimental spin-tune map has been taken with the full-fledged magnet system of the \SI{120}{keV} cooler, which we attempted to approximate by a simple solenoid S$_2$. The results presented in Table\,\ref{tab:rescaling-fit-parameters} do clearly show that the cross talk parameter $F$ is $10\,\sigma$ away from zero. The allowance for this cross-talk term $F$ entails a substantial improvement of the fit quality of the measured spin-tune map from $\chi^2/N_{\text{dof}} = 226.4/56$  down to  $\chi^2/N_{\text{dof}} =137.5/55$. We also observe a simultaneous reduction of the magnitude of the fitted rescaling parameters $K_{1,2}^{\text{fit}}$.

We interprete the above comparison as a significant hint that the magnet system of the \SI{120}{keV} electron cooler, consisting of solenoids, toroids and steerers, is not yet properly implemented into COSY-Infinity. The approximation that this magnetic system can be treated as a simple solenoid is most likely in part responsible for the discrepancy between the simulated and empirically-determined results for $K_{1,2}$. Another part may stem from the fact that the simulations using COSY-Infinity do not take into account neither the finite emittance of a stored cooled beam (of about 1 to \SI{2}{\micro.m}\,\cite{PhysRevSTAB.18.020101}), nor the non-vanishing momentum dispersion $\delta p/p \simeq \num{e-4}$, nor the stochastic heating process used for extraction, and the associated gradual removal of particles from the periphery of the beam in the target. Nevertheless, the comparison of simulated and empirically-determined results provides an important insight into the significance of solenoid misalignment effects. 

In the future, it will be possible to go beyond the present interim solution. Recently, it was shown that the troublesome electron cooler magnets, including the involved steerer magnets, can be switched off on flattop without beam loss. 

\section{Interpretation of the results and possible applications of the spin tune mapping technique }
\label{sec:V}
The spin tune mapping determines the parameters $a_\pm^*$ [Eq.\,(\ref{eq:master-formula})]. The pitfalls of the makeshift two solenoid scheme contributes systematic uncertainties to the interpretation of those $a_\pm^*$. To get rid of these uncertainties, it is imperative to have an alignment of solenoids S$_{1,2}$ as ideal as possible, such that they do not disturb the beam orbit. 

According to the simulations based on COSY-Infinity, it is of prime importance to eliminate the vertical steering effect of the solenoids, {\it i.e.}, to keep the beam orbit a planar one. The parameters $a_\pm^*$ are projections of the stable spin axis $\vec c$ onto a plane spanned by the vectors $\vec n_1$ and ${\vec{n}_2}^{\,\text{r}}$.  This plane is very close to the ring plane, and the stable spin axis points in a direction very close to the normal of this plane, \textit{i.e.}, along the direction of $[\vec n_1 \times {\vec{n}_2}^{\,\text{r}}]$.  

We estimate the accuracy to which the projections $c_{x,z}$ onto the ring plane can be controlled, using the approximation of Eq.\,(\ref{eq:2C.5}), which entails
\begin{equation}
\begin{split}
(\vec c \cdot \vec n_1) &\approx c_z\,, \\
(\vec c \cdot \vec n_2^r) &\approx\cos(\pi\nu_s^0) c_z  - \sin(\pi\nu_s^0)c_x \, ,\\
a_\pm^* &\approx\cos(\pi\nu_s^0) c_z  - \sin(\pi\nu_s^0)c_x \pm c_z\,.
\end{split}
\end{equation}
Solving the last equation for $c_{x,z}$, we obtain
\begin{widetext}
\begin{align}
\delta c_z & \approx  \frac{1}{2}\Big\{(\delta {a_+^*})^2 +(\delta a_-^*)^2\Big\}^{\frac{1}{2}} && =  \num{0.7e-6}\, ,\\
\delta c_x & \approx  \frac{1} {|\sin(2\pi\nu_s^0)|}  \Big\{[1- \cos(\pi\nu_s^0)]^2 (\delta a_+^*)^2 +[1+\cos(\pi\nu_s^0)]^2 (\delta a_-^*)^2\Big\}^{\frac{1}{2}}&& =  \num{1.7e-6}\,,
\label{eq:orientation-errors}
\end{align}
\end{widetext}
where the $\delta {a_\pm^*}$ denote the uncertainties of ${a_\pm^*}$, as listed in Table\,\ref{tab:rescaling-fit-parameters}. Our principal finding can be summarized by stating that the angular orientation of the stable spin axis with respect to the plane defined by its normal vector $[\vec n_1 \times {\vec{n}_2}^{\,\text{r}}]$ can be determined to a {\it statistical} accuracy better than \SI{1.7}{\micro.rad}.

A slight trouble  with the two-solenoid scheme is that the exact orientation of the normal vector $[\vec n_1 \times {\vec{n}_2}^{\,\text{r}}]$ cannot be determined to such a high precision, because besides the uncertainties of the solenoid axes, the vector ${\vec{n}_2}^{\,\text{r}}$ depends also on the imperfection content of $\vec m_1$, the spin rotation axis in arc A$_1$. This is an intrinsic feature of the two-solenoid scheme used in the experiment based on the makeshift devices that were already available at COSY. 

In our derivation of the combined AI, we could equally have arranged for the transport of the spin rotation in solenoid S$_2$ over the arc A$_2$ [see Eq.\,(\ref{eq:2C.5})]. Consequently, the orientation of the stable spin axis can be controlled in both straight sections.

This complication could have been avoided, if we had arranged for the local AI supplementing a solenoid S$_1$ by a static Wien filter, generating a horizontal magnetic field, as discussed briefly in Appendix \ref{sec:appendixG}.

The above shortcoming of the two-solenoid scheme does not present a major impediment to some of the future applications of the spin tune mapping technique. For instance, recall the driven rotations of the particle spins in an RF WF. Here, the attainable spin rotation angle is proportional to the spin coherence time $\tau_{\text{SCT}}$. The JEDI collaboration has already achieved very long spin coherence times $\tau_{\text{SCT}}\geq \SI{1000}{s}$\,\cite{Guidoboni:2016bdn}. Evidently, one can only take full advantage of the large spin coherence time, if the spin resonance condition $f_{\text{WF}} = f_s$ is maintained, and the experimentally observed walk of the spin tune is compensated for during times $t \simeq \tau_{\text{SCT}}$. 

One possibility is to let the spin tune drift in the cycle and to adjust $f_{\text{WF}}$ accordingly to match the resonance condition. The other possibility is to keep $f_{\text{WF}}$ fixed in the cycle and to maintain the resonance condition $f_s = f_{\text{WF}}$ in the cycle by adjusting one solenoid field. This way, spin tune mapping becomes a tool to maintain the resonance condition. Both approaches demand for a continuous determination of the spin tune. The corresponding experimental technique has already been developed by the JEDI collaboration (see\,\cite{Eversmann:2015jnk}). In order to measure the spin tune, one needs a horizontal polarization, and it would be appropriate to observe the buildup of a vertical polarization component in the beam as function of time starting with the particle spins precessing in the horizontal plane.

The second point is that by fine tuning the spin rotations in the two solenoids, one can bring the stable spin axis of the ring, including the solenoids themselves, to a desired direction with the above stated angular precision. An illustration of such a precision alignment of the stable spin axis is presented in Appendix\,\ref{sec:appendixE}.

\section{Summary and Outlook}
\label{sec:VI}
We reported about the first ever attempt for the {\it in situ} determination of the spin stable  axis of polarized particles in a storage ring. The experiment, carried out by the JEDI Collaboration in September 2014 at COSY, was  motivated by the search for electric dipole moments (EDMs) of protons and deuterons using a storage ring\,\cite{srEDM-collaboration,jedi-collaboration}. On a purely statistical basis, a sensitivity to the proton and deuteron EDMs at the level of $ \sigma |d_{p,d}| < \SI{e-29}{e.cm}$ looks feasible\,\cite{doi:10.1063/1.4967465}. Such an upper bound on the $CP$- and time reversal invariance violating EDM would be 15 orders of magnitude smaller than the magnetic dipole moment, allowed by all symmetries. Correspondingly, one needs to eliminate spurious effects from the interactions of the MDM of particles with the magnetic fields in a storage ring. The issue becomes extremely acute for the methodical precursor experiments planned at the all-magnetic storage ring COSY, which, in a first step, will make use of an RF Wien filter. This calls for pushing the frontiers of precision spin dynamics at storage rings.

The principal aim of the present experiment was to explore the imperfection magnetic field content of the COSY ring. Our approach was to probe the integral effect of the imperfections acting in the ring plane using a modulation of the spin tune of the stored particles by tunable spin rotators inserted in the ring. The point is that the spin tune can be utilized as a high-precision diagnostics tool, as it can be measured to a precision of nine decimal places for $\SI{100}{s}$ cycles and still higher precision for longer measurement cycles.

In the present exploratory study, the drift (and compensation) solenoids of the two electron coolers installed in COSY have been used as two makeshift spin rotators. An encouraging point is that already these two AIs offer the possibility to fully control the angular orientation of the stable spin axis at two locations in the ring. Our principal conclusion is that the spin tune mapping emerges as a very useful tool to control the spin closed orbit with an accuracy, never achieved before. We uncovered several systematic effects which need further scrutiny, but these do not compromise the fundamentals of the spin tune mapping technique. 

Specifically, we demonstrated that with the interim setup presently available at COSY, the orientation of the stable spin axis $\vec c$ can be determined to a statistical accuracy  $\delta c_{x,z} \sim \SI{1.7}{\micro.rad}$, and eventually to an even higher precision. There are reasons to anticipate a further reduction of $\chi^2/N_{\text{dof}}$ with custom-tailored solenoids. In the meantime, the results of Fit\,2, given in Table\,\ref{tab:rescaling-fit-parameters}, suggest the scaling factor $S = \sqrt{137.5/55} \sim 1.6$. Including this scaling factor, our final estimate for the accuracy of the angular orientation of the stable spin axis is $\delta  c_{x,z} = \SI{2.8}{\micro.rad}$ or better.  

In the future, it will be possible to substantially reduce the systematic effects by employing dedicated solenoids during the spin tune mapping measurements, while the electron cooler magnets and involved steerers are switched off. We mentioned also the use of double-helix magnets, which can simultaneously produce both a longitudinal and horizontal magnetic field. In the latter case, such a device must be complemented with a static electric field in order to operate it in the Wien filter mode. We anticipate that the spin tune mapping technique will prove most useful for the calibration of various devices to be employed in high-precision EDM searches at all magnetic and hybrid magnetic-electric storage rings.

An analysis of the data taken within a very limited scope of the exploratory beam time at COSY has identified certain systematic background effects to the EDM signal. The AI-induced distortions of the beam orbit emerge as the most unwanted one. This makes it imperative to orient the AI field in future investigations to ensure orbit-distortion-free operation, and to also upgrade the beam position monitors. 

The initial motivation for the study was to identify the background from imperfection magnetic fields in view of the precursor experiment searching for the EDM using an RF Wien filter. As a spin-off, we identified that the orientation of the stable spin axis constitutes yet another \textit{static} observable which is also sensitive to the EDM.   As for COSY, a tentative accuracy for the deuteron EDM at $T = \SI{270}{MeV}$,
\begin{equation}
\sigma(d) = \frac{G q}{\beta m_d} \delta c_x \approx 
\SI{e-20}{e.cm}\,, \label{bound-on-D}
\end{equation}
is feasible.

The static and RF WF approaches both suffer from the same systematic background from imperfection fields. As we
have seen above, the spin rotation signal in the RF WF approach is suppressed by the weak spin kick in the WF.
Nevertheless, the importance of the planned RF WF  experiment stems from the point that it provides a testing ground for further perfectioning of the technique by measuring tiny spin rotations of the kind to be measured in the ultimate EDM experiments at future dedicated EDM storage rings.

Although COSY was never intended to be used as an EDM ring, our findings will serve as a plumb line for an upgrade of COSY and even the modest constraints on the proton and deuteron EDMs would be an indispensable step towards the development of dedicated high-precision EDM storage rings.

\begin{acknowledgments}
This work is supported by an ERC Advanced-Grant of the European Union (proposal number 694340). N.N.\ Nikolaev and A.\ Saleev are supported by a Grant from the Russian Science Foundation (grant number RNF-16-12-10151). A.\ Saleev  gratefully acknowledges his support by an FFE grant of Forschungszentrum J\"ulich [grant number 42028695 (COSY-125)].
\end{acknowledgments}

\appendix

\section{Parametric EDM resonance in the RF Wien filter}
\label{sec:appendixB}
\subsection{On-resonance case}
\label{sec:appendixB-1}
The FT-BMT equation [given in Eqs.\,(\ref{eq:spin-precession}), (\ref{eq:Omega_s})] constitutes a homogeneous and linear equation and the EDM resonance can only be a parametric one.  The RF excitation of collective betatron oscillations of the beam is a no go for a precision experiment, because such collective effects produce unwanted and hard to control systematic errors with respect to the determination of the EDM signal. The Wien filter condition for vanishing Lorenz force, given in Eq.\,(\ref{eq:vanishing-lorentz-force}), makes the RF WF entirely EDM transparent. But it leaves a non-vanishing sum of the $\vec B_{\text{WF}}$ and the motional cross product $\vec \beta \times \vec E_{\text{WF}}$, described by Eq.\,(\ref{eq:WFprecession}). The Wien filter axis is denoted by the unit vector $\vec w$, which points in the direction of $\vec B_{\text{WF}}$. 

According to the FT-BMT equation, the RF WF generates a spin rotation of the MDM of a particle around $\vec w$\,\cite{PhysRevSTAB.16.114001} with the spin transfer matrix, given by 
\begin{equation}
\begin{split}
\matr{t}_{\text{WF}}(t)  =  \cos\frac{1}{2}\chi_{\text{WF}}(t) - i (\vec{\sigma} \cdot \vec{w} )\sin\frac{1}{2}\chi_{\text{WF}}(t) \, .  
\label{eq:App-B.3}
\end{split}
\end{equation}
The corresponding spin rotation angle amounts to [see Eq.\,(\ref{eq:WFprecession})]
\begin{equation}
\begin{split}
&\chi_{\text{WF}}(t) =\\
& = -\frac{L_{\text{WF}}}{\beta} \cdot \frac{q  E_{\text{WF}} }{m \beta} \cdot \frac{G+1}{\gamma^2} 
{\cos \left(2\pi f_{\text{WF}}t + \Delta_{\text{WF}}\right)} \\
 & = \chi_{\text{WF}} \cos\left(2\pi  f_{\text{WF}} t + \Delta_{\text{WF}}\right)\, ,  
\label{eq:App-B.2}
\end{split}
\end{equation}
where $L_{\text{WF}}$ is the length of the  RF WF, $E_{\text{WF}}$ is the electric field amplitude, $f_{\text{WF}}$ is the RF frequency. In addition, an allowance is made for the phase shift $\Delta_{\text{WF}}$ with respect to the phase of the spin precession $\theta_s(n)=   2\pi \nu_s f_{\text{R}} t = 2\pi\nu_s n$, where $n$ is the number of revolutions.  Wherever appropriate, we work to the lowest order in a small parameter $\chi_{\text{WF}} \ll 1$. In the ideal case, the Wien filter axis $\vec{w}=(w_x,w_y,w_z)=(0,1,0)$ points along the vertical direction. 

\begin{figure}[tb]
\includegraphics[width=0.8\columnwidth]{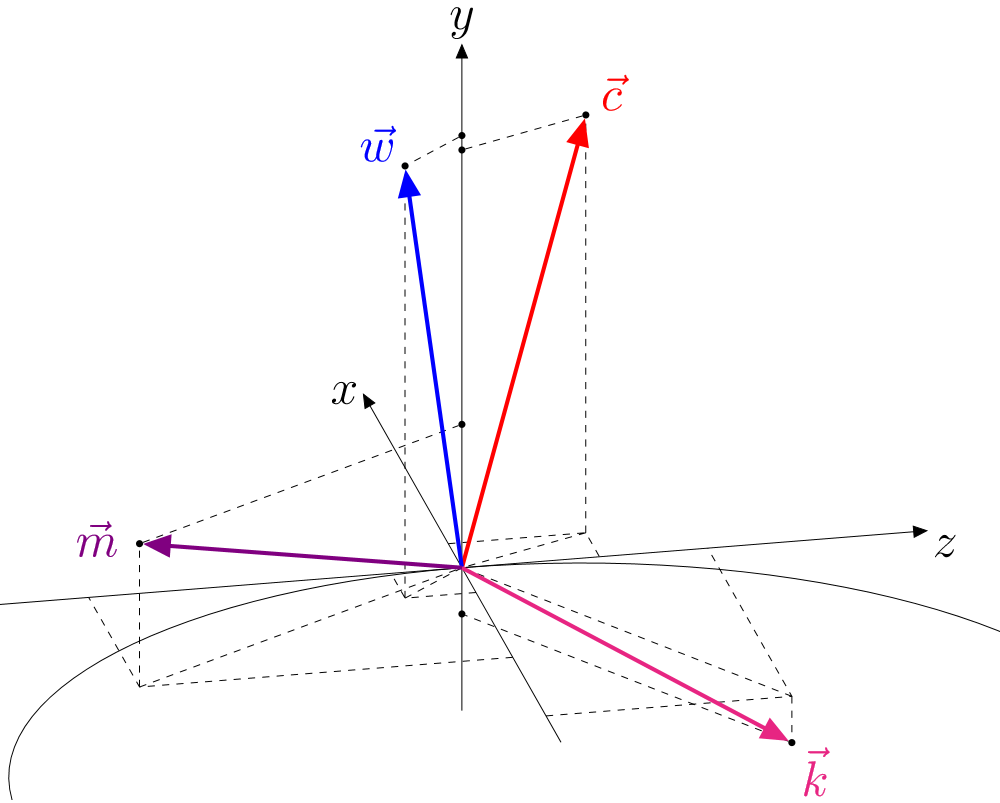}
\caption{\label{fig:scotimesrf} Relative alignment of the spin closed orbit vector $\vec c$, the RF field vector $\vec{w}$, $\vec k \propto (\vec c \times \vec w)$ and $\vec m \propto (\vec k \times \vec c)$, upstream of the RF Wien filter.}
\end{figure}

The evolution of the spinor wave function $\psi$ of the stored particle per turn is described by the one turn map
\begin{equation}
\psi(n+1) = \matr{t}_{\text{WF}}(n+1) \matr{T}\psi(n)\, ,   \label{eq:App-B.4}
\end{equation}
where $\matr{T}$ is the spin transfer matrix of the ring including the AIs, if they are switched on [see Eq.\,(\ref{eq:T})]. We factor out the rapid precession of the spin around the $\vec{c}$-axis.\footnote{We reiterate, that the  $\vec c$-axis is defined for a static ring before the RF spin rotators were activated. For the sake of brevity, we omit the argument $\chi_{\text{WF}}$ of $\vec c(\chi_{\text{WF}})$.} Passing to the conventional interaction representation $\psi(n) = \matr{T} ^n \eta(n)$, where $\eta(n)$ describes the envelope over the rapid oscillations of the spin, $\eta(0)=\psi(0)$. The evolution equation for $\eta(n)$ is given by
\begin{equation}
\begin{split}
\eta(n) = & \matr{T}^{-n} \matr{t}_{\text{WF}}(n) \matr{T}^n \eta(n-1) \\
          & = \exp \left\{-  \frac{i}{2} \vec{\sigma} \cdot \vec{U}(n) \right\}\eta(n-1)\, ,
\end{split}
\label{eq:Evolution}
\end{equation}
where 
\begin{equation}
\begin{split}
\vec U (n) =  & 2 \sin \left( \frac{1}{2} \chi_{\text{WF}}(n)\right) \\
              & \times \Big\{ \phantom{-}\cos\theta_s(n) \Big[\vec w - (\vec c \cdot \vec w)\vec c \Big]  \\ 
              & \quad\;\;\, -  \sin\theta_s(n) \vec c \times \vec w + (\vec c \cdot \vec w)\vec c \Big\}
\label{eq:App-B.6}
\end{split}
\end{equation}
is the instantaneous spin rotation axis in the rotating frame. Here the three vectors,
\begin{equation}
\begin{split}
& \vec c\,,\\
& \vec k = \frac{\vec c \times \vec w}{\sqrt{1- (\vec c \cdot \vec w)^2}}\,, \text{ and }\\  
& \vec m =\frac{(\vec c \times \vec w) \times \vec c}{\sqrt{1 - (\vec c \cdot \vec w)^2}} = \frac{\vec w -(\vec c \cdot \vec w) \vec c}{\sqrt{1 - (\vec c \cdot \vec w)^2}}\, ,  
\label{eq:vector-k-from-c-and-omega}
\end{split}
\end{equation}
form an orthonormal set.  Schematically, the interplay of the above introduced vectors is shown in Fig.\,\ref{fig:scotimesrf}.

Equation Eq.\,(\ref{eq:Evolution}) has a formal solution 
\begin{equation}
\eta(n) = T_n \exp \left\{-  \frac{i}{2} \sum_{k=1}^n \vec{\sigma} \cdot \vec{U}(k) \right\}\psi(0) \, ,	
\label{eq:App-B.5}
\end{equation}
where $T_n$ denotes $n$ ordering. In the resonance regime of $f_{\text{WF}} = f_{\text{R}} (\nu_s + K)$, where the integer $K$ is the harmonic number, the large-$n$ behavior of $\eta(n)$ is evaluated using the Bogolyubov-Krylov-Mitropolsky averaging method\,\cite{Bogolyubov}. It amounts to keeping in the sum $\sum_{k=1}^n \vec{\sigma} \cdot \vec{U}(k)$  only the linearly rising terms
\begin{equation}
\begin{split}
\sum_{k=1}^{n} 2 \chi_{\text{WF}}(k) \cos \theta_s(k)  & \simeq \phantom{-}n \cos \Delta_{\text{WF}} \,, \text{ and} \\
\sum_{k=1}^n 2 \chi_{\text{WF}}(k) \sin \theta_s(k)  & \simeq -n \sin \Delta_{\text{WF}} \,,
\end{split}
\label{eq:App-B.8}
\end{equation}
and neglecting the oscillating terms. The result is 
\begin{equation}
\eta(n) = \exp\left(-\frac{i}{2} n \epsilon {\vec \sigma}\cdot \vec{u}\right)\psi(0)\,,
\label{eq:App-B.7}
\end{equation} 
where 
\begin{equation} 
\vec{u} = \cos \Delta_{\text{WF}} \vec{k} + \sin \Delta_{\text{WF}} \vec{m} \label{eq:App-B.9}
\end{equation}
denotes the stable spin axis in the rotating frame. The envelope $\vec{S}_{\text{env}}(n)$ is given by
\begin{widetext}
\begin{equation}
\vec{S}_{\text{env}}(n) = \frac{1}{2} \Tr \left\{ \exp\left[+\frac{i}{2} n \epsilon \left({\vec \sigma} \cdot \vec{u}\right)\right]\vec{\sigma}\exp\left[-\frac{i}{2} n \epsilon \left({\vec \sigma}\cdot \vec{u}\right)\right]\left[\vec{\sigma}\cdot \vec{S}(0)\right]\right\}\, ,
\label{eq:enevelope}
\end{equation}
\end{widetext}
where $\vec{S}(0)$ is the initial polarization vector.

The spin resonance strength $\epsilon$, given by Eq.\,(\ref{eq:2B.1}), is the product of the MDM spin rotation in the RF WF ($\chi_{\text{WF}}$) and the sine of the angle $\xi_{cw}$ between the stable spin axis $\vec{c}$ and the RF WF axis $\vec{w}$. It should be noted that $\epsilon$ is independent of the phase shift $\Delta_{\text{WF}}$. The generic solution for the spin evolution as function of time,  however,  depends on $\Delta_{\text{WF}}$,  and we shall comment on that below.

\subsection{Off-resonance case}
\label{sec:appendixB-2}
The off-resonance case is of practical interest, if the RF frequency of the WF $f_{\text{WF}}$ does not exactly follow the spin-precession frequency $f_s$ [Eq.\,(\ref{eq:3A.1})], for instance, because
of the spin tune walk in runs with long spin coherence time. 
We parameterize the fractional mismatch of the two frequencies via 
\begin{equation}
\delta_{\text{WF}} = \frac{1}{4\pi} \frac{f_{\text{WF}} - f_s}{f_{\text{R}}} \,.   
\label{eq:App-D.1}
\end{equation}

It is convenient to reabsorb the mismatch effect into the spin transfer matrix of the RF WF,
\begin{equation}
\begin{split}
&\matr{t}_{\text{WF}}(n) \\
& =  \left( 1 + \frac{i}{4}(\vec \sigma \cdot {\vec c\,}) \delta_{\text{WF}} \right) 
                 \left(1 - \frac{i}{2}(\vec \sigma \cdot \vec w )\chi_{\text{WF}}(t) \right)\, \\
 & \simeq  1 - \frac{i}{4} \vec \sigma \cdot \Big(2\chi_{\text{WF}}(t) \vec w - \delta_{\text{WF}} \vec c\,  \Big) \, . 
\end{split}
\label{eq:App-D.2}
\end{equation}
Repeating the analysis described in Appendix\,\ref{sec:appendixB-1}, one finds for the spin rotation axis in the rotating frame
\begin{equation}
\begin{split}
\vec u  = &  \frac{\chi^0_{\text{WF}} \Big( \cos \Delta_{\text{WF}} \vec k + \sin \Delta_{\text{WF}} \vec m \Big) - \delta_{\text{WF}} \vec c}{\sqrt{ {\chi^0_{\text{WF}}}^2 + {\delta_{\text{WF}}}^2}} \\
        = & \cos\rho \Big( \cos \Delta_{\text{WF}} \vec k + \sin \Delta_{\text{WF}} \vec m \Big) -\sin\rho \vec c \, , 
\label{eq:App-D.3}
\end{split}
\end{equation}
where $\chi^0_{\text{WF}} = \chi_{\text{WF}} \sqrt{1 - (\vec c \cdot \vec w)^2 }$ and the angle $\rho$ provides a convenient parameterization for the frequency mismatch,
\begin{equation}
 \sin\rho = \frac{ \delta_{\text{WF}}}{\sqrt{ {\chi^0_{\text{WF}}}^2 + {\delta_{\text{WF}}}^2}}.
\end{equation}
The corresponding resonance strength is given by
\begin{equation}
\epsilon(\chi^0_{\text{WF}}, \delta_{\text{WF}})  =  \frac{1}{2} \sqrt{ {\chi^0_{\text{WF}}}^2 + {\delta_{\text{WF}}}^2} \, . 
\label{eq:App-D.4}
\end{equation}

The generic solution for the evolution of the envelope of the rapid oscillations of the polarization vector $\vec S(n)$ as a function of turn number $n$ (suppressing the arguments of $\epsilon$) reads
\begin{widetext}
\begin{equation}
\vec S_{\text{env}}(n) =  \vec u \left( \vec u \cdot \vec S(0) \right) 
	  +  \left\{ 1 -(\vec u \cdot \vec S(0))^2 \right\}^{1/2} \left[\vec n_a \cos( \epsilon n) + \vec n_b \sin( \epsilon n) \right]\, , 
\label{eq:App-D.SPIN}
\end{equation}
\end{widetext}
where
\begin{equation}
\begin{split}						
\vec n_a = & \left[ \vec S(0) -\vec u \left( \vec u \cdot \vec S(0) \right)\right] \left[1 -\left(\vec u \cdot \vec S(0)\right)^2\right]^{-1/2}\, \\
\vec n_b = & \left[\vec u \times \vec S(0)\right] \left[1 - \left(\vec u \cdot \vec S(0)\right)^2\right]^{-1/2}\, . \label{eq:plane}
\end{split}
\end{equation}
The projection of the envelope $\vec S_{\text{env}} (n)$ onto the spin precession axis $\vec u$ is conserved, $ \vec u \cdot \vec S_{\text{env}}(n) =  \vec u \cdot \vec S(0)$. The spin envelope rotates in the plane defined by two unit vectors $\vec n_{a,b}$. The orientation of this plane depends on both the relative phase $\Delta_{\text{WF}}$ of the RF field and the spin rotation phase, and the initial orientation of the polarization $\vec S(0)$. 

This result generalizes the considerations given in\,\cite{lee1997spin} (see also the more recent analysis in\,\cite{Silenko:2015taa}). Below we illustrate the salient features of this solution for the off-resonance case by two typical examples.

\subsubsection{Evolution of polarization starting with initial spin along the stable spin axis of a ring}
When initially the spins are oriented along the (approximately vertical) stable spin axis $\vec c$, {\it i.e.},  $\vec S(n=0) = \vec c $, the solution  for the envelope is described by 
\begin{widetext}
\begin{equation}
 \begin{split}
  \vec S_{\text{env}}(n)  = &  \sin\rho \left\{\sin\rho \vec c -
	\cos\rho \left[\cos \Delta_{\text{WF}}\vec k + \sin\Delta_{\text{WF}}\vec m \right]\right\} \\ 
	& + \cos\rho \left\{\cos\rho \vec c +\sin\rho \left[\cos \Delta_{\text{WF}}\vec k + \sin\Delta_{\text{WF}}\vec m\right]\right\} \cos \left(\epsilon n\right) \\
	& + \cos\rho \left\{\cos \Delta_{\text{WF}}\vec m - \sin\Delta_{\text{WF}}\vec k \right\}\sin \left(\epsilon n\right)\, .
	\label{eq:Svertical}
 \end{split}
\end{equation}
\end{widetext}
The up-down oscillations of polarization along the $\vec{c}$-axis are described by
\begin{equation}
 \begin{split}
 \left( \vec S_{\text{env}}(n)\cdot \vec c\right)   =  \sin^2\rho + \cos^2\rho \left(\epsilon n\right)\,.
	\label{eq:OscillationsOffset1}
	 \end{split}
\end{equation}

Whether $\vec S(n)$ would cross the ring plane into the lower hemisphere or not, depends on the angle $\rho$.  
\begin{itemize}
 \item[(i)] If $|\rho| < \pi/4$, the up-down oscillation amplitude exceeds the offset value and the polarization will flip the sign. 
 \item[(ii)]  If   $\cos^2 \rho <\sin^2 \rho$, vertical component of the spin  does not flip the sign.
\item[(iii)] Far away off the resonance, {\it i.e.} if $\cos^2 \rho \ll \sin^2 \rho$, the axis of the driven spin motion approaches $\vec{c}$ and the driven motion reduces to weak nutations.
\end{itemize}

For the sake of completeness we cite the other two projections of the polarization envelope
\begin{widetext}
\begin{equation}
 \begin{split}
  \vec S_{\text{env}}(n)\cdot \vec k   &= 	-\sin\rho \cos\rho \cos \Delta_{\text{WF}}  + \sin\rho \cos\rho \cos \Delta_{\text{WF}}\cos \left(\epsilon n\right) -
	\cos\rho \sin \Delta_{\text{WF}}\sin \left(\epsilon n\right)\, , \\
	\vec S_{\text{env}}(n)\cdot \vec m   &= 	-\sin\rho \cos\rho \sin \Delta_{\text{WF}}  + \sin\rho \cos\rho \cos \Delta_{\text{WF}}\sin \left(\epsilon n\right) -
	\cos\rho \cos \Delta_{\text{WF}}\sin \left(\epsilon n\right)\, .
	\label{eq:Horizontal-Oscillations-c}
 \end{split}
\end{equation}
\end{widetext}

\subsubsection{Evolution of polarization starting with initial spin perpendicular to the stable spin axis of a ring}
Here one starts with the spin in the plane spanning the vectors $\vec k$ and $\vec m$. The prime signal of the RF-driven spin rotations is the buildup of a (vertical) polarization along the stable spin axis $\vec c$ of the ring [see Eqs.\,(\ref{eq:2A.3},\ref{eq:spin-tune-without-solenoid})]. For instance, for $\vec{S}(0) =\vec k$ [given by Eq.\,(\ref{eq:vector-k-from-c-and-omega})], the evolution of the spin envelope is described by
\begin{widetext}
\begin{equation}
 \begin{split}
  \vec S_{\text{env}}(n)  = &  \cos\rho \cos \Delta_{\text{WF}}
	\left\{
	\cos\rho \left[\cos \Delta_{\text{WF}}\vec k  + \sin\Delta_{\text{WF}}\vec m \right]  -\sin\rho \vec c\right\} \\
	&+ \left\{\left[1-\cos^2\rho \cos ^2\Delta_{\text{WF}}\right]\vec k -\cos^2 \rho \cos \Delta_{\text{WF}}\sin \Delta_{\text{WF}}\vec m + \cos\rho\sin\rho\cos\Delta_{\text{WF}}\vec c\right\} \cos \left(\epsilon n\right) \\
	       & +\left\{ \cos\rho \sin \Delta_{\text{WF}}\vec c + \sin\rho\vec m\right\}\sin \left(\epsilon n\right)\, .\\
  \vec S_{\text{env}}(n)\cdot \vec c  = &   -\cos\rho \sin\rho  \cos \Delta_{\text{WF}} +  \sin\rho \cos\rho \cos \Delta_{\text{WF}}\cos \left(\epsilon n\right) 
	+\cos\rho \sin \Delta_{\text{WF}}\sin \left(\epsilon n\right) \, , \\
		\vec S_{\text{env}}(n)\cdot \vec k   = &	\cos^2\rho \cos ^2\Delta_{\text{WF}}  + (\cos^2\rho \cos ^2\Delta_{\text{WF}} +\sin^2\rho ) \cos \left(\epsilon n\right)
		+	\cos\rho \sin \Delta_{\text{WF}}\sin \left(\epsilon n\right)\, , \\
	\vec S_{\text{env}}(n)\cdot \vec m   =& \cos^2\rho \sin \Delta_{\text{WF}} \cos \Delta_{\text{WF}} -\cos^2\rho \sin \Delta_{\text{WF}} \cos \Delta_{\text{WF}} 
	\cos \left(\epsilon n\right)
		+	\sin\rho \sin \left(\epsilon n\right)\, .
			\label{eq:Horizontal-Oscillations-c2}
 \end{split}
\end{equation}
\end{widetext}
The vertical polarization buildup is suppressed by a factor $\cos\rho$, while its polarization offset is suppressed by a further factor $\sin\rho$. The offset of the polarization oscillations is manifest in the other two projections as well.  

\subsubsection{Spin motion frequency spectrum and utility of the phase of the RF Wien Filter}
In the real experiment one determines the the up-down and left-right asymmetries in the scattering of beam 
particles extracted onto the carbon target. The time dependence of these asymmetries is an interplay of 
the RF driven rotation of the envelope of the polarization with the resonance strength $\epsilon$ and the idle precession
with the spin tine $\nu_s$:
\begin{equation}
 \begin{split}
\psi(n)=& \matr{T}(n)\psi(0)\, , \\
\matr{T}(n) = & \exp\left[-i\pi n\nu_s \left( \vec \sigma \cdot \vec c \right)\right] \exp\left[-\frac{i}{2}n\epsilon\vec\sigma \cdot \vec u \right]\, , \\
\vec{S}(n) = & \frac{1}{2} \Tr \left\{ \matr{T}^{\dagger}(n)\vec{\sigma}\matr{T}(n)\left(\vec{\sigma}\cdot \vec{S}(0)\right)\right\}
 \end{split}
\end{equation}
For the sake of completeness, we give here the generic evolution law
\begin{widetext}
\begin{equation}
 \begin{split}
  \vec S(n)  = \,& \vec c \left(\vec S(0)\cdot \vec u\right) (\vec c\cdot \vec u) +
	\vec c \left[\left(\vec S(0)\cdot \vec c\right)-\left(\vec S(0)\cdot \vec u\right) (\vec c \cdot\vec u)\right] \cos \left(\epsilon n\right)+
	\vec c \left(\vec S(0)\cdot\left[ \vec u \times \vec c\right]\right) \sin \left(\epsilon n\right)\\
	&+ \left[\vec u-\vec c (\vec c \cdot \vec u) \right] \left(\vec S(0)\cdot \vec u\right) \cos \left(2\pi\nu_s n\right)
	+\left[\vec c\times \vec u\right]  \left(\vec S(0)\cdot \vec u\right ) \sin\left(2\pi\nu_s n\right)\\
	&+ \left[\vec S(0) - \vec u \left(\vec S(0)\cdot \vec u\right)-\vec c \left(\vec S(0)\cdot \vec c\right)+
	\vec c \left(\vec S(0)\cdot \vec u\right) (\vec c \cdot\vec u)\right] \cos \left(2\pi\nu_s n\right)\cos \left(\epsilon n\right)\\
	&+\left\{\left[\vec S(0) \times \vec u \right] -\vec c  \left(\vec S(0)\cdot\left[ \vec u \times \vec c\right]\right)\right\}
	 \cos \left(2\pi\nu_s n\right)\sin \left(\epsilon n\right)\\
	&+ \left\{\left[\vec S(0) \times \vec c \right] -\left[\vec c \times \vec  u\right]\left(\vec S(0)\cdot \vec u\right) \right\}  
	 \sin \left(2\pi\nu_s n\right)\cos \left(\epsilon n\right) \\
	&+\left[\vec S(0)(\vec c\cdot \vec u)-\vec u  \left(\vec S(0)\cdot \vec c\right)\right]  \sin \left(2\pi\nu_s n\right)\sin \left(\epsilon n\right)\, .
\end{split}
\end{equation}
\end{widetext}
A comprehensive analysis of this evolution law goes way beyond the scope of this publication, we only mention its salient features:
\begin{itemize}
\item[(i)] Besides the obvious frequencies $\nu_s f_{\text{R}}$ and $\epsilon f_{\text{R}}/2\pi$, the full frequency spectrum of the spin motion includes the side bands $(\nu_s \pm \epsilon/2\pi)f_{\text{R}}$.
\item[(ii)] In the presence of imperfection fields and in the off-resonance regime, the Fourier spectrum of the up-down oscillations shall contain the frequencies $\epsilon f_{\text{R}}/2\pi $ and $(\nu_s \pm \epsilon/2\pi)f_{\text{R}}$.
\item[(iii)] All Fourier amplitudes shall exhibit a non-trivial dependence on the phase shift $\Delta_{\text{WF}}$, which can be utilized as a cross check of the RF WF operation.  
\end{itemize}

\section{Spin transfer matrix in an imperfection-loaded ring}
\label{sec:appendixA}
Spin-wise the MDM interaction with the imperfection magnetic fields mimics the EDM interaction with the motional electric field in the FT-BMT equation, given in Eqs.\,(\ref{eq:spin-precession}, \ref{eq:Omega_s}). The equation for the spinor wave function of the stored particle $\Psi(\theta)$ is
\begin{equation}
\frac{d\Psi(\theta)}{d\theta} = -\frac{i}{2} \Big\{ 
G\gamma \sigma_y -F_x(\theta)\sigma_x -F_z(\theta)\sigma_z \Big\}\Psi(\theta)\, ,
\label{eq:App-A.1}
\end{equation}
where $F_{x,z} \propto B_{x,z}/B_y \ll 1$, and the EDM interaction with the motional $\vec{E}$-field enters under the umbrella of $F_x$. We proceed to the customary interaction representation in which
\begin{equation} 
\Psi(\theta) = \exp\left[-\frac{i}{2}  G\gamma\theta\sigma_y\right]u(\theta) = \matr{t}_{\text R}(\theta) u(\theta)\, .
\label{eq:App-A.2}
\end{equation}
Here $\matr{t}_{\text R}(\theta)$ is the spin transfer matrix without ring imperfections,  and $u(\theta)$ satisfies the equation  
\begin{equation}
\begin{split}
\frac{du}{d\theta} & =  \frac{i}{2} \matr{t}_{\text R}(-\theta) \Big[ F_x(\theta)\sigma_x +F_z(\theta)\sigma_z \Big]
\matr{t}_{\text R}(\theta)u(\theta)\\
& = \frac{i}{2} \vec{\sigma}\cdot \vec{a}(\theta) u(\theta)\, , 
\label{eq:App-A.3}
\end{split}
\end{equation}
where 
\begin{equation}
\begin{split}
\vec{a}(\theta) = \phantom{+}  &\Big[ F_x(\theta) \cos\left( G\gamma\theta\right) - F_z\sin \left(G\gamma\theta \right) \Big] \vec{e}_x  \\
  +  &\Big[ F_z(\theta) \cos \left( G\gamma\theta \right) + F_x\sin \left( G\gamma\theta \right) \Big] \vec{e}_z\, 
\label{eq:App-A.4}
\end{split}
\end{equation}
is the imperfection field in the reference frame which rotates with the ideal spin tune frequency $G\gamma f_{\text{R}}$.

A formal solution of Eq.\,(\ref{eq:App-A.3}) is $u(\theta) = \matr{t}_{\text{R}}^{\text{imp}}(\theta) \Psi(0)$, where $\matr{t}_{\text{R}}^{\text{imp}}(\theta)$ is given by the $\theta$-ordered exponential
\begin{equation}
\matr{t}_{\text{R}}^{\text{imp}}(\theta) = T_{\theta} \exp\left\{ \frac{i}{2} \int_0^{\theta} d\theta_1 \vec{\sigma}\cdot \vec{a}(\theta_1) \right\}\, .
\label{eq:App-A.5}
\end{equation}
To the second order in the imperfection fields, the spin transfer matrix per turn is given by
\begin{equation}
\begin{split}
\matr{t}_{\text{R}}^{\text{imp}}(2\pi) & = T_{\theta} \exp\left\{ \frac{i}{2} \int_0^{2\pi} d\theta_1 \vec{\sigma}\cdot \vec{a}(\theta_1) \right\} \\
                      & = 1+\frac{i}{2} \vec{\sigma}\cdot \vec{b} + \frac{1}{2!} \left(\frac{i}{2} \vec{\sigma}\cdot \vec{b}\right)^2 \, , \\
\end{split}
\label{eq:App-A.6}
\end{equation}
where $\vec{b} = \vec{b}(2\pi)$, with the components
\begin{equation}
\begin{split}
b_{x,z}(\theta)   = &\int_0^{\theta} d\theta_1 \vec{a}_{x,z}(\theta_1)\, ,\\
b_y(\theta)       = &\frac{1}{2} \int_0^{\theta} d\theta_1 \Big[ a_x(\theta_1) b_z(\theta_1) - a_z(\theta_1) b_x(\theta_1) \Big] \, .
\end{split}
\label{eq:App-A.7}
\end{equation}
Here, $b_y(\theta)$ comes from the non-commuting spin rotations around the horizontal and longitudinal imperfection fields. An extension to higher orders is straightforward and is redundant for the purposes 
of the present paper.

The total spin transfer matrix per turn can be cast as
\begin{equation}
\matr{T} = \matr{t}_{\text R}(2\pi)  \matr{t}_{\text{R}}^{\text{imp}}(2\pi) = \exp \Big[-i \pi\nu_s (\vec{\sigma} \cdot \vec{c}) \Big] \,,\label{eq:App-A.8}
\end{equation}
with the spin tune $\nu_s$, given by 
\begin{equation}
\begin{split}
 &\cos(\pi\nu_s)  \\
 & = \left[1-\frac{1}{8} (b_x^2+b_z^2)\right] \cos(\pi G\gamma) \\
 & +\frac{1}{2} b_y \sin(\pi G\gamma) \,,
 \label{eq:spin-tune-formula}
\end{split}
\end{equation}
and the stable spin axis $\vec{c}$,
\begin{widetext}
\begin{equation}
\begin{split}
c_x\sin \left( \pi\nu_s \right) & = -\frac{1}{2} \left[b_x \cos \left(\pi G\gamma\right) + b_z \sin\left(\pi G\gamma\right)\right] \,, \\
c_y\sin \left( \pi\nu_s \right) & = \phantom{-}\frac{1}{2} \left[ -b_y \cos\left(\pi G\gamma\right) + \left[1-\frac{1}{8} (b_x^2+b_z^2)\right]\sin\left(\pi G\gamma\right)\right]\,, \\
c_z\sin \left(\pi\nu_s\right)   & = -\frac{1}{2}\left[b_z \cos\left(\pi G\gamma\right) - b_x \sin\left(\pi G\gamma\right)\right]\, . 
\end{split}
\label{eq:App-A.9}
\end{equation}
\end{widetext}
It should be noted that the correction to the spin tune starts to the second order in the imperfection field. In an imperfection-free ring
\begin{equation}
\vec{c}=(\sin\xi_{\text{EDM}}, \cos\xi_{\text{EDM}}, 0) \,, \label{eq:App-A.10}
\end{equation}
while in an imperfection-loaded ring 
\begin{equation}
c_x = c_x(\text{MDM})+\sin \xi_{\text{EDM}}\,. 
\end{equation}

\section{Cooler solenoids as artificial imperfections at COSY}
\label{sec:appendixF}
Here we present technicalities of the derivation of Eq.\,(\ref{eq:2C.3}) and its accuracy. The preliminary ideas have already been exposed in Sec.\,\ref{sec:II.C.2}. 

The exact formula for the spin tune modified by the AI reads
\begin{widetext}
\begin{equation}
\begin{split}
  \cos\left(\pi \left[\nu_s^0 + \Delta\nu_s(\chi_1,\chi_2)\right]\right)  = & \frac{1}{2}\Tr \matr{T} = \frac{1}{2}\Tr \left( \matr{t}_{\text{R}} \matr{t}_{\text{AI}} \right)  
  =  \cos\left(\pi\nu_s^0\right) \cos \left(\frac{1}{2}\chi_1 \right) \cos \left(\frac{1}{2}\chi_2 \right) \\ 
  &- (\vec{c} \cdot \vec{n}_1)\sin \left( \frac{1}{2}\chi_1 \right) \cos \left( \frac{1}{2}\chi_2 \right) - (\vec{c} \cdot {\vec{n}_2}^{\,\text{r}}) \cos \left( \frac{1}{2}\chi_1 \right) \sin\left( \frac{1}{2}\chi_2 \right) \\
 & - \left\{ \cos \left(\pi \nu_s^0 \right) ({\vec{n}_2}^{\,\text{r}} \cdot \vec n_1) + \sin \left( \pi \nu_s^0 \right) \left(\vec c \cdot \left[{\vec{n}_2}^{\,\text{r}} \times \vec n_1\right]\right) \right\}\sin \left( \frac{1}{2}\chi_1 \right) \sin \left(  \frac{1}{2}\chi_2 \right)
\,,
\end{split}
\label{eq:App-F-3}
\end{equation}
\end{widetext}
where $\matr{t}_{\text{AI}}$ is given by Eq.\,(\ref{eq:t_AI}). 

We have several small imperfection parameters in the problem. First of all, the departure of the stable spin axis $\vec c$  from the exact vertical orientation  is described by the non-vanishing imperfection parameters $c_x$ and $c_z$. Similar imperfections arise from the non-vanishing imperfection components $m_{1x}$ and $m_{1z}$ of the axis $\vec m_1$ of the spin rotation in the arc A$_1$. A possible misalignment of the magnetic field axes of solenoids $\text{S}_{1,2}$ with respect to the beam axis is of similar magnitude, and the spin rotation angles in the solenoids $\text{S}_{1,2}$ are in the same ballpark as well. 

Now we argue that the coefficient $E$ in front of the quadratically small product $\sin \left(\frac{1}{2}\chi_1 \right) \sin \left(\frac{1}{2}\chi_2\right)$ is approximately unity,
\begin{equation}
\begin{split}
E = & \cos \left(\pi \nu_s^0\right) \left(\vec n_2^{\, \text{r}} \cdot \vec n_1\right) \\
    & + \sin  \left( \pi \nu_s^0 \right) \left(\vec c \cdot \left[\vec n_2^{\, \text{r}} \times \vec n_1\right) \right] \simeq 1\, .
\end{split}
\label{eq:App-F-4}
\end{equation}
Indeed, it can be evaluated to the zeroth order in the above used small imperfection parameters. Specifically, we can take the solenoid axes $\vec n_1 \simeq \vec n_2 \simeq \vec e_z$. To the same accuracy, the stable spin axis $\vec c$ and the spin rotation axis $\vec m_1$ in the arc $A_1$ can be approximated as  $\vec c \simeq \vec m_1 \simeq \vec e_y$. Both arcs rotate the beam momentum by an angle $\pi$, and to the same accuracy to which $\vec m_1 \simeq \vec e_y$, we have  $\theta_1 \simeq \pi \nu_s^0$. Lumping all these approximations together, we find
\begin{equation}
\begin{split}
(\vec n_2^{\, \text{r}} \cdot  \vec n_1) &\simeq \cos \left( \pi \nu_s^0 \right)\, , \\
\vec c \cdot [\vec n_2^{\, \text{r}} \times \vec n_1) ] &\simeq \sin \left( \pi \nu_s^0 \right)\,.
\end{split}
\end{equation}
This entails $E \simeq 1$ and completes the derivation of Eq.\,(\ref{eq:2C.3}). The omitted terms are of the fourth order, an  example is $\sim c_j^2 \chi_i^2 $.

\section{Error analysis of spin tune jumps} 
\label{sec:appendixC}
Each time interval $\Delta T_i$ ($i=1, 2, 3$) is analyzed using first a first guess of the spin tune $q_{s_i}$. Subsequently, keeping this input $q_{s_i}$ fixed, we allow for the additional time-dependent variation of the phase of the  spin precession, which is monitored as function of turn number $n$,
\begin{equation}
\Phi_i(n) = 2\pi q_{s_i} n +   \varphi_{s_i} (n) \,.
\label{eq:App-C.1}
\end{equation}
For each macroscopic time interval representing about $\Delta n =\SI{e6}{turns} \approx \SI{1.3}{s}$, one phase value $\varphi_{s_i}(n)$ is determined based on the available statistics corresponding to $\approx 5000$ events. The time walk of the spin tune in each interval $i$ is then given by  
\begin{equation}
\nu_{s_i}  = q_{s_i} + \frac{1}{2\pi}\cdot \frac{\partial \varphi_{s_i}(n)}{\partial n} \, .
\label{eq:App-C.2}
\end{equation}

Following the findings described in\,\cite{Eversmann:2015jnk}, an allowance is made for a linear drift of the spin tune, \textit{i.e.}, a parabolic fit of the phase is performed using
\begin{equation}
\begin{split}
\varphi_{s_i}(n)  = & 2\pi a_i + 2\pi N_i b_i x(n) \\
                    & + \pi N_i c_i \left(x(n)^2 - \frac{1}{3}\right)\, , \\
 \nu_{s_i}(n)     = & q_{s_i} + b_i + c_i x(n) \, .
\end{split}
\label{eq:App-C.3}
\end{equation}
Here $x(n) = (n - n_{0_i})/{N_i}$, and $2N_i$ represents the number of particle turns in each of the time intervals $\Delta T_i$, and $n_{0_i}$ refers to the midpoint of each interval, so that $-1 < x < 1$. Such an expansion in the basis of orthogonal functions ensures a minimal correlation between the expansion parameters $b_i$ and $c_i$. 

\begin{table}[tb]
\renewcommand{\arraystretch}{1.15}
\begin{ruledtabular}
\begin{tabular}{cccc}
Parameter                   & \multicolumn{3}{c}{Time intervals} \\ 
& $\Delta T_1$              & $\Delta T_2$              & $\Delta T_3$   \\ 
                            & ($i = 1$)                 & ($i = 2$)         & ($i = 3$)      \\\hline
$\sigma_{b_i}/\sigma_{a_i}$ & $1.744 \pm 0.027$         & $1.743 \pm 0.025$ & $1.709 \pm 0.037$     \\
                            & \multicolumn{3}{c}{$\braket{\sigma_{b_i}/\sigma_{a_i}} = 1.737 \pm 0.016$} \\\hline
$\sigma_{c_i}/\sigma_{a_i}$ & $6.835 \pm 0.113$         & $6.738 \pm 0.131$ & $6.489 \pm 0.204$     \\
                            & \multicolumn{3}{c}{$\braket{\sigma_{c_i}/\sigma_{a_i}} = 6.748 \pm 0.079$} \\\hline                           
$\sigma_{b_i}\,[\num{e-10}]$ & $\phantom{-}7.9 \pm 1.1$  & $5.3 \pm 0.7$     & $\phantom{-}4.3 \pm 0.9$  \\\hline
$c_i/\sigma_{c_i}$          & $-0.1 \pm 1.3$            & $0.8 \pm 2.0$     & $-0.3 \pm 1.8$ \\
\end{tabular}
\end{ruledtabular}
 \caption{\label{tab:all-abc} Uncertainties of the fitted linear phase-parameters $\sigma_{b_i}/\sigma_{a_i}$, $\sigma_{c_i}/\sigma_{a_i}$, $b_i$, and $c_i/\sigma_{c_i}$ for the three time intervals $\Delta T_i$, averaged over all 359 cycles. The corresponding distribution for $c_2/\sigma_{c_2}$ is shown in Fig.\,\ref{fig:c2-over-sigma-c2}, the distributions of $\sigma_{b_i}$ are shown in Fig.\,\ref{fig:sigma-b1-sigma-b2-sigma-b3}. }
\end{table}

In the absence of such correlations, the standard deviations of the fitted parameters in each time interval $i=1,2,3$ are expected to satisfy 
\begin{equation}
\begin{split}
\sigma_{a_i} \, : \, \sigma_{b_i} \, : \, \sigma_{c_i}  & = \, 1 \, : \, \sqrt{3} \, :\, \sqrt{45} \\ 
                                                        & = 1\,: 1.732\, : \, 6.708\, , 
\label{eq:App-C.4}
\end{split}
\end{equation}
which is perfectly confirmed by the ratio of fitted results, listed in Table\,\ref{tab:all-abc},
\begin{equation}
\begin{split}
&\sigma_a \, : \, \sigma_b \, : \, \sigma_c \\
&= \, 1 \, : \, (1.737 \pm 0.016) \, :\, (6.748 \pm 0.079) \,.
\end{split}
\label{eq:App-C.5}
\end{equation}

\begin{figure}[tb]
\includegraphics[width=\columnwidth]{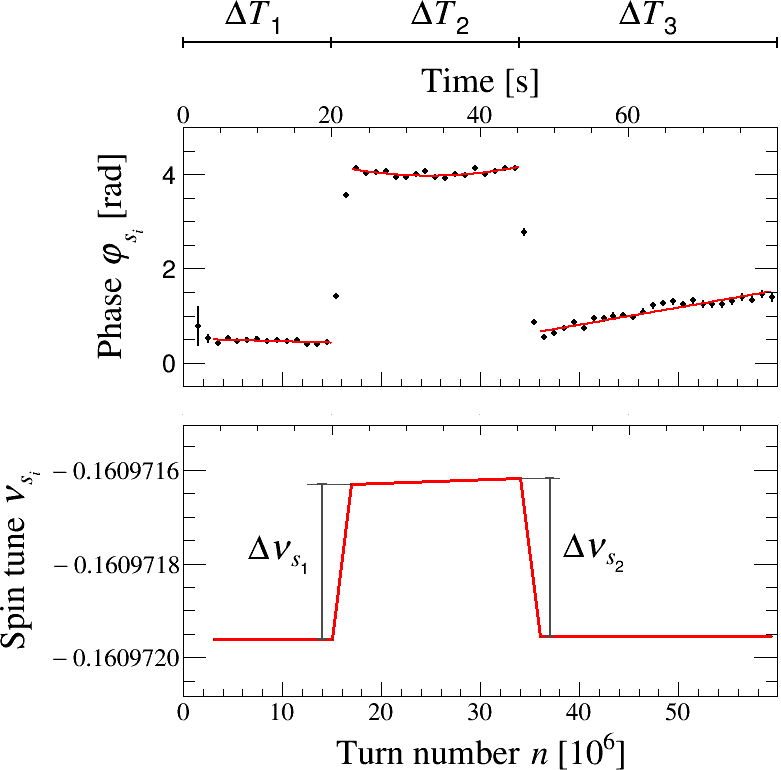}
\caption{\label{fig:phasejump} Upper panel: Spin phase $\varphi_{s_i}$ as function of the number of turns $n$ for the three time intervals $i=1,2,3$ of one particular cycle. In red, the fitted result using Eq.\,(\ref{eq:App-C.3}) is indicated, exhibiting a pronounced non-linearity in the time interval $\Delta T_2$. Bottom panel: Spin tunes $\nu_{s_i}$, calculated using Eq.\,(\ref{eq:App-C.3}), and spin tune jumps $\Delta \nu_{s_{1,2}}$, calculated from Eq.\,(\ref{eq:spin-tune-jumps-1-and-2}).}
\end{figure}
Figure\,\ref{fig:phasejump} shows the measured dependence of the phase $\varphi_{s_i}$ as function of turn number $n$ during one particular cycle for the three time intervals $\Delta T_i$. The chosen example is one of the few cases, where $c_2$ differs from $0$ in the time interval $\Delta T_2$. 

\begin{figure}[t]
\includegraphics[width=\columnwidth]{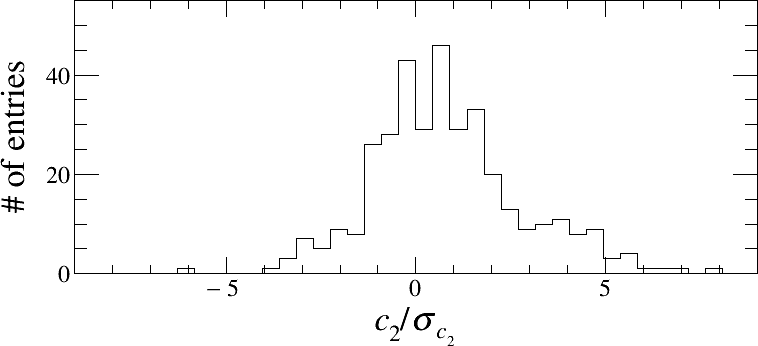}
\caption{\label{fig:c2-over-sigma-c2} Statistical distribution of $c_2/\sigma_{c_2}$, the ratio of the quadratic parameter $c_2$ to its error, for the second time interval $\Delta T_2$ for all $\SI{359}{cycles}$, which yields $c_2/\sigma_{c_{2}} = 0.8 \pm 2.0 $.}
\end{figure}

For a total of $\SI{359}{cycles}$ considered in the analysis, the calculated ratio of $c_2/\sigma_{c_{2}} = 0.85 \pm 2.00 $ indicates that the quadratic term is small and that its consideration in the data analysis is statistically not justified (see Fig.\,\ref{fig:c2-over-sigma-c2}). For the solenoid-off time intervals $\Delta T_1$ and $\Delta T_3$, the ratios $c_1/\sigma_{c_{1}}$ and $c_2/\sigma_{c_{2}}$ are found to be entirely negligible (see Table\,\ref{tab:all-abc}). 

The magnitude of the spin tune jump $\Delta \nu_s$ is thus determined from 
\begin{equation}
\begin{split}
\Delta\nu_{s_1} = \nu_{s_2} - \nu_{s_1} & =  q_{s_2} + b_2 - q_{s_1} - b_1 \,,\\
\Delta\nu_{s_2} = \nu_{s_2} - \nu_{s_3} & =  q_{s_2} + b_2 - q_{s_3} - b_3   \,,
 \end{split}
 \label{eq:spin-tune-jumps-1-and-2}
\end{equation}
where the quadratic phase parameters $c_i$ have been neglected because of their their statistical insignificance. An example is shown in Fig.\,\ref{fig:phasejump} (bottom panel). The spin tune jump $\Delta \nu_s$ is computed from the average of the two spin tune jumps $\Delta\nu_{s_1}$ and $\Delta\nu_{s_2}$,
\begin{equation}
\begin{split}
 \Delta \nu_s & = \frac{\Delta\nu_{s_1} + \Delta\nu_{s_2}}{2}  \\
              & = \frac{2q_{s_2} - q_{s_1} - q_{s_3} + 2 b_2  -b_1 -b_3}{2}\,.
\end{split}
\label{eq:statistical-error-of-spin-tune-jump}
\end{equation}

\begin{figure}[tb]
\includegraphics[width=\columnwidth]{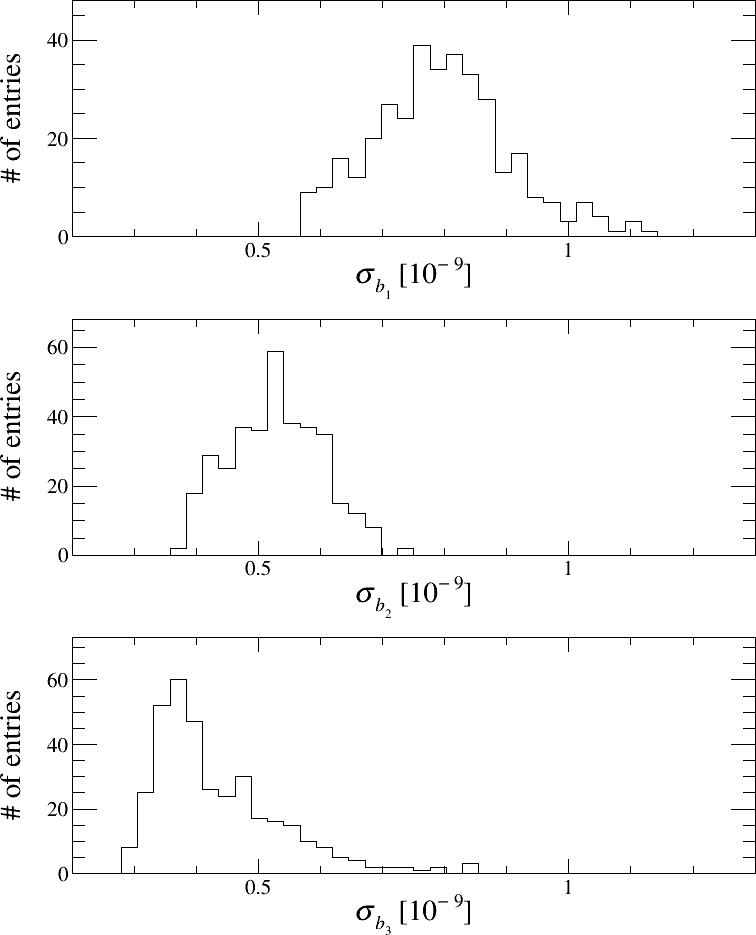}
\caption{\label{fig:sigma-b1-sigma-b2-sigma-b3} Distribution of the statistical errors $\sigma_{b_1}$,  $\sigma_{b_2}$ and $\sigma_{b_3}$ of the linear parameters $b_1$, $b_2$, and $b_3$ of Eq.\,(\ref{eq:App-C.3}). The mean values of statistical errors are listed in Table\,\ref{tab:all-abc}.}
\end{figure}

In Fig.\,\ref{fig:sigma-b1-sigma-b2-sigma-b3}, the distributions of $\sigma_{b_i}$ for the three time intervals are shown for all $\SI{359}{cycles}$. The fitted parameters of the baseline spin tune in each of the three time intervals are given in Table\,\ref{tab:all-abc}. Using all $\SI{359}{cycles}$, the parameters $b_i$ are determined with a statistical uncertainty in the range $\numrange[range-phrase = -]{4e-10}{8e-10}$.
\begin{figure}[htb]
\includegraphics[width=\columnwidth]{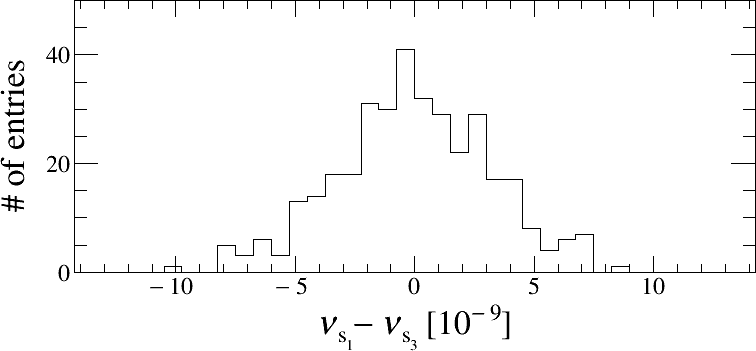}
\caption{\label{fig:dnu13} Distribution of the difference of the baseline spin tunes between time intervals $\Delta T_1$ and $\Delta T_3$. The RMS value of this distribution is used as an estimate of the systematic error of the spin tune jumps, yielding $\delta \Delta\nu_s^{\text{syst}} = \num{3.23e-9}$.}
\end{figure}

Since the $q_{s_i}$ in Eq.\,(\ref{eq:statistical-error-of-spin-tune-jump}) carry no uncertainty, using the $\sigma_{b_i}$ given in Table\,\ref{tab:all-abc}, the statistical error of $\Delta\nu_s$ is given by
\begin{equation}
\begin{split}
 \delta\Delta\nu_s^{\text{stat}} = & \sqrt{\left(\frac{1}{2} \right)^2 \sigma_{b_1}^2 + \sigma_{b_2}^2 + \left(\frac{1}{2} \right)^2 \sigma_{b_3}^2}\\
 = & \num{7.0e-10}\,.
\end{split}
\end{equation}
Thus, the spin tune jumps for runs consisting of 6 cycles can be determined to a \textit{statistical} precision of $\delta\Delta \nu_s^{\text{stat}} = \num{7.0e-10}$. 

In order to estimate the \textit{systematic} error of the spin tune jumps, the distribution of the difference between the baseline spin tunes $\nu_{s_1} - \nu_{s_3}$ in the time intervals $\Delta T_1$ and $\Delta T_3$ is used (see Fig.\,\ref{fig:dnu13}). The width of this distribution is used as an estimate of the systematic error of the spin tune jumps, yielding $\delta \Delta\nu_s^{\text{syst}} = \num{3.23e-9}$.

\section{Static Wien Filter option for local artificial imperfection}
\label{sec:appendixG}
Here we briefly mention the static Wien filter as an option for a local AI which generates simultaneously horizontal and longitudinal magnetic fields. This can be achieved by using a double helix solenoid\,\cite{Ball:2007zzb} which can readily be fit into the ring. The Lorentz force from the horizontal magnetic field must be compensated for by a corresponding vertical electric field. 

The results of the present study call for a longitudinal magnetic field integrals up to $\int B_z dz = 10 - \SI{15}{T.mm}$. For a $\SI{0.5}{m}$ long solenoid with $B_x= \SI{0.03}{T}$, this calls for a corresponding electric field $E_y = \beta B_x = \beta \cdot \SI{10}{MV \per \meter}$. Such strong electric fields will be a challenge but are still in the admissible ballpark. 

The demand on the electric field can be relaxed, though, if such a static Wien filter with a double helix solenoid could be just a supplement for fine tuning the two main solenoids. Hopefully, the intrinsic imperfections of the COSY ring can be further reduced after the precision geodetic survey of the ring magnetic elements has been completed, and the magnetic elements have been aligned more precisely. 

\section{Artificial imperfections as a tool to align the stable spin axis}
\label{sec:appendixE}
We demonstrate the possibility to align the stable spin axis by an artificial imperfection, using a model defined by Eqs.\,(\ref{eq:spin-tune-without-solenoid}) and (\ref{eq:2C.1}). Let the axis of the AI point along $\vec k$ in the $xz$ plane. We decompose the stable spin axis of the ring without artificial imperfections into components along the vertical direction $c_y \vec{e}_y$ and the in-plane component $\vec c_\|= c_x\vec e_x + c_z \vec e_z $. With AI switched on, we would like to align the stable spin axis of the ring along the vertical direction $\vec e_y$. 

Let $\vec{c}_\|$ be determined by spin tune mapping. We demand that the total spin rotation matrix $\matr{T}$ of the ring with the AI switched ON [see Eq.\,(\ref{eq:T})] yields a vanishing in-plane component of the stable spin axis, thus 
\begin{equation}
\begin{split}
\vec{c}_{\matr{T}} = &\sin\left( \frac{1}{2} \chi_{\text{AI}}\right) \\
        & \times \left[ \cos \left( \pi \nu_s^0\right)  \vec k  +  c_y \sin \left( \pi \nu_s^0 \right) \left( \vec e_y \times \vec k \right) \right] \\ 
        & + \sin \left( \pi \nu_s^0\right) \cos \left( \frac{1}{2} \chi_{\text{AI}}\right) \vec{c}_\| =0\,. 
\end{split}
\label{eq:App-E.1}
\end{equation}
One can readily solve this equation for $\chi_{\text{AI}}$ and the orientation of the AI axis $\vec k$. Upon some algebra we find
\begin{equation}
\begin{split}
\tan\left( \frac{1}{2} \chi_{\text{AI}}\right) \vec{k} & = - \frac{\sin\left( \pi \nu_s^0\right)}{D} \matr{A} \vec{c}_\| \, , \text{ where} \\
D & = \sqrt{ \cos^2 \left(\pi\nu_s\right) + {c_y}^2\sin^2\left(\pi \nu_s\right)}\,.
\end{split}
\end{equation}
$\matr{A}$ denotes the rotation matrix of unit determinant,
\begin{equation}
\matr{A} = \frac{1}{D} \left(
\begin{matrix}
\cos\left(\pi\nu_s\right) & -c_y\sin\left(\pi\nu_s\right) \\
c_y \sin\left(\pi\nu_s\right) & \cos\left(\pi\nu_s\right) 
\end{matrix}
\right) \,.
\end{equation}

The AI axis $\vec k$ must point counter to the imperfection vector $\vec c_\|$ rotated by an angle $\theta_p$ such that
\begin{equation}
\tan \theta_p = c_y \tan\left(\pi\nu_s\right)\, .
\label{eq:App-E.4}
\end{equation}
In the approximation of $\vec c = \vec e_y$, this corresponds to a rotation by an angle $\theta_p=\pi\nu_s$. The AI must be run at a spin kick angle
\begin{equation}
\tan \left( \frac{1}{2} \chi_{\text{AI}}\right)  = - \frac{\sin \left( \pi \nu_s \right) }{D} \left| \vec{c}_\| \right|\,.
\label{eq:App-E.5}
\end{equation} 

\bibliographystyle{apsrev4-1}
\bibliography{spintunemapping_23.01.2017}
\end{document}